\documentclass[fleqn,usenatbib]{mnras}

\usepackage{newtxtext, newtxmath}

\usepackage[T1]{fontenc}

\DeclareRobustCommand{\VAN}[3]{#2}
\let\VANthebibliography\thebibliography
\def\thebibliography{\DeclareRobustCommand{\VAN}[3]{##3}\VANthebibliography}

\usepackage{graphicx}	
\usepackage{amsmath}	
\graphicspath{{figs/}}

\title[Non-linear SR Transients under Wide Doppler Broadening]{Transient Structure in the Non-linear Superradiance Regime of Widely Doppler Broadened Media}

\author[C. Wyenberg et al.]{C. Wyenberg,$^{1}$\thanks{E-mail: cwyenber@uwo.ca} F. Rajabi,$^{2}$ M. Chamma,$^{1}$ A. Kumar,$^{1}$ and M. Houde$^{1}$\thanks{E-mail: mhoude2@uwo.ca}\\
$^{1}$Department of Physics and Astronomy, The University of Western Ontario, 1151 Richmond Street, London, Ontario N6A 3K7, Canada\\
$^{2}$Perimeter Institute for Theoretical Physics, Waterloo, ON N2L 2Y5, Canada}

\date{}

\pubyear{2022}

\begin{document}
\label{firstpage}
\pagerange{\pageref{firstpage}--\pageref{lastpage}}
\maketitle

\begin{abstract}

We investigate transient radiation processes in the non-linear superradiance (SR) regime of the Doppler broadened Maxwell-Bloch equations when the velocity distribution is of total bandwidth greatly exceeding that of the transient process itself. We demonstrate the formation of global polarisation phase correlation and the quenching of temporal structure if a smooth distribution is inverted above the critical threshold required to enter the non-linear SR regime. We propose candidate stochastic velocity distributions capable of sustaining finite temporal structure in the non-linear emission process. We develop a novel algorithm for simulating the Doppler broadened Maxwell-Bloch equations which is $\mathcal{O}\left(n\right)$ complex in the number of velocity channels $n$ whenever the emerging polarisation correlation is of moderate bandwidth, and we apply it to a stochastic velocity distribution in order to demonstrate sustained delay and duration of peak intensity in the widely Doppler broadened limit. We discuss the transverse inversion process and recognise an autoregulation mechanism on the number of molecules cooperatively participating in SR emission. This mechanism has the effect of limiting the temporal duration of the intensity pulse to a lower bound proportional to the length of the sample, which we confirm through simulation.

\end{abstract}


\section{Introduction}\label{sec:intro}

In the quantum phenomenon of superradiance (SR) a population of two-level molecules evolves, through its interaction with the quantized radiation field, into a composite excitation state possessing a high degree of entanglement. This highly entangled state couples strongly to the radiation field and produces cooperative emission of much greater intensity than would otherwise be generated by the independent spontaneous emission of the constituent molecules. SR was first described theoretically by \citet{Dicke1954} and was verified experimentally in an HF gas by Skribanowitz et al. in 1973 \citep{Skribanowitz1973}. It has more recently been applied to astrophysical processes \citep{Rajabi2016a, Houde2018a,Houde2019, Rajabi2020}, including the description of flux transient features observed in maser-harbouring regions \citep{Rajabi2016b, Rajabi2017, Rajabi2019, Rajabi2020b} where velocity coherence conditions within the environment foster evolution into the entangled state prerequisite to SR emission.

Indeed, velocity coherence is of paramount importance to the development of SR. A collection of stationary two-level molecules is able to evolve into a state of high entanglement because of the molecules' interactions with a common radiation field; molecules in motion, however, interact only with those field excitation states near their Doppler shifted spontaneous emission modes. Molecules of excessive velocity differences thus do not see a common radiation field and cannot be a priori assumed to evolve into a highly entangled state. In the large velocity separation limit molecules radiate as completely independent, statistically uncorrelated emitters.\footnote{Such independence can be easily demonstrated analytically in the Schr\"{o}dinger picture of two-level molecules \citep{Benedict1996}, but we choose to demonstrate it in Section \ref{subsec:swept-two} within the framework of SR suited to the physical configurations of interest to this paper.}

Velocity coherence plays a central role in astrophysical applications of SR, but it has thus far been introduced in the literature at only a high level based upon back-of-the-envelope physical arguments. Previous studies have reduced the populations of molecules proposed to interact coherently to those sub-populations of the full velocity distribution which share a neighbourhood of bandwidth established by the SR transient pulse duration. Despite the fact that such a velocity width represents an exceedingly small fraction of the total population, the large astronomical scales (which are many orders of magnitude larger than those realised in typical laboratory experiments) have been considered sufficient to produce such a large total number of molecules that the sub-population of a narrow velocity slice can feasibly exceed the critical inverted column density threshold required to initiate SR.

In order to distinguish the present work from existing results in the literature, let us quantify the extent of Doppler broadening in astrophysical SR processes. The coherent SR emission transients described in \citet{Rajabi2019}, for example, were generated over a timescale $T \sim 5 \times {10}^8 \textrm{ s}$ by molecules possessing a natural emission frequency $f_0 = 6.7 \textrm{ GHz}$. The spectrum of the SR process therefore covered a bandwidth $\Delta f$ on the order of $1/T$, such that the electric field generated would interact with molecules over a velocity range $\Delta v \sim c \Delta f / f_0 \sim {10}^{-8} \textrm{ m/s}$. Any realistic velocity distribution is certainly many orders of magnitude wider than such a velocity coupling width, and numerical simulations of such systems must therefore invoke extremely fine velocity channel discretisation in order to capture the important physics of the problem. Were the channel discretisation width in a simulation of the aforementioned sample to exceed $10^{-6}\textrm{ m/s}$, for example, all channels would be a priori decoupled and the response would be the trivial uncorrelated combination of independent SR processes occurring within each velocity channel.

If a sample's Doppler broadening is multiple orders of magnitude wider than the velocity width defined by the SR timescale, we refer to it as widely Doppler broadened (WDB); conversely, if a sample's Doppler broadening is on the order of the velocity width defined by the SR timescale, we refer to it as narrowly Doppler broadened. We refer to a collection of molecules entirely on resonance (without velocity extent) as a resonant sample. Numerical simulations of SR under Doppler broadening exist in the literature, but have thus far been restricted to narrowly broadened processes. Such a restriction is of course sufficient for laboratory realisations of SR; \citet{MacGillivray1976}, for example, modelled SR processes in an HF gas of velocity width $10^4 \textrm{ cm/s}$, occurring over a duration of about $10^{-7} \textrm{ s}$. Such a duration corresponds to a velocity spread of about $3 \times 10^3 \textrm{ cm/s}$ for HF emission. The narrowly broadened system demonstrated SR with minor quantitative deviations from the peak intensity and timescales of the resonant case.

The objective of the present work is to model WDB distributions of two-level molecules with such high column densities that even narrow slices of the velocity distribution can possess population inversion levels sufficient to initiate non-linear SR processes. Such degrees of saturation differ dramatically from those of typical laboratory experiments \citep{Skribanowitz1973} or of existing analytical solutions \citep{Benedict1996, Gross1982, MacGillivray1976} (as discussed in more detail in Section \ref{subsec:theory-existing}) and are found, through numerical simulation, to yield rich features beyond minor corrections to the resonant theory. Whereas one might naturally expect the emergence of multiple locally coherent groups of velocity channels in some globally decoherent combination, we show that certain inversion processes lead to the development of wide velocity bandwidth polarisation correlation and the consequent quenching of temporal features. We demonstrate that a WDB sample responding to a transverse inversion can, however, sustain SR temporal structure if its length exceeds a critical threshold; while one responding to a swept inversion may sustain SR temporal structure only if its distribution obeys certain statistical properties.

\subsection{Structure of the paper}\label{subsec:struct}

The paper is structured as follows. In Section \ref{sec:theory} we summarise the theoretical models used to simulate WDB SR processes and place the regime of the present work in the context of existing theoretical and experimental results for Doppler broadened SR at laboratory scales. We discuss numerical complexity scaling with distribution size and introduce efficient solution methods. One of these methods is unique to the present work; however, we summarise only its main features in Section \ref{sec:theory} while leaving its mathematical derivation and details concerning its numerical implementation to Appendix \ref{app:sf_algorithm}.

We commence our numerical study of WDB SR processes by first simulating transversely inverted samples in Section \ref{sec:transverse}, whose responses are found to most closely resemble that of a resonant sample (though with some important modifications). We argue that a transverse pumping mechanism may inhibit the formation of wide bandwidth polarisation phase correlation via the so-called Arrechi-Courtens condition, whereby we recognise an autoregulation mechanism on the number of molecules cooperatively participating in the SR emission process during the transition to the WDB limit. We confirm finite temporal duration in the transversely pumped WDB SR regime via simulation, and we demonstrate it to be proportional to the sample's length.

Conversely, a WDB SR process initiated by a swept inversion mechanism is found to deviate substantially from the resonant case; in light of its nuanced response, we therefore analyse a swept SR process in two steps. First, we simulate in Section \ref{sec:swept-disc} the swept inversion of simple (though physically unrealistic) discrete distributions; second, we simulate in Section \ref{sec:swept-cont} the more complicated (and physically realistic) case of a continuum distribution. This order of presentation allows us to build important intuition from simple models before presenting the unexpected response of a real system. We demonstrate in Section \ref{subsec:cont-polphases} the formation of wide bandwidth polarisation phase correlation in the continuum limit which we quantify as a function of the initial population inversion of the sample. We introduce in Section \ref{subsec:cont-candidates} an intuitive picture for the formation of correlation which motivates our study in Section \ref{subsec:cont-noisysims} of the relationship between the stochastic characteristics of a velocity distribution and the emergence of SR temporal structure in the WDB limit.

Finally, we conclude in Section \ref{sec:summary} by placing these results in the context of realistic astrophysical processes. We emphasize that the order of magnitude of an SR process timescale dictates its ability to demonstrate finite temporal features in the non-linear SR regime. We recommend candidate astrophysical phenomena for future research including, for example, events within turbulent environments or fast radio bursts.


\section{Theory and Numerical Methods}\label{sec:theory}

\subsection{The Maxwell-Bloch equations}\label{subsec:theory-mbeqns}

Superradiance may be described by a semi-classical set of partial differential equations called the (Doppler broadened) Maxwell-Bloch (MB) equations. The MB equations may be derived from the Schr\"{o}dinger picture density operator analysis of a collection of two-level molecules or from the Heisenberg picture of polarisation and inversion operators evolving in the presence of a semi-classical electromagnetic field. The MB equations describe the evolution of the expectation values of the coarse-grained molecular population inversion operator $\hat{N}$, the coarse-grained raising (lowering) operators $\hat{P}^+$ ($\hat{P}^-$), and the photon annihilation (creation) operators $\hat{E}^+$ ($\hat{E}^-$). In the classical limit such operators correspond to, respectively, the population inversion density, the forward (reverse) rotating parts\footnote{If $f\left(z,t\right)=\int_{-\infty}^{+\infty}\tilde{f}\left(z,\omega\right)e^{i \omega t} \mathrm{d}\omega$, then $f^{\pm}\left(z,t\right)$ are defined as $f^{-}\left(z,t\right)=\int_{-\infty}^{0}\tilde{f}\left(z,\omega\right)e^{i \omega t} \mathrm{d}\omega$ and $f^{+}\left(z,t\right)=\int_{0}^{+\infty}\tilde{f}\left(z,\omega\right)e^{i \omega t} \mathrm{d}\omega$.} of the polarisation, and the forward (reverse) rotating parts of the electric field. For a detailed derivation, see \citet{Gross1982}.

For simplicity and without loss of generality, we refer in all that follows to a medium interacting with the electromagnetic field through an electric dipole transition. We assume that the molecular dipole moments, the molecular polarisations, and the electric field polarisation are all aligned in a shared direction perpendicular to the $z$ axis of a one-dimensional sample. The MB equations across a velocity distribution then read \citep{Gross1982}
\begin{align}
    \left[\frac{\partial}{\partial t} + v \frac{\partial}{\partial z}\right]N_{v} &= \frac{i}{\hbar} \left(E^{+}+E^{-}\right) \left(P_{v}^{+} - P_{v}^{-}\right) \label{eq:MBE_Inv} \\
    \left[\frac{\partial}{\partial t} + v \frac{\partial}{\partial z} \right] P_{v}^{+} &= i \omega_{0} P_{v}^{+} + 2i \frac{d^{2}}{\hbar} E^- N_v \label{eq:MBE_Pol} \\
    \left[\frac{\partial^{2}}{\partial t^{2}} - c^{2} \frac{\partial^{2}}{\partial z^{2}}\right] E^{+} &= -\frac{1}{\epsilon_{0}} \int \mathrm{d} v' F_{v'} \frac{\partial^2 P_{v'}^{-}}{\partial t^2}, \label{eq:MBE_Field}
\end{align}
where we have partitioned molecules according to their velocity $v$ within a distribution $F_v$ defined such that the fraction of molecules of velocity between $v$ and $v+dv$ is $F_v dv$ (with $\int F_v dv = 1$), and where $d$ is the molecular dipole moment and $\omega_0$ the angular frequency of spontaneous emission from a single molecule. Operator ``hats'' have been removed from all quantities above, since the MB equations describe the evolution of expectation values of all operators. Note that the formal definition of $\hat{N}$ (omitted above for brevity) implies that the expectation value $N_v$ corresponds, physically, to half the population inversion density.

We make a change of variables to the retarded time $\tau=t-z/c$ and factor the polarisations and electric field functions with envelopes (Doppler shifted for the polarisations) as
\begin{align}
    P_{v}^{\pm}\left(z,\tau\right) & =\bar{\mathcal{P}}'^\pm_v \left(z,\tau\right)e^{\pm i\omega_{0}\left(1+v/c\right)\tau} \label{eq:p_doppler_defn}\\
    E^{\pm}\left(z,\tau\right) & =\mathcal{E}'^\pm \left(z,\tau\right)e^{\mp i\omega_{0}\tau}\label{eq:Efactorisation}
\end{align}
(the primes to be removed when we normalise momentarily). The conventional theory is to eliminate the fast-rotating terms $E^+ P^-_v$ and $E^- P^+_v$ (the so-called ``rotating wave approximation'') in equation (\ref{eq:MBE_Inv}) and recognise that $\partial/\partial z\ll\omega_{0}/c$ and $\partial/\partial \tau \ll \omega_0$ when acting on the envelope functions, to arrive at the so-called slowly-varying envelope approximation (SVEA) of the MB equations \citep{Gross1982},
\begin{align}
    \frac{\partial N_v}{\partial\tau} &= \frac{i}{\hbar} \left(\bar{\mathcal{P}}'^+_v \mathcal{E}'^+ e^{i\omega_0\frac{v}{c}\tau} - \bar{\mathcal{P}}'^-_v \mathcal{E}'^- e^{-i\omega_0\frac{v}{c}\tau}\right) \nonumber \\ &\qquad -\frac{N_v - N_0}{T_{1}} + \Lambda^{(N)} \label{eq:MB_TD-1} \\
    \frac{\partial\bar{\mathcal{P}}'^+_v}{\partial\tau} &= i \frac{2 d^2}{\hbar} \mathcal{E}'^- N_v e^{-i\omega_0\frac{v}{c}\tau} - \frac{\bar{\mathcal{P}}'^+_v}{T_2} + \Lambda^{(P')} \label{eq:MB_TD-2} \\
    \frac{\partial\mathcal{E}'^+}{\partial z} &= i \frac{\omega_0}{2\epsilon_0 c} \int \mathrm{d} v' F_{v'} \mathcal{\bar{P}}'^-_{v'} e^{-i\omega_0\frac{v'}{c}\tau}, \label{eq:MB_TD-3}
\end{align}
for a molecular number density $2 N_0$, where we have introduced population inversion and polarisation pumping sources $\Lambda^{(N)}\left(\tau\right)$ and $\Lambda^{(P')}\left(\tau\right)$, respectively, as well as relaxation and dephasing time scales $T_1$ and $T_2$, respectively. Additionally, we made the approximation $\left(1+v'/c\right) \approx 1$ in our derivation of \eqref{eq:MB_TD-3}.

We may simplify equations \eqref{eq:MB_TD-1}--\eqref{eq:MB_TD-3} by recognising an intrinsic SR timescale
\begin{equation}
    T_\textrm{R} = \frac{2 \epsilon_0 \hbar c}{d^2 2 N_0 \omega_0 L} = \tau_\textrm{sp} \frac{8\pi}{3 \lambda^2 2 N_0 L} \label{eq:TR_def}
\end{equation}
for an initial population inversion density $2 N_0$, where $\tau_\textrm{sp}$ is the spontaneous emission timescale of a free molecule and $L$ the length of the sample. It can be shown \citep{Benedict1996, Gross1982} that, at least in the resonant case, both the delay $\tau_\textrm{D}$ and duration $\tau_\textrm{P}$ of peak SR emission intensity are proportional to $T_\textrm{R}.$ Furthermore, if we work in the dimensionless quantities
\begin{equation}
    Z_\omega = \frac{1}{2 N_0} N_\omega \textrm{, }
    \bar{\mathcal{P}}^\pm_\omega = \frac{1}{d 2 N_0} \bar{\mathcal{P}}'^\pm_\omega \textrm{, and }
    \mathcal{E}^\pm = \frac{d}{\hbar} T_\textrm{R} \mathcal{E}'^\pm \textrm{;} \label{eq:dimless_defns}
\end{equation}
whose velocity channels are now indexed by angular frequencies $\omega = \omega_0 v / c$, then the MB equations read
\begin{align}
    \frac{\partial Z_\omega}{\partial\tau} &= \frac{i}{T_\textrm{R}} \left(\bar{\mathcal{P}}^+_\omega \mathcal{E}^+ e^{i\omega\tau} - \bar{\mathcal{P}}^-_\omega \mathcal{E}^- e^{-i\omega\tau}\right) - \frac{Z_\omega - 1}{T_1} + \Lambda^{(Z)} \label{eq:MB_TD_norm-1} \\
    \frac{\partial\bar{\mathcal{P}}^+_\omega}{\partial\tau} &= \frac{2i}{T_\textrm{R}} \mathcal{E}^- Z_\omega e^{-i\omega\tau} - \frac{\bar{\mathcal{P}}^+_\omega}{T_2} + \Lambda^{(P)} \label{eq:MB_TD_norm-2} \\
    \frac{\partial\mathcal{E}^+}{\partial z} &= \frac{i}{2 L} \int \mathrm{d} \omega' F_{\omega'} \mathcal{\bar{P}}^-_{\omega'} e^{-i\omega'\tau}. \label{eq:MB_TD_norm-3}
\end{align}

For a long cylindrical sample of cross-sectional area $A$ and length $L$ having fresnel number $A/(L\lambda)$ on the order of unity, the resonant case of equations \eqref{eq:MB_TD_norm-1}--\eqref{eq:MB_TD_norm-3} can be solved analytically \citep{Benedict1996}. The resulting theoretical transient intensity response to an initially entirely inverted total number of molecules $n_\textrm{tot}$ is characterised by peak intensity bursts of decreasing amplitude. In what follows we will scale our intensity plots to the theoretical peak intensity $I_\textrm{p}$ of the first burst in said analytical response given by \citep{MacGillivray1976, Feld1980}
\begin{equation}
    I_\textrm{p} = \frac{4 n_\textrm{tot} \hbar \omega_0 / (A T_\textrm{R})}{\left|\ln \left[\theta_0/(2\pi)\right]\right|^2}, \label{eq:Ip}
\end{equation}
where $\theta_0 = 2/\sqrt{n_\textrm{tot}}$ as discussed further in Section \ref{subsec:theory-initialqm}.

\subsection{The present regime in the context of existing results}\label{subsec:theory-existing}

This paper investigates Doppler broadening of the MB equations, but in fact any inhomogeneous broadening mechanism\footnote{That is, any mechanism by which a distribution of spontaneous emission frequencies exists within the sample.} is described by equations \eqref{eq:MB_TD_norm-1}--\eqref{eq:MB_TD_norm-3}, for which analytical solutions as well as numerical results exist in the literature for a variety of regimes. It is important to place the present work in the context of these known results.

One may distinguish between the linear and non-linear regimes of SR. Below a certain population inversion density threshold a sample evolves with only negligible loss of population inversion, such that the $Z_\omega$ may be considered constants with respect to time. For such a process the MB equations are linearised and the sample is said to be in the linear regime. Conversely, if the initial population inversion density is sufficient to cause a loss of population inversion (potentially after some delay) the system is said to be in the non-linear SR regime. 

We are interested in conditions within which a WDB sample may generate high intensity emission over a finite timescale via non-linear SR processes. As investigated in \citet{Rajabi2019} and \citet{Feld1980}, the quasi-steady state regime of the MB equations describes maser processes, while the fast flux rises and transient structures modelled in \citet{Rajabi2017, Rajabi2019} result from the non-linear SR regime of the MB equations. Although there exist analytical solutions to equations \eqref{eq:MB_TD_norm-1}--\eqref{eq:MB_TD_norm-3} for an inhomogeneously broadened system in the linear regime \citep{Benedict1996}, the non-linear case must be treated numerically.

Let us make our introductory comments of Section \ref{sec:intro} regarding the column densities per velocity interval within our WDB astronomical samples more precise in the context of existing results in the literature. We imagine commencing the construction of a WDB sample with a rather narrow velocity extent $\delta v$ and an initial population inversion defining a characteristic SR timescale $T_\textrm{R, i}$ much less than the characteristic inhomogeneous broadening time $T^*_\textrm{2, i} = 2 \pi c / \left(\omega_0 \delta v\right) $. This initial configuration is able to generate a transient SR intensity pulse of some delay $\tau_\textrm{d, i}$ and duration $\tau_\textrm{p, i}$, but as of yet represents only a negligible fraction of the full WDB velocity extent.

A conventional study of inhomogeneous broadening \citep{Benedict1996} and its effect upon laboratory systems would proceed, from this starting configuration, to redistribute the entire population of molecules over a wider $\Delta v > \delta v$ corresponding to a smaller ${T^*_2}'$. As ${T^*_2}'$ is reduced below $T_\textrm{R, i}$, both a reduction in peak intensity and an elongation of timescales to $\tau_\textrm{d}'>\tau_\textrm{d, i}$ and $\tau_\textrm{p}' > \tau_\textrm{p, i}$ are then observed. See, for example, \citet{Benedict1996, Gross1982}.

In our present investigation, however, we build upon the initial distribution of extent $\delta v$ by adding molecules further and further from resonance. This addition of molecules decreases the formal quantity $T_\textrm{R}$ with the increase in $N_0$ (as per equation \eqref{eq:TR_def}); however, we are interested in how the observed timescales $\tau_\textrm{d}$ and $\tau_\textrm{p}$ respond to the simultaneous reduction of $T^*_2$ (caused by the widening of the distribution) and of $T_\textrm{R}$ (caused by the increase in total number of initially inverted molecules). While continuously adding molecules to the outer ends of the velocity distribution, the development of correlation between velocity channels deep inside the distribution will tell us the degree to which a sample's SR temporal features will be reduced in the WDB limit, and will help us make realistic quantitative conclusions regarding SR emission intensities from full distributions.

\subsection{Numerical complexity}\label{subsec:theory-complexity}

Equations \eqref{eq:MB_TD_norm-1}--\eqref{eq:MB_TD_norm-3} in the time domain are $\mathcal{O}\left(n^2\right)$ complex in the number of velocity channels $n$ simulated. As the velocity channel count increases, the temporal step size must concurrently decrease in order to avoid aliasing of the oscillating exponentials $e^{\pm i \omega_0 \frac{v}{c} \tau}$. For a detailed discussion see \citet{Wyenberg2021}. We herein refer to the simulation of the MB equations via time stepping\footnote{Via a fourth order Runge-Kutta scheme.} of the unmodified equations \eqref{eq:MB_TD_norm-1}--\eqref{eq:MB_TD_norm-3} as the conventional time domain (CTD) method.

There exist, however, (at least) two algorithms for solving equations \eqref{eq:MB_TD_norm-1}--\eqref{eq:MB_TD_norm-3} which are $\mathcal{O}\left(n\right)$ complex under certain approximations. The first, developed in \citet{Wyenberg2021}, is known as the integral Fourier (IF) algorithm and is summarised in Section \ref{subsec:theory-ifalg}. The second is unique to the present work and offers improved performance over the IF algorithm for the SR regimes investigated by this paper; it is referred to as the supplementary fields (SF) algorithm and is introduced in Section \ref{subsec:theory-sfalg} but derived in Appendix \ref{app:sf_algorithm}. The two algorithms share a fundamental physical approximation which enables their advantageous complexity scaling, but each has its advantages in differing applications. We will only briefly describe the IF algorithm for the purpose of identifying its advantageous regimes; however, only the CTD and SF algorithms will be used throughout this paper.

\subsection{The integral Fourier algorithm}\label{subsec:theory-ifalg}

The IF algorithm established in \citet{Wyenberg2021} is based upon a Fourier series representation of the (time) integral equations equivalent to equations \eqref{eq:MB_TD_norm-1}--\eqref{eq:MB_TD_norm-3}, and is able to accurately model transient processes described by the MB equations (a novel feature over other Fourier domain methods \citep{Menegozzi1978} which are suited to quasi-steady state modelling). We briefly summarise the IF algorithm's main features now, leaving its full expression to Appendix \ref{app:if_representation}. Its detailed derivation may be found in \citet{Wyenberg2021}. The purpose of this brief introduction is only to place the IF algorithm in the context of the SF algorithm to be developed in Section \ref{subsec:theory-sfalg}, in order to guide future research in selecting the most efficient algorithm for a particular regime of WDB SR under investigation.

In an IF simulation of temporal duration $T$, the population inversion, the polarisation envelopes, the pumping sources, and the field envelopes are expanded in temporal Fourier series of mode separation $d\omega=2\pi/T$. The velocity distribution is partitioned with a granularity $dv$ of the equivalent Doppler shift $d\omega$ between adjacent channels (namely, $dv=c d\omega / \omega_{0}$), such that each velocity slice of the material quantities ($\mathcal N$ and  $\bar{\mathcal{P}}^\pm$) possesses its own Fourier series expansion. The population inversion of the sample is described by a spatially dependent two-dimensional array $\mathbb{N}_{p,m}\left(z\right)$, where $p$ indexes the velocity channel and $m$ its Fourier expansion coefficient. Similarly, the polarisation is described by a spatially dependent two-dimensional array $\bar{\mathbb{P}}_{p,m}^{\pm}\left(z\right)$. The electric field, on the other hand, is not a velocity-partitioned material quantity and is defined by a one-dimensional array of Fourier expansion coefficients $\mathbb{E}_{m}^{\pm} \left(z\right)$. The precise Fourier expansions of $\mathcal N$,  $\bar{\mathcal{P}}^\pm$, and $\mathcal E^\pm$ in terms of the coefficient arrays $\mathbb{N}_{p,m}$, $\bar{\mathbb{P}}_{p,m}^{\pm}$, and $\mathbb{E}_{m}^{\pm}$ may be found in Appendix \ref{app:if_representation}.

The IF simulation of the MB equations propagates solutions for the Fourier coefficients down the length of the sample in two looping steps: (1) the $\mathbb{E}_{m}^{\pm}$ array at a given $z$ establishes, for each velocity channel, a system of linear equations to be solved in $\mathbb{N}_{p,m}$ and $\bar{\mathbb{P}}_{p,m}^{\pm}$; and (2) said solution is used to propagate forward $\mathbb{E}_{m}^{\pm}$ to $z+dz$. In practice, Step 2 is executed with a fourth-order Runge-Kutta scheme.

The computational scaling advantage of the IF method is achieved through truncation of interactions between material quantities over a finite neighbourhood of electric field modes $\mathbb{E}_{m}^{\pm}$ centred upon each velocity channel's natural frequency; such neighbourhoods are set to a fixed bandwidth, regardless of the total number of velocity channels added to the simulation. This interaction bandwidth truncation is known as the local mode interaction ($\textrm{LMI}^\textrm{IF}$) approximation, and a very similar (though slightly distinct) approximation will be made in the SF algorithm (introduced in Section \ref{subsec:theory-sfalg}) and referred to as the $\textrm{LMI}^\textrm{SF}$ approximation. For physical motivation and numerical validation of the $\textrm{LMI}^\textrm{IF}$ approximation, as well as detailed discussion of the IF algorithm's implementation, see \citet{Wyenberg2021}.

In Sections \ref{subsec:swept-comb} and \ref{subsec:cont-polphases} we will observe extremely wide polarisation correlation across a Doppler broadened SR system, over a bandwidth many times greater than that of the transient SR pulse which a resonant sample in the same non-linear SR regime would produce. The IF algorithm under its $\textrm{LMI}^\textrm{IF}$ approximation is in fact able to recover wide correlation, even if that correlation bandwidth exceeds the $\textrm{LMI}^\textrm{IF}$ approximation truncation width used. The physical correlation of far distant frequency modes is a higher-order effect of the MB equations, and the IF algorithm is immune to the removal of formally manifest mathematical coupling between said modes. Conversely, the algorithm introduced next in Section \ref{subsec:theory-sfalg} must be executed with full fidelity to model such systems, which renders the IF algorithm potentially advantageous in such circumstances. We therefore compare in more detail the performance of all algorithms as function of the distribution size and emergent correlation bandwidth in Appendix \ref{app:alg_comparisons}.

\subsection{The method of supplementary fields}\label{subsec:theory-sfalg}

We derive in Appendix \ref{app:sf_algorithm} a novel algorithm built upon the physical principles of the IF algorithm (namely, the concept of local spectral interaction between the field and the medium) but operating entirely within the time domain, which we call the method of supplementary fields (SF). Its name is rooted in our introduction of an array of electric field envelopes
\begin{equation}
    \bar{\mathcal{E}}^\pm_\omega \equiv \mathcal{E}^\pm e^{\pm i \omega \tau}, \label{eq:sf_envelopes}
\end{equation}
such that equations \eqref{eq:MB_TD_norm-1}--\eqref{eq:MB_TD_norm-3} may be expressed as
\begin{align}
    \frac{\partial Z_\omega}{\partial\tau} &= \frac{i}{T_\textrm{R}} \left(\bar{\mathcal{P}}^+_\omega \bar{\mathcal{E}}^+_\omega - \bar{\mathcal{P}}^-_\omega\bar{\mathcal{E}}^-_\omega\right) - \frac{Z_\omega - 1}{T_1} + \Lambda^{(N)} \label{eq:mb_sf-1} \\
    \frac{\partial\bar{\mathcal{P}}^+_\omega}{\partial\tau} &= \frac{2i}{T_\textrm{R}} \bar{\mathcal{E}}^-_\omega Z_\omega - \frac{\bar{\mathcal{P}}^+_\omega}{T_2} + \Lambda^{(P)} \label{eq:mb_sf-2} \\
    \frac{\partial\bar{\mathcal{E}}^+_\omega}{\partial z} &= \frac{i}{2 L} \int \mathrm{d} \omega' F_{\omega - \omega'} \mathcal{\bar{P}}_{\omega - \omega'}^- e^{i\omega'\tau} s_\delta\left(\omega'\right). \label{eq:mb_sf-3}
\end{align}
The function $s_\delta \left(\omega'\right)$ is referred to as the local mode interaction kernel. If $s_\delta = 1$ everywhere, then equations \eqref{eq:mb_sf-1}--\eqref{eq:mb_sf-3} are mathematically equivalent to equations \eqref{eq:MB_TD_norm-1}--\eqref{eq:MB_TD_norm-3}; however, we permit $s_\delta$ to be a function non-vanishing on only a finite domain $|\omega'| \lessapprox \delta/2$. We refer to this act, which effectively limits the convolution bandwidth in equation \eqref{eq:mb_sf-3}, as the local mode interaction approximation in the SF representation (abbreviated as the $\textrm{LMI}^\textrm{SF}$ approximation). The $\textrm{LMI}^\textrm{SF}$ approximation is similar in character to the $\textrm{LMI}^\textrm{IF}$ approximation of the IF algorithm \citep{Wyenberg2021} as described previously in Section \ref{subsec:theory-ifalg}, but with some important distinctions detailed in Appendix \ref{app:alg_comparisons}.

Equations \eqref{eq:mb_sf-1}--\eqref{eq:mb_sf-3} together form an $\mathcal{O}\left(n\right)$ complex algorithm in the number $n$ of velocity channels. The problematic aliasing of fast oscillating exponentials is removed by effectively truncating integration ranges over a finite velocity bandwidth via the finite width of $s_\delta$, such that additional velocity channels may be indefinitely added to the system without concurrently reducing the temporal step size in a Runge-Kutta temporal propagation. Although equation \eqref{eq:mb_sf-3} has a convolution structure (which suggests greater than $\mathcal{O}\left(n\right)$ complexity), the convolution is in fact $\mathcal{O}\left(n\right)$ since the kernel $e^{i\omega'\tau} s_\delta \left(\omega'\right)$ does not increase in width with increasing channel count. In fact, we demonstrate in Appendix \ref{app:sf_algorithm} that a rectangular kernel has the special status of circumventing the convolution calculation altogether; however, we detail also in Appendix \ref{app:sf_algorithm} a fast Fourier transform (FFT) method which performs optimally in the regimes of interest to this paper and is therefore used wherever the SF algorithm is applied in what follows.

\subsection{The initial quantum mechanical stage and considerations for widely Doppler broadened superradiance}\label{subsec:theory-initialqm}

The MB equations describe the self-consistent evolution of an invertible, polarisable medium exposed to and interacting with an electromagnetic field; however, they do not describe quantum fluctuations in any of these quantities. Such fluctuations are critical to the initial stages of SR. In the fully quantum mechanical analysis of the combined molecular and radiative system, an initially inverted, non-polarised, zero photon state is not an eigenstate of the Hamiltonian. Such an initial state therefore evolves very quickly, as described by perturbation theory, into a state of non-zero photons and non-zero polarisation.

One can show \citep{Gross1982} that the important transient features of SR--including the time delay to peak emission intensity, the scaling of peak intensity with the square of the initial population inversion, and the duration of peak intensity--are all modelled by the MB equations, so long as they are prescribed initial conditions which capture the initial stages of quantum fluctuations. The initial condition prescription is derived from a statistical analysis of the expectation values of the initial evolution of the population inversion, the polarisation, and the electric field. The derivation finds that, at a certain time $\tau_\textrm{classical}$, commutators associated with the classical limit become negligible and the system can thereafter be evolved from a statistical distribution of states at $\tau_\textrm{classical}$ along trajectories for $\tau>\tau_\textrm{classical}$ defined by the MB equations.

\begin{figure}
    \centering
    \includegraphics[width=.8\columnwidth, trim=0cm 0cm 0cm 0cm, clip]{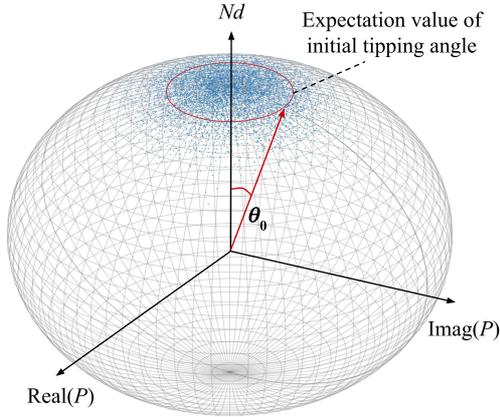}
    \caption{The Bloch sphere with example statistical distribution (scatter points) of initial tipping angles and polarisation phases. Expectation value shown in solid. Note that this example is greatly exaggerated in its standard deviation and in its expectation value of $\theta_0$; the initial tipping angles are highly localised near $\theta=0$ (the top pole) for all simulations of this paper.}
    \label{fig:bloch_sphere}
\end{figure}

The statistical distribution of states at $\tau_\textrm{classical}$ is prescribed as follows. Suppose the system is first prepared at $\tau=0$ with an initial inversion $N_0$ and with no polarisation. The state of the inversion and the polarisation may be plotted in three dimensional space, such that the $x$ and $y$ axes correspond to the real and imaginary parts of the polarisation and the $z$ axis to the dipole moment $d$ times the inversion $N$. The conventional ``Bloch sphere'' is defined as a sphere of radius $N_0 d$ centred at the origin, and the initial state of the system is located at the North pole of the Bloch sphere. A detailed quantum mechanical analysis \citep{Gross1982} demonstrates that the state of the system quickly tips, at a small time $\tau=\tau_\textrm{classical}$, into a statistical distribution of states on the surface of the Bloch sphere. Specifically, the distribution is uniform about the azimuthal axis (the polarisation phase) and Gaussian about a mean polar angle (the angle between $N d$ and $|P^+|$) of $\theta_0 = 2/\sqrt{n_\textrm{tot}}$, where $n_\textrm{tot}$ is the total number of interacting atoms in the sample. A visualisation of an initial tipping angle distribution on the Bloch sphere is shown in Figure \ref{fig:bloch_sphere}.

The initial Bloch tipping angle affects the transient system behaviour, but its value is not well-defined in the WDB limit. The mean value of $\theta_0 = 2/\sqrt{n_\textrm{tot}}$ is usually (in resonant or narrowly Doppler broadened samples) computed from the total number of atoms in the sample; however, the quantum mechanical derivation of this result \citep{Gross1982} assumes that all $n_\textrm{tot}$ interact with a common quantized radiation field. Conversely, in a widely Doppler broadened sample we expect some degree of independence between distant velocity channels. It is beyond the scope of this paper to generalise the quantum mechanical calculation of \citet{Gross1982} to the complicated problem of an ensemble of molecules possessing a continuum velocity distribution; instead, we make the ``best guess'' that $\theta_0 = 2/\sqrt{n_{\textrm{atoms, }\Delta \omega}}$, where $n_{\textrm{atoms, }\Delta \omega}$ is the number of atoms inside the bandwidth $\Delta \omega$ of the transient SR pulse generated by the simulation. This definition may appear circular; however, it is in fact well-defined, as the timescale of the SR transients generated by WDB samples are found to be affected by $\theta_0$ only if it is modified by multiple orders of magnitude.

\subsection{Velocity channel discretisation}\label{subsec:theory-chdisc}

We are investigating the effect of a continuum velocity distribution upon transient SR features. In addition to the initial Bloch tipping angle prescription, we must also choose a discretisation to our velocity distribution which sufficiently captures the physical coupling of channels in the continuum while remaining numerically manageable. If the velocity channel separation is too coarse, the physics of the continuum will be lost and the channels will behave as physically inaccurate, individually resonant Dirac delta-like distributions; on the other extreme, excessively fine discretisation comes at the cost of numerical complexity.

In all simulations we establish our distribution discretisation from the simulation duration $T$; i.e., we separate velocity channels by $dv = 2 \pi c / \left(\omega_0 T\right)$. Achieving a physically accurate discretisation under this prescription requires that transient SR processes evolve over a period less than $T$. We typically meet this constraint by commencing with a resonant sample demonstrating SR transient timescales less than $T$; we then regulate initial population inversion levels as the Doppler broadened system is constructed, ensuring that transient features never exceed a moderate fraction of $T$. In fact, since in the resonant case $T_\textrm{R}$ is inversely proportional to the total initial population inversion density $N_0$, we expect transient SR timescales to only possibly reduce during the addition of molecules to the ends of our distribution (though the exact degree of that reduction is as of yet unknown for the broadened system). We may therefore assume that our starting duration $T$ for a resonant sample yields physically accurate velocity discretisation granularity under our Doppler broadened distribution construction method described near the end of Section \ref{subsec:theory-existing}.


\section{Transverse Inversion Processes}\label{sec:transverse}
The WDB SR processes analysed in this paper are initiated by the sudden inversion of a sample via some pumping process. Mathematically speaking, the most simple\footnote{That is, in our retarded time representation of the MB equations \eqref{eq:MB_TD_norm-1}---\eqref{eq:MB_TD_norm-3}.} of such processes is a swept inversion, which corresponds to the simultaneous onset of population inversion everywhere at retarded time $\tau=\tau_0$. A swept inversion is realised by a pumping action propagating at the speed of light $c$ down the length of the sample such as, for example, the sudden introduction of transition-inducing radiation at $z=0$ in the direction of the sample's longitudinal axis. Alternatively, radiation incident in a direction perpendicular to the longitudinal axis of a sample may initiate a population inversion everywhere at simultaneous non-retarded time $t=t_0$; such a process is conventionally referred to as a transverse inversion.

If the length of the sample is small in comparison to $c$ times any relevant timescales, a transverse inversion will yield the same response as a swept inversion. Conversely, if the length of the sample is greater than the distance which light can travel during the characteristic SR timescale, then the type of pumping process will dramatically alter the intensity transient at the end of the sample. In such a configuration, transversely inverted molecules separated by a distance greater than $c \tau_\textrm{d}$ (for $\tau_\textrm{d}$ the delay to peak SR emission) will not become correlated before the onset of peak SR emission. The number of molecules cooperatively contributing to the SR intensity pulse generated by a transverse inversion will thus be limited to a fraction of the total population within the so-called cooperation length $L_\textrm{c} \propto c \tau_\textrm{d}$, and non-linear SR emission may only develop after transverse inversion if the number of molecules within $L_\textrm{c}$ exceeds the critical threshold for that SR regime. This requirement, known as the Arecchi-Courtens condition \citep{Arecchi1970}, causes the response to a WDB transverse inversion to more closely resemble the resonant case than does the response to a WDB swept inversion.

The Arecchi-Courtens condition inhibits the formation of global polarisation phase correlation during a transverse WDB SR process via an autoregulation mechanism on the number of molecules acting cooperatively; we justify this claim via the following thought experiment. Suppose that a sample of length $L$ possesses a smooth distribution of some initial total velocity bandwidth $\Delta v_i$, such that the resulting SR endfire intensity transient possesses a peak duration $\tau_\textrm{p, i} \approx L/c$. This duration is inversely proportional to the total number of molecules partaking in the SR process \citep{Gross1982} which is, at present, the population of the entire velocity distribution. Imagine that this distribution is now widened by adding molecules adjacent to its lower and upper bounds. In a short sample of length $L \ll L_\textrm{c}$---or in response to a swept inversion---the added molecules can correlate with the existing distribution and reduce the peak intensity duration to some $\tau'_\textrm{p} <\tau_\textrm{p, i}$.\footnote{In Sections \ref{subsec:swept-comb} and \ref{sec:swept-cont} we will in fact demonstrate that the duration of a swept SR process reduces arbitrarily as molecules are added to the distribution wings.} In a transverse SR process of length $L\approx L_\textrm{c}$, however, if the added velocity population were to reduce the SR timescale, then the cooperation length $L_\textrm{c}$ would simultaneously reduce to $L_textrm{c} < L$. This shorter cooperation length would, in turn, contain a smaller total number of molecules and thereby increase the SR timescale and decrease the velocity bandwidth of molecules cooperatively participating in the SR process. This feedback loop forms an autoregulation mechanism which inhibits arbitrarily wide bandwidth correlation and acts to sustain finite temporal structure. If a sample's length is on the order of or exceeds $L_\textrm{c}$, then, we expect its transversely initiated SR intensity response to retain finite temporal features in the WDB limit.

\begin{figure}
    \centering
    \includegraphics[width=1.\columnwidth, trim=.4cm .4cm .4cm .4cm, clip]{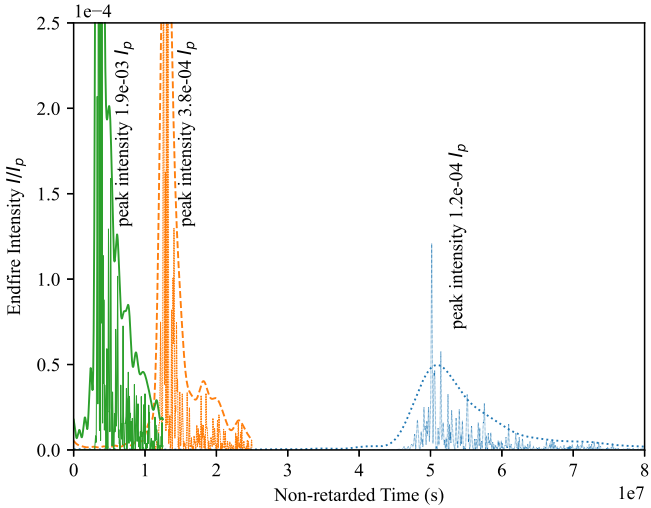}
    \caption{Endfire intensity transients resulting from the sudden transverse inversion at $t=0$ of three samples of lengths $c T / 32$ (solid, leftmost), $c T / 8$ (dashed), and $c T /2$ (dotted, rightmost) for a simulation of duration $T=10^8 \textrm{ s}$. All transients normalised to $I_p$ with peak values provided in text overlays. See accompanying text for discussion of filtered overlays. Velocity extent of $4{\small,}095$ channels total, simulated with the CTD algorithm. We observe transient structure in the WDB limit proportional to the length of the transversely inverted sample.}
    \label{fig:transverse}
\end{figure}

We show in Figure \ref{fig:transverse} the transient endfire intensity responses to the transverse inversion of three samples of differing lengths and identical column densities, such that only the final $10\%$ of their lengths enter the non-linear SR regime. For a simulation duration $T=1\times {10}^{8} \textrm{ s}$, the samples of lengths $c T / 32$, $c T / 8$, and $c T /2$ are suddenly inverted everywhere at simultaneous non-retarded time $t=0$. Each sample possesses a Gaussian velocity distribution of $4{\small,}095$ channels spanning $\pm \sqrt{2} \sigma$ for a Gaussian characteristic width $\sigma = 2{\small,}047 d\omega / \sqrt{2}$. The SR emission reaches peak intensity approximately after the delay required for radiation to propagate down the length of each sample.

Figure \ref{fig:transverse} presents both the raw intensity transients generated by simulation (jagged plots) as well as filtered plots which more accurately represent the intensity measured by a real observation. Although the number of channels simulated is large relative to the fundamental frequency differential $1/T$, the total velocity extent remains small relative to the bandwidth of the observing instrument. A real telescope would gather incident radiation generated by the decoherent combination of many versions of the above transient across the observing bandwidth. Furthermore, we expect a real astrophysical sample to be comprised of a large number of unresolved independent cylindrical samples generating SR radiation. More precisely, the filtered plots of Figure \ref{fig:transverse} were constructed by (1) taking the full FFT of the signal; (2) masking out a 20-channel band at the lower end of the spectrum, inverting back to the time domain, and computing its intensity; and (3) repeating Step 2 for a sliding band across the entire FFT, and averaging all intensity transients so generated.

Each transient of Figure \ref{fig:transverse} contains a finite duration intensity pulse, demonstrating that transverse inversion indeed sustains temporal structure in the WDB limit; additionally, the duration of a pulse is observed to be proportional to the sample's length. This proportionality can be explained by the autoregulation mechanism as follows. Suppose we construct a WDB distribution by appending velocity channels to the boundaries of an initially narrow velocity extent. As the distribution is widened, the SR emission duration and the cooperation length $L_\textrm{c}$ decrease as more molecules enter the SR process, but $L_\textrm{c}$ at first remains above the sample length $L$. Eventually, however, $L_\textrm{c}$ reduces below the order of the length of the sample. At this point the temporal duration attains a critical value $\tau_\textrm{p, crit}$ proportional to the length of the sample. If the velocity distribution is widened further, then any potential reduction in temporal duration and corresponding cooperation length would also decrease the number of molecules participating in the SR process, thereby contradictorily extending the temporal duration. The reduction in SR intensity duration is thus arrested at $\tau_\textrm{p, crit} \propto L$.

\begin{figure}
    \centering
    \includegraphics[width=1.\columnwidth, trim=.4cm .4cm .4cm .4cm, clip]{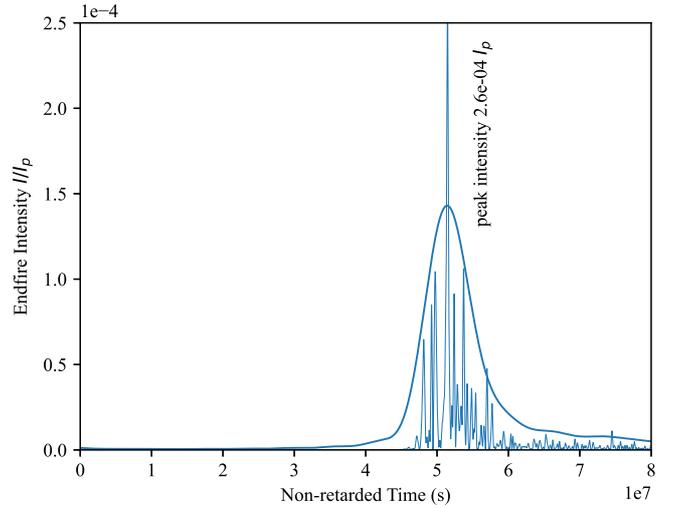}
    \caption{Endfire intensity transient resulting from the sudden transverse inversion at $t=0$ of a sample of length $c T /2$ for a simulation of duration $T=10^8 \textrm{ s}$. Normalised to $I_p$ with peak value provided in text overlay. Velocity extent of $2{\small,}047$ channels total, simulated with the CTD algorithm; cf. the rightmost intensity transient of Figure \ref{fig:transverse}. See accompanying text for discussion of filtered overlays. The transient duration is immune to the change in total velocity channel count (cf. Figure \ref{fig:transverse}).}
    \label{fig:transverse_redwidth}
\end{figure}

We can verify that the finite temporal duration $\tau_\textrm{p}$ of, for example, the rightmost transient intensity pulse in Figure \ref{fig:transverse} is indeed a critical limiting value by repeating the simulation at a revised total velocity width. The $4{\small,}095$ velocity channel extent is in fact an order of magnitude greater than the critical velocity extent for that sample, so we choose for numerical convenience to reduce the width of the distribution by a factor of two, to span $2{\small,}047$ channels total. The resulting transient intensity pulse, shown in Figure \ref{fig:transverse_redwidth}, indeed possesses the same temporal duration as its earlier counterpart in Figure \ref{fig:transverse}.

In the opening to this section we discussed the role of the Arecchi-Courtens condition in a transverse inversion SR process, where we preemptively stated an upcoming result of Sections \ref{sec:swept-disc} and \ref{sec:swept-cont}; namely, that a swept inversion leads to extremely wide bandwidth polarisation phase correlation and the quenching of temporal structure in the WDB limit. For a pump propagating with speed $c$ at an angle $\theta$ to the longitudinal axis of the sample, the SR emission duration $\tau_\textrm{p}$ of a WDB sample is only at present known in two cases; i.e.,
\begin{equation}
    \tau_\textrm{p}\left(\theta\right) \approx
    \begin{cases}
        \tau_\textrm{p, coh}   &\theta=0 \\
        \textrm{max}\left(\alpha \frac{L}{c} \textrm{, } \tau_\textrm{p, coh}\right)    &\theta=\pi/2
    \end{cases} \label{eq:tau_p_twoangles}
\end{equation}
where $\tau_\textrm{p, coh}$ is the duration of SR emission that a resonant ensemble of the same total initial population inversion would generate ($\tau_\textrm{p, coh} \rightarrow 0$ in the infinitely WDB limit) and $\alpha$ is a calibration factor dictated by the detailed workings of the transverse autoregulation mechanism. We estimate from Figure \ref{fig:transverse} that $\alpha \approx 0.2$.

We may make an educated guess as to the generalisation of expression \eqref{eq:tau_p_twoangles} to intermediate angles $0<\theta<\pi/2$ if we first recognise the quantity $L/c$ appearing in the transverse case ($\theta=\pi/2$) as the amount of time required for the initial onset of inversion to propagate back up the length of the sample in the retarded frame. For an arbitrary $\theta$, the duration of said propagation becomes $(L/c) \sin \theta$; it is therefore natural to propose that
\begin{equation}
    \tau_\textrm{p} \left(\theta\right) = \textrm{max}\left(\alpha \frac{L}{c} \sin\theta \textrm{, } \tau_\textrm{p, coh}\right).
\end{equation}

\section{Swept Inversion Processes Part I: \\ \hspace{11 pt} Discrete Distributions}\label{sec:swept-disc}

We now shift our attention to swept inversion processes; unless otherwise mentioned, every WDB SR process in this section as well as in Section \ref{sec:swept-cont} will be initiated by a swept inversion occurring at simultaneous retarded time $\tau$. Before investigating the development of WDB SR transients in response to the swept inversion of a continuum velocity distribution, it is important to first quantify the degree of interaction between discrete velocity populations at various Doppler offsets. In this section we simulate distributions comprised of multiple Dirac delta-like contributions at varying velocity offsets and degrees of SR saturation. The so-called spectral interaction distance evaluated in this section will guide our physical intuition of the continuum in later sections and will serve three quantitative purposes: first, it will determine the required fidelity of our later $\textrm{LMI}^\textrm{SF}$ approximations; second, it will allow us to quantify the degree of SR saturation in the continuum; and third, it will determine characteristic scales of stochastic distribution variations which become important in Section \ref{subsec:cont-noisysims}.

\subsection{Two discrete samples and spectral interaction distance}\label{subsec:swept-two}

\begin{figure}
    \centering
    \includegraphics[width=.9\columnwidth, trim=0cm 0cm 0cm 0cm, clip]{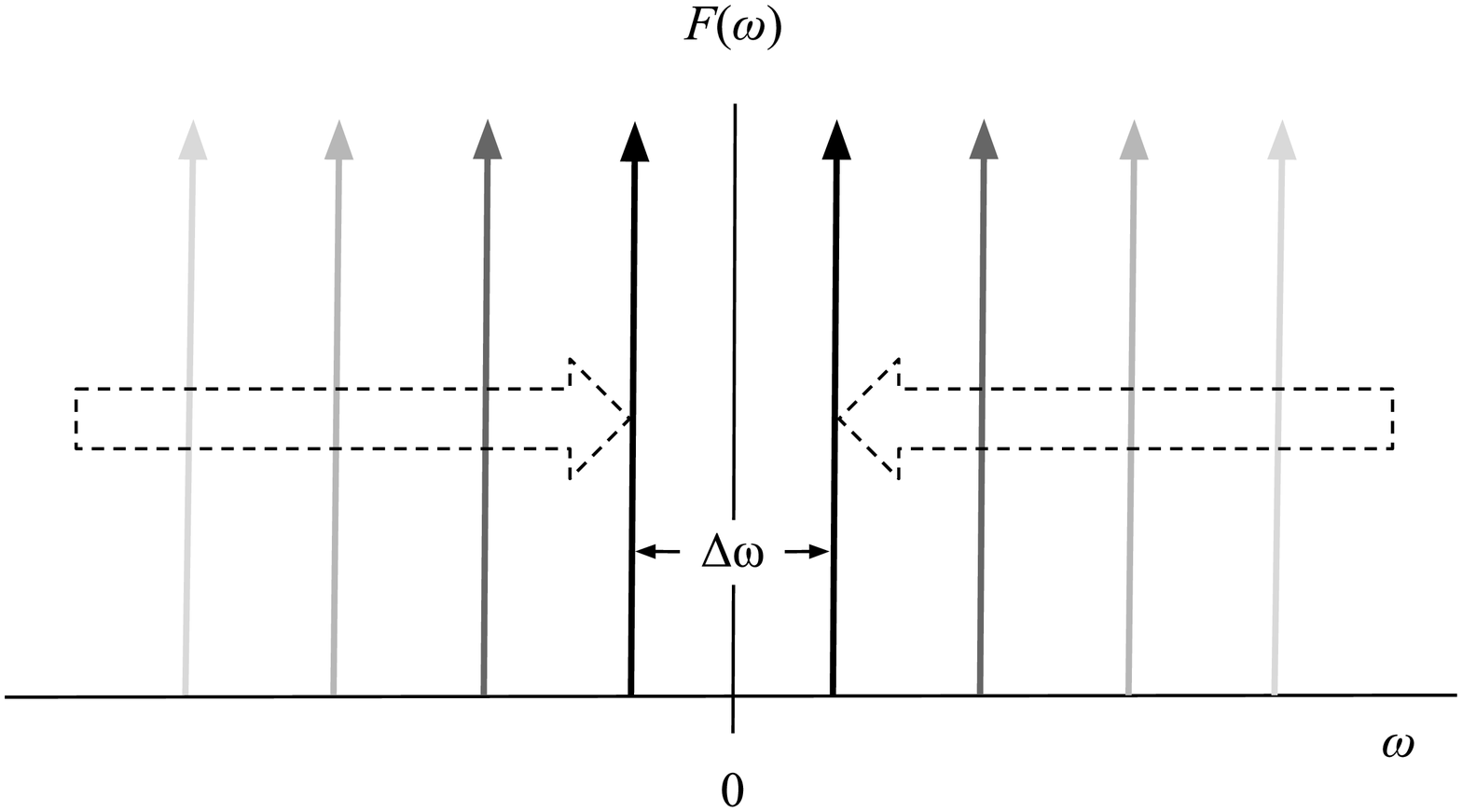}
    \caption{Velocity distributions of successive simulations indicated by increasingly dark shades. Each simulation involves a pair of Dirac delta-like distributions at equal and opposite velocity offsets from the central envelope carrier frequency.}
    \label{fig:two_distrns}
\end{figure}

We now execute a series of simulations of varying distributions, each comprised of two Dirac delta sub-populations at successively reducing Doppler offsets as depicted in Figure \ref{fig:two_distrns}. Our simulations merge the two sub-populations together while evaluating, for each separation, the degree of interaction between them. Our goal in this exercise is to quantify the relationship between velocity separation and SR coherence modelled by the MB equations.

In order to establish the dependency of SR coherence upon velocity separation, we must first establish a meaningful metric for the degree of SR cooperation between two populations. Recalling that SR is characterised by an $n^2$ peak intensity dependency upon population size $n$ (see equation \ref{eq:Ip} and note there that $T_\textrm{R}$ is inversely proportional to $n$), it may be tempting to simply evaluate peak intensity as a function of velocity separation; however, such measurements are sensitive to the beating effect of interfering SR radiation from two carrier frequencies. The time of occurrence of peak intensity in the SR transient does not necessarily coincide, for different velocity separations, with a repeatably constructive or destructive phase of the two populations' interference pattern. Peak SR intensity therefore does not vary smoothly with the velocity separation between two samples.

A better metric for the degree of interaction between two samples is, instead, the total energy released by the system during the transient process. The justification of such a metric is not a priori apparent from simple physical intuition regarding a small sample SR system, since the total energy released in such a sample scales linearly with its population;\footnote{Although the peak intensity of a small sample scales quadratically with its population, the intensity timescale concurrently reduces inversely \citep{Gross1982}.} however, in the large sample case the degree of SR saturation determines the length fraction entering the non-linear regime (see, for example, \citealt{Rajabi2020}) and thereby converting its initial population inversion into radiated energy. The total energy released by the sample is dominated by that length fraction which loses its inversion and is, therefore, greater at higher SR saturations. If two large sample populations are interacting during their emission of radiation, then, we should observe greater total radiated energy over the transient duration than in their far-separated limit.

\begin{figure}
    \centering
    \includegraphics[width=1.\columnwidth, trim=.4cm .4cm .3cm .4cm, clip]{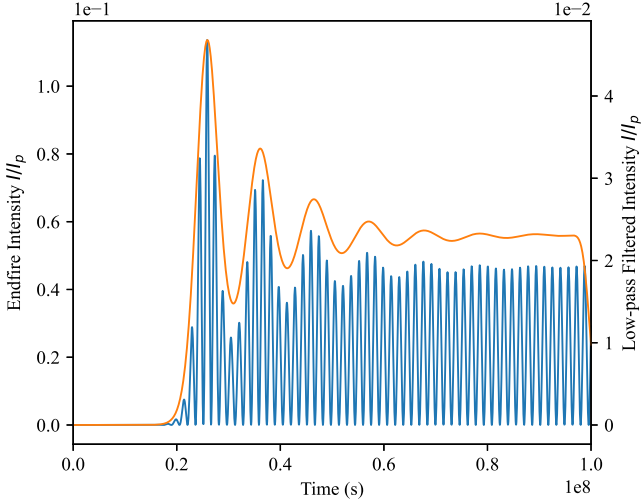}
    \caption{Normalised endfire intensity from two far-separated samples, $v_\textrm{sep} = 64 dv$. Fast-varying curve: endfire intensity; slowly-varying curve: low-pass filtered intensity. Normalised to $I_\textrm{p}$. Two independent transient SR processes are present, with a beating pattern equal to the natural frequency offset between the two samples (as expected).}
    \label{fig:two_distrns_vsep64}
\end{figure}

We simulate the CTD representation of the MB equations \eqref{eq:MB_TD_norm-1}–\eqref{eq:MB_TD_norm-3} for a one-dimensional sample of methanol molecules ($\omega_0 = 2 \pi \times 6.7 \text{ GHz}$, $d=0.7\text{ D}$) over a duration $T={10}^8 \text{ s}$, having length $L=2 \times {10}^{13} \text{ m}$, radius $w=5.4 \times {10}^{5} \text{ m}$, population inversion relaxation time constant $T_1=1.64 \times {10}^{7} \text{ s}$, and polarisation dephasing time constant $T_2=1.55 \times {10}^{6} \text{ s}$. This system is similar to that of \citet{Rajabi2020}, but differs in its dimensions and–most importantly–in that we now simulate two velocity distributions each with initial population inversions of $2 N_0=3.0 \times {10}^{-6} \text{ m}^{-3}$ at $\tau=0$. At present, the value of $2 N_0$ is considered to represent the total number of molecules inside the bandwidth of the transient process generated which is found, momentarily (see Figure \ref{fig:two_distrns_vsep64}), to be on the order of ${10}^{-7} \textrm{ Hz}$. Extrapolating this to the population of a sample spanning, say, $1 \textrm{ km/s}$, this value of $2 N_0$ would represent a total Doppler-broadened population of $2 N_0 (1 \textrm{ km/s}) (6.7 \textrm{ GHz}) / \left[c ({10}^{-7} \textrm{ Hz})\right] = 6.7 \times {10}^{5} \textrm{ m}^{-3}$. We apply a constant restoring population inversion pump equal to the relaxation rate; i.e., $\Lambda^{\left(N\right)} \left(\tau\right) = 2 N_0/ \left(2 T_1\right)$.\footnote{Recall that $N_v$ of equation \eqref{eq:dimless_defns} is half the population inversion, hence the factor of $1/2$ in the restoring pump.} Note that these dimensions correspond to a Fresnel number $\pi w^2 / L \lambda$ of unity.

The velocity distribution is, for each simulation, a pair of Dirac delta functions separated by some $v_\textrm{sep}$; i.e.,
\begin{equation}
    F_v = \frac{1}{2} \delta(v+v_\textrm{sep}/2) + \frac{1}{2} \delta(v-v_\textrm{sep}/2).
\end{equation}
Note that we choose to normalise $F_v$ which implies that, actually, $2 N_{-v_\textrm{sep}/2}(\tau=0,z) = 2 N_{v_\textrm{sep}/2}(\tau=0,z) = 6.0 \times {10}^{-6} \text{ cm}^{-3}$ in order for each of the two sub-samples to correspond to an initial population inversion density
\begin{equation}
    \int_{\pm v_\textrm{sep}-\epsilon}^{\pm v_\textrm{sep}+\epsilon} \textrm{d}v 2 N_v(\tau=0, z) = 3.0 \times {10}^{-6} \text{ cm}^{-3}.
\end{equation}
Defining the fundamental velocity step $dv = (c/\omega_0) 2\pi/T$, we commence our series of simulations with two (thus far assumed, but shortly verified) independent samples at $v_\textrm{sep} = 64 dv$. The resulting radiation intensity at the endfire ($z=L$) is shown in Figure \ref{fig:two_distrns_vsep64}.

The endfire intensity is, as expected, an interference of two Doppler offset SR processes, each precessing positively or negatively relative to the reference envelope frequency. The transient profile of the underlying SR processes occurring in either independent sample may be visualised by applying a low-pass filter: the orange slowly-varying curve superimposed on Figure \ref{fig:two_distrns_vsep64} is the result of convolution with a Gaussian of width $T/40$ and clearly demonstrates the salient SR transient features of a delay to peak intensity and a ringing thereafter. Note that this filter is distinct from those used in previous figures (Figure \ref{fig:transverse}, for example), and is chosen in order to visualise the transient profile undergoing beating (rather than to represent real observation, as was previously the case).

These results are not surprising, but they do demonstrate two physically accurate properties of the MB equations; first, that the SVEA of the MB equations is not preferential to the central carrier frequency--the MB equations under the SVEA  model SR equally well anywhere within the velocity distribution, and not only at the central factorisation frequency $\omega_0$. The choice of $\omega_0$ introduces a visualisation asymmetry between velocities, but not a physical one. Second, the independence of sufficiently Doppler-offset SR processes is properly modelled by the MB equations.

\begin{figure}
    \centering
    \includegraphics[width=1.\columnwidth, trim=.4cm .4cm .3cm .4cm, clip]{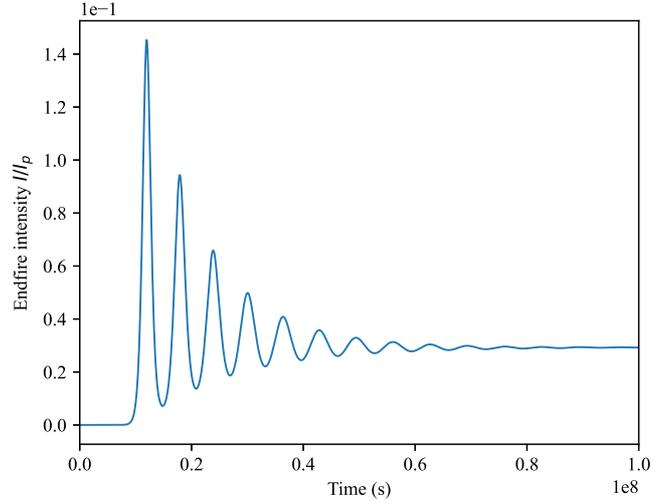}
    \caption{Endfire intensity from two superimposed samples, $v_\textrm{sep} = 0$. Normalised to $I_p$. The Doppler broadened MB equations now describe interaction between the two samples and properly generate a transient response equivalent to that of a higher-saturated single sample.}
    \label{fig:two_distrns_vsep0}
\end{figure}

\begin{figure}
    \centering
    \includegraphics[width=1.\columnwidth, trim=1cm 1cm 1cm 1cm, clip]{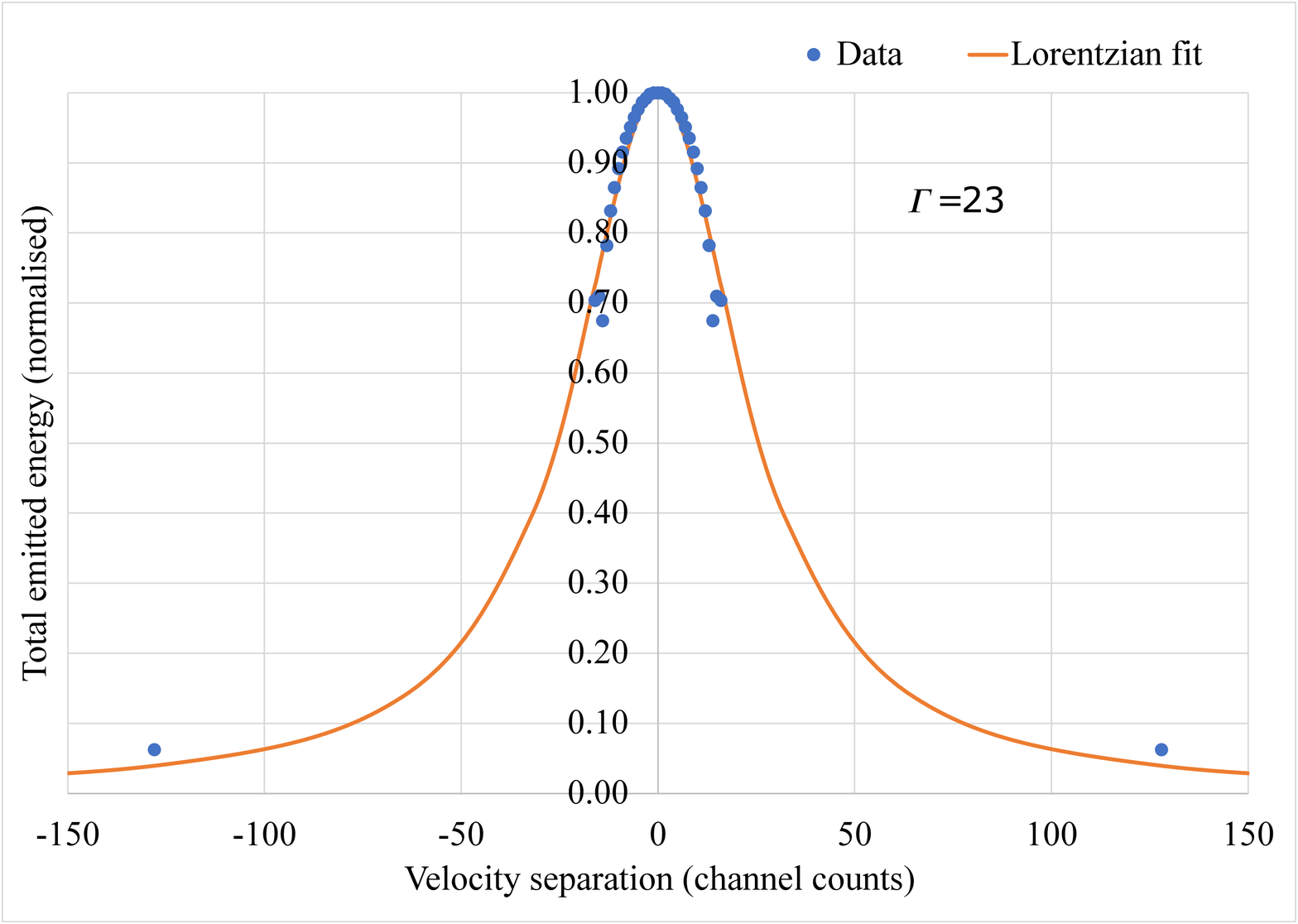}
    \caption{Normalised total SR energy emitted as a function of two-sample velocity separation (separation in counts of the fundamental velocity differential $dv$). Orange curve: least-squares Lorentzian fit. The Doppler broadened MB equations are seen to model spectral interaction as expected.}
    \label{fig:two_distrn_merger}
\end{figure}

We now proceed to merge the two Dirac delta velocity sub-components by simulating at separations of $v_\textrm{sep} / dv = \pm \left\{64, 32, 16, 15, 14, 13, 12, \dots 1, 0\right\}$. The terminal case $v_\textrm{sep}=0$ is shown in Figure \ref{fig:two_distrns_vsep0}. For each separation we integrate the intensity for the first quarter of the simulation duration, which roughly covers the transient SR pulse visible in the first $2.5\times{10}^7$ s of Figure \ref{fig:two_distrns_vsep0}. We subtract from this the total energy emitted in the independent limit $v_\textrm{sep} \rightarrow \infty$ (practically established from $v_\textrm{sep} = 256 dv$) and normalise to the zero separation case. The resulting dependency upon velocity separation is plotted in Figure \ref{fig:two_distrn_merger}.

Some spurious features are visible in Figure \ref{fig:two_distrn_merger}. Around $v_\textrm{sep} = 15 dv$ the total energy dependency varies erratically. Near such separations, the two channels remain locked in phase during the SR transient phase for $\tau \lesssim 2.5\times {10}^7 \textrm{ s}$ but separate immediately thereafter, at which point they engage in constructive or destructive interference for $\tau > 2.5\times {10}^7 \textrm{ s}$. The total integrated intensity therefore varies quickly with velocity separation as the constructive or destructive phases traverse the exit of the integration window. Such behaviour is similar to the issues alluded to earlier regarding peak intensity, where constructive or destructive phasing relative to the time of peak onset corrupts the smooth dependency of peak intensity upon velocity separation. In the present case, however, this spurious effect is minor and a Lorentzian-like profile is clearly apparent in the data.

We perform a least-squares Lorentzian fit on the central $\pm 16$ velocity separations (in units of $dv$), under the assumption that the central lobe behaviour is most representative of the interaction between the two-samples. The resulting fit shown in Figure \ref{fig:two_distrn_merger} has $\Gamma=23$ for a functional form
\begin{equation}
    G\left(v_\textrm{sep}\right) = \frac{\Gamma^2}{\left(v_\textrm{sep}/dv\right)^2+\Gamma^2}.
\end{equation}
For the sample of the present simulation we therefore a identify the so-called spectral interaction distance as $v_\textrm{int} = \Gamma dv = 23 dv$.

The factor of $23$ should not be surprising. In Figure \ref{fig:two_distrns_vsep64} the fastest transient SR feature (the first peak of the low-pass filtered SR transient pulse) generated by either separate sub-sample occurs over a duration $T_\textrm{d} \approx T/23$. We expect the electric field generated by a sample to interact with channels Doppler shifted by $23 d\omega$, an angular frequency extent which corresponds to 23 velocity channels under our discretisation scheme. The present result is the quantitative verification of this physical intuition that two samples should interact only if they are Doppler offset by a velocity on the order of their transient spectral extent. Though unsurprising, this spectral interaction distance result was not mathematically guaranteed in the non-linear regime of the MB equations.

\subsection{Comb distributions}\label{subsec:swept-comb}

We now conduct three experiments on so-called comb distributions, which elucidate the connection between discrete distributions and continuum distributions. We define a comb distribution as a chain of discrete channels of adjacent separation $\Delta v_\textrm{sep}$, and we identify an experiment as a series of simulations of progressively smaller $\Delta v_\textrm{sep}$ for a fixed initial population inversion. In all simulations the channel polarisation phases and initial Bloch tipping angles are initiated in accordance with the discussion of Section \ref{subsec:theory-initialqm}.

\begin{figure}
    \centering
    \includegraphics[width=1.\columnwidth, trim=0cm 0cm 0cm 0cm, clip]{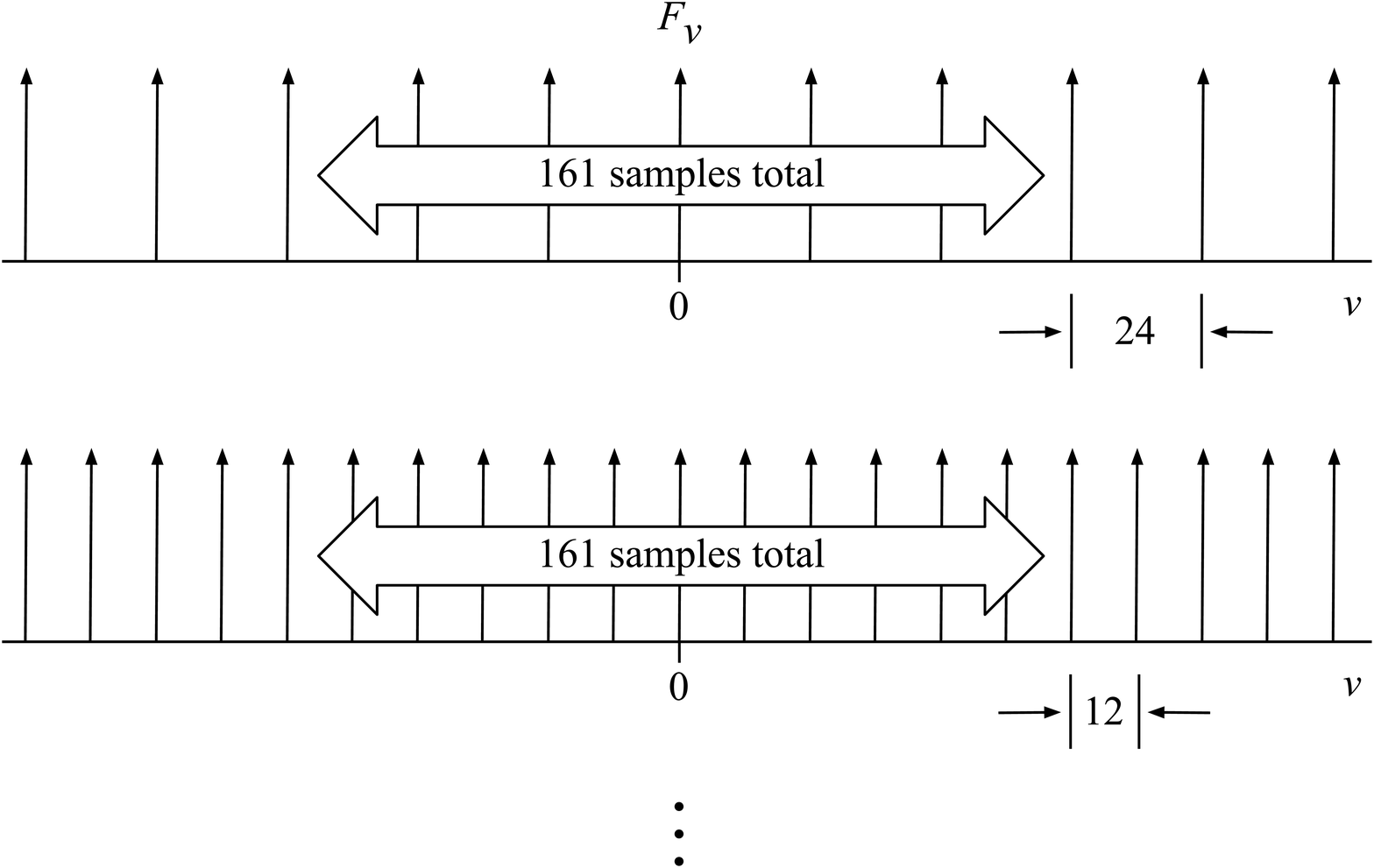}
    \caption{Successive comb distributions for a given comb experiment, which proceeds multiple iterations further than depicted and terminates at $\Delta v_\textrm{sep}=1dv$.}
    \label{fig:combs}
\end{figure}

The three experiments differ only in their total initial population inversions $N_0$, and each experiment proceeds as follows. For a given $N_0$ we simulate 161 channels separated by successively decreasing $\Delta v_\textrm{sep}$. We commence with a simulation of $\Delta v_\textrm{sep} = 24 dv$, and proceed to compress the comb distribution through simulations of finer $\Delta v_\textrm{sep}$ until the terminal case of $\Delta v_\textrm{sep} = 1 dv$. The first two iterations of the process are depicted in Figure \ref{fig:combs}.

We classify each experiment relative to the so-called ``critical threshold,'' identified as follows. In the case of a single resonant velocity channel, the critical initial population inversion threshold $N_{0 \textrm{ crit (coh)}}$ is defined as that value of $N_0$ above which the inversion at the end of the sample is lost (i.e., $N(z=L)\rightarrow 0$ and the system therefore attains the non-linear regime) at some time during the transient response of the system. Similarly, there should exist some critical per-channel initial population inversion threshold $N_{0 \textrm{ crit (non-coh)}}$ above which a continuum velocity distribution will lose its inversion at $z=L$. If the transient response of a resonant sample at the critical initial population inversion level has bandwidth $\Delta f$, then we roughly expect that
\begin{equation}
    N_{0 \textrm{ crit (non-coh)}} \approx N_{0 \textrm{ crit (coh)}} \frac{dv}{c \Delta f / f_0}
\end{equation}
for a natural emission frequency $f_0$; however, we practically identify $N_{0 \textrm{ crit (non-coh)}}$ by experiment.

\subsubsection{First experiment: initial population inversion well above the critical threshold}\label{subsubsec:swept-comb-hisat}

\begin{figure}
    \centering
    \includegraphics[width=1.\columnwidth, trim=.7cm .8cm .5cm .4cm, clip]{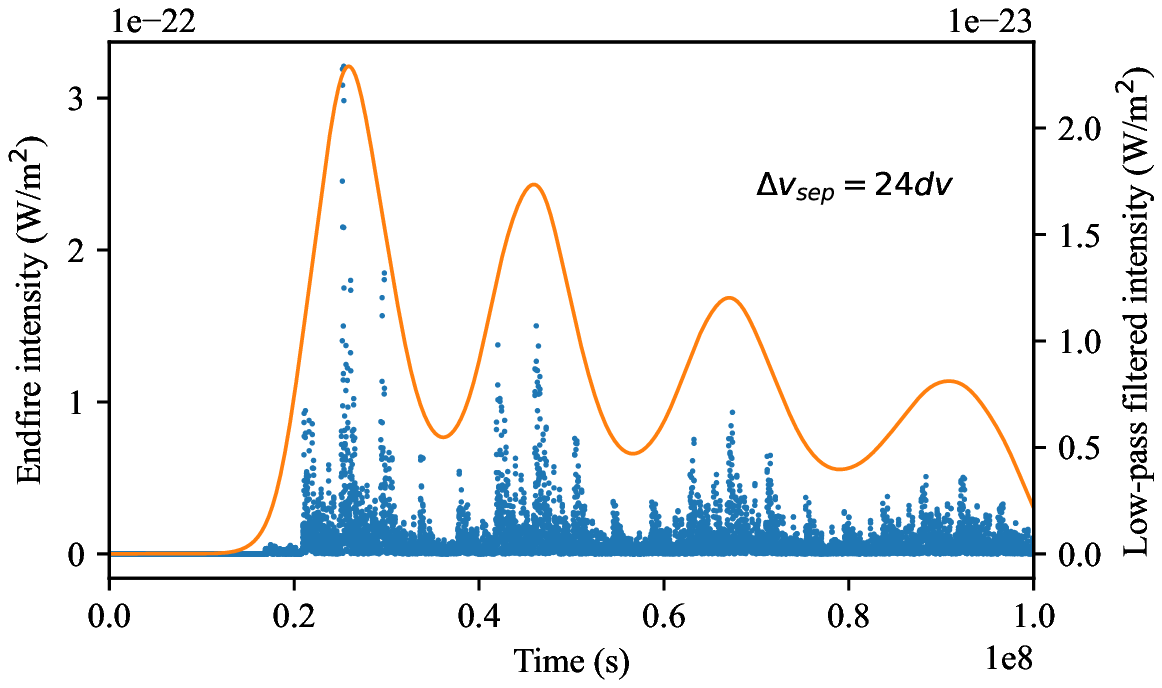}
    \includegraphics[width=1.\columnwidth, trim=.4cm .8cm .5cm .4cm, clip]{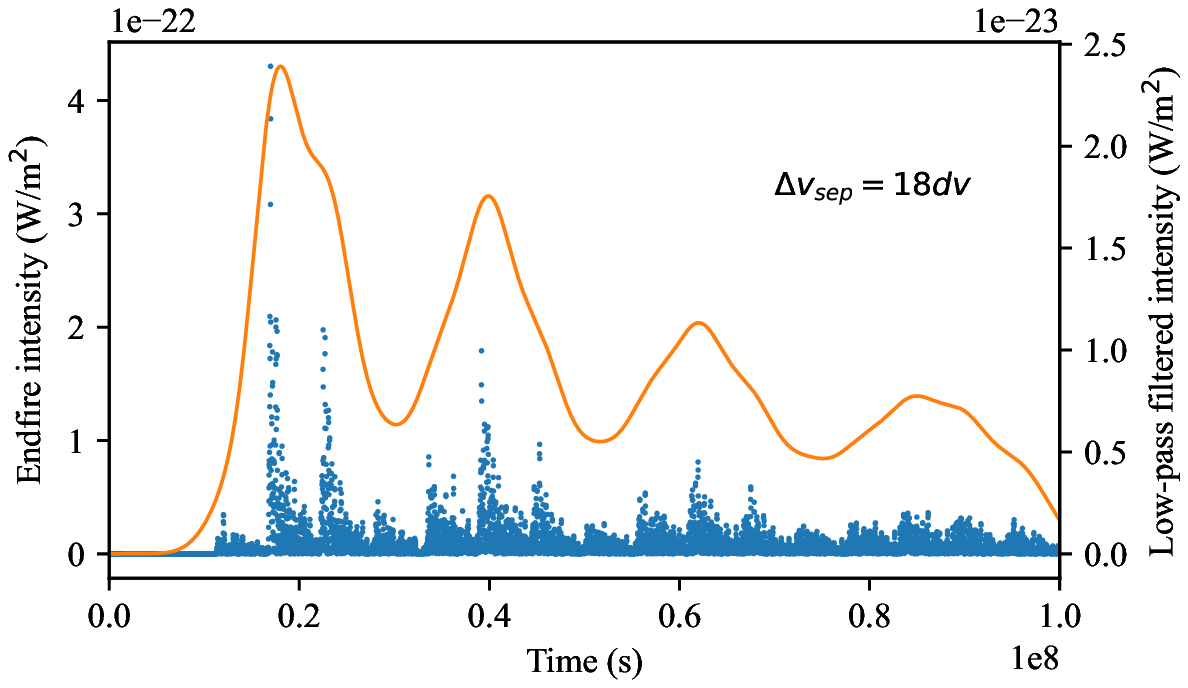}
    \includegraphics[width=1.\columnwidth, trim=.4cm .8cm .5cm .4cm, clip]{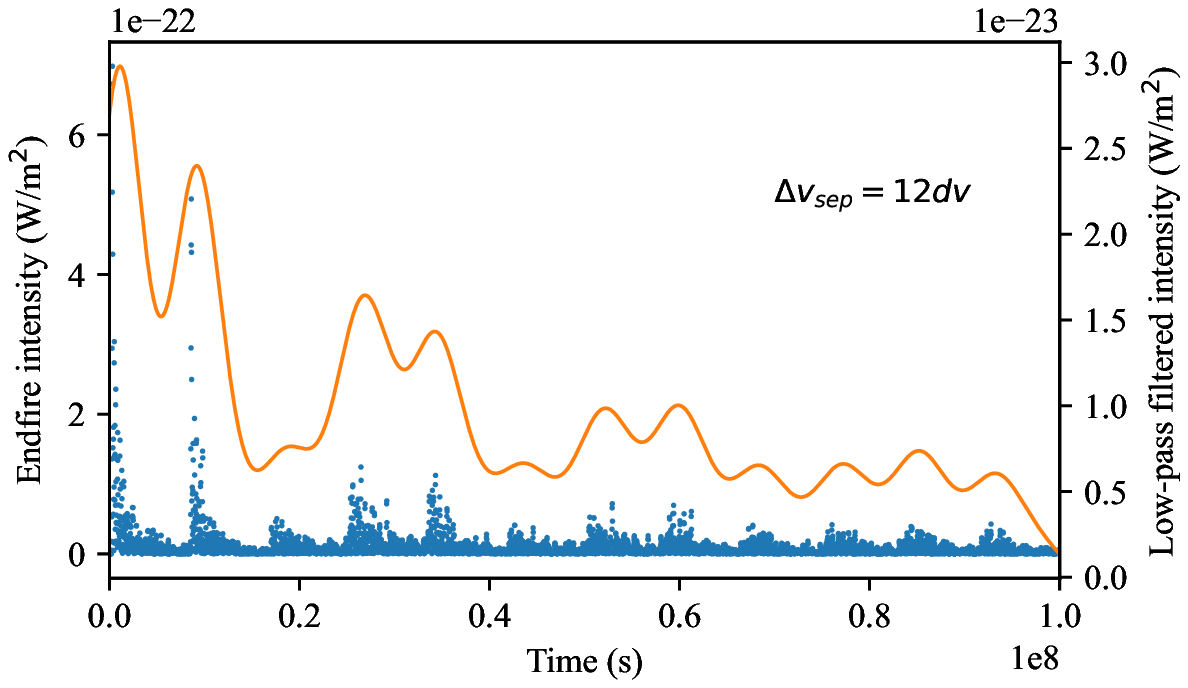}
    \includegraphics[width=1.\columnwidth, trim=.4cm .5cm .5cm .4cm, clip]{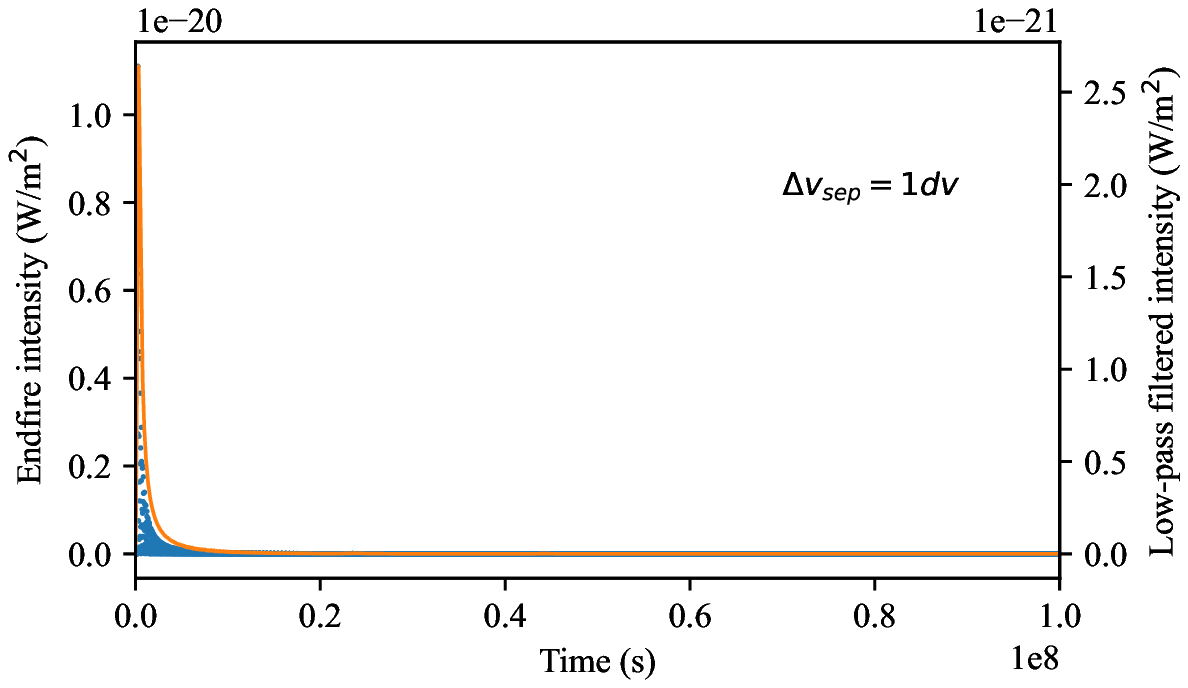}
    \caption{Intensity transients at the endfire of a comb distribution of 161 channels and initial population inversion $N_{0,\textrm{ high}} = 8.0\times {10}^{-7} \textrm{ m}^{-3}$ for velocity separations $\Delta v_\textrm{sep}/dv=24,\:18,\:12,\textrm{ and } 1$ (top to bottom). Filters of top three panels equal to those described in text accompanying Figure \ref{fig:transverse} (sliding 20-channel FFT band mask); bottom panel filter modified to 80-channel FFT mask for resolving faster temporal response. Wide spectral correlation is found to develop before reaching the final separation $\Delta v_\textrm{sep} = dv$.}
    \label{fig:comb_161channels_toothsepdecreasing}
\end{figure}

Our first experiment begins with 161 channels, each possessing an initial population inversion of $N_{0,\textrm{ high}} = 8.0\times {10}^{-7} \textrm{ m}^{-3}$ and separated by $\Delta v_\textrm{sep} = 24 dv$. Such a population inversion is well above the critical threshold in each channel (more precisely, $50\%$ of the sample length loses its population inversion during the SR process) and $\Delta v_\textrm{sep}$ is sufficiently wide such that each of the 161 channels evolves independently. The transient response of the top panel of Figure \ref{fig:comb_161channels_toothsepdecreasing} is, therefore, the decoherent combination of 161 SR transients, each modulated by an individual channel's natural frequency. Note that the intensity peaks of periodicity $T/24$ in the top panel of Figure \ref{fig:comb_161channels_toothsepdecreasing} are spurious features of the artificial comb distribution which would not be generated by a real continuum distribution; however, their emergence indicates a small degree of correlation between neighbouring channels. For further discussion and quantification of this remark, see Appendix \ref{app:comb_spurious}.

We proceed to simulate combs of decreasing velocity separations, each for the same initial population inversion $N_{0,\textrm{ high}}$. The intensity responses at the end of the sample are shown in Figure \ref{fig:comb_161channels_toothsepdecreasing} for $\Delta v_\textrm{sep} / dv=24,\:18,\:12,\textrm{ and } 1$ from top to bottom, respectively. In each panel we overlay a low-pass filtered intensity generated by convolution against a Gaussian of $\sigma$-width $T/20$ (top three panels) or $T/200$ (bottom panel).

The initial simulation (Figure \ref{fig:comb_161channels_toothsepdecreasing}, top) generates a transient SR pulse of full-width-half-maximum (FWHM) duration in its first lobe equal to $\tau_\textrm{p} \approx T/10$. We therefore expect channel coupling when $\Delta v_\textrm{sep}$ reduces below the order of $\sim \! 10 dv$. In the second panel of Figure \ref{fig:comb_161channels_toothsepdecreasing}, $v_\textrm{sep}$ has been reduced to $18dv$ and demonstrates only minor deviation from the independent channel limit; however, at $\Delta v_\textrm{sep}\approx 12 dv$ (Figure \ref{fig:comb_161channels_toothsepdecreasing}, third panel from top) we observe the onset of an initial intensity pulse near $\tau=0$.

As the comb is compressed further, such that $\Delta v_\textrm{sep} < 12 dv$, groups of adjacent channels couple together to form more populated SR ensembles. This cooperative effect is a non-linear one, as the larger populations of these groups will generate SR processes of shorter $\tau_\textrm{p}$ \citep{Gross1982}, and thus of wider spectral interaction width; this wider spectral interaction width in turn couples more channels together, reducing $\tau_\textrm{p}$ even further. In the terminal case $\Delta v_\textrm{sep} = 1 dv$ shown in the fourth panel of Figure \ref{fig:comb_161channels_toothsepdecreasing}, the population inversion is lost extremely quickly and the temporal duration of the peak intensity collapses to a very short pulse near $\tau=0$.

We numerically demonstrated the formation of wide velocity bandwidth correlation in a highly saturated sample ($N_0 \gg N_{0 \textrm{ crit (non-coh)}}$), which we intuitively understood via a theoretical estimate of the total number of molecules $n_\textrm{tot}$ participating in the SR process. This estimate assumed $\tau_\textrm{p}$ to be inversely proportional to $n_\textrm{tot}$; however, such a statement is theoretically derived \citep{Gross1982} for molecules of identical velocities. We thus consider this picture an intuitive one only, and one which does not a priori predict WDB SR intensity behaviour at all degrees of saturation. We proceed next to determine numerically if the wide velocity coherence demonstrated in Figure \ref{fig:comb_161channels_toothsepdecreasing} (bottom) is realised for lower initial population inversions $N_0$, or if $N_0$ may be tuned to sustain finite SR temporal structure.

\subsubsection{Second experiment: initial population inversion slightly above the critical threshold}\label{subsubsec:swept-comb-medsat}

We are interested in determining if a swept WDB sample is able to sustain an SR process of finite temporal structure. The initial population inversion used in the previous experiment of Section \ref{subsubsec:swept-comb-hisat} was found to introduce coupling between channels when $\Delta v_\textrm{sep}$ crossed below a critical threshold of $\sim\!12 dv$, such that further reduction of $\Delta v_\textrm{sep}$ down to the continuum case $\Delta v_\textrm{sep} = 1 dv$ resulted in a wide coupling of velocity channels and the quenching of temporal structure. We now use this $\Delta v_\textrm{sep} = 12dv$ coupling threshold of the previous experiment to inform our attempt to configure a sample which may sustain finite temporal SR structure down to the continuum limit, by setting our initial population inversion to $N_{0,\textrm{ mod}} = N_{0,\textrm{ high}}/12 = 6.7\times {10}^{-8} \textrm{ m}^{-3}$.

\begin{figure}
    \centering
    \includegraphics[width=1.\columnwidth, trim=.4cm .8cm .5cm .4cm, clip]{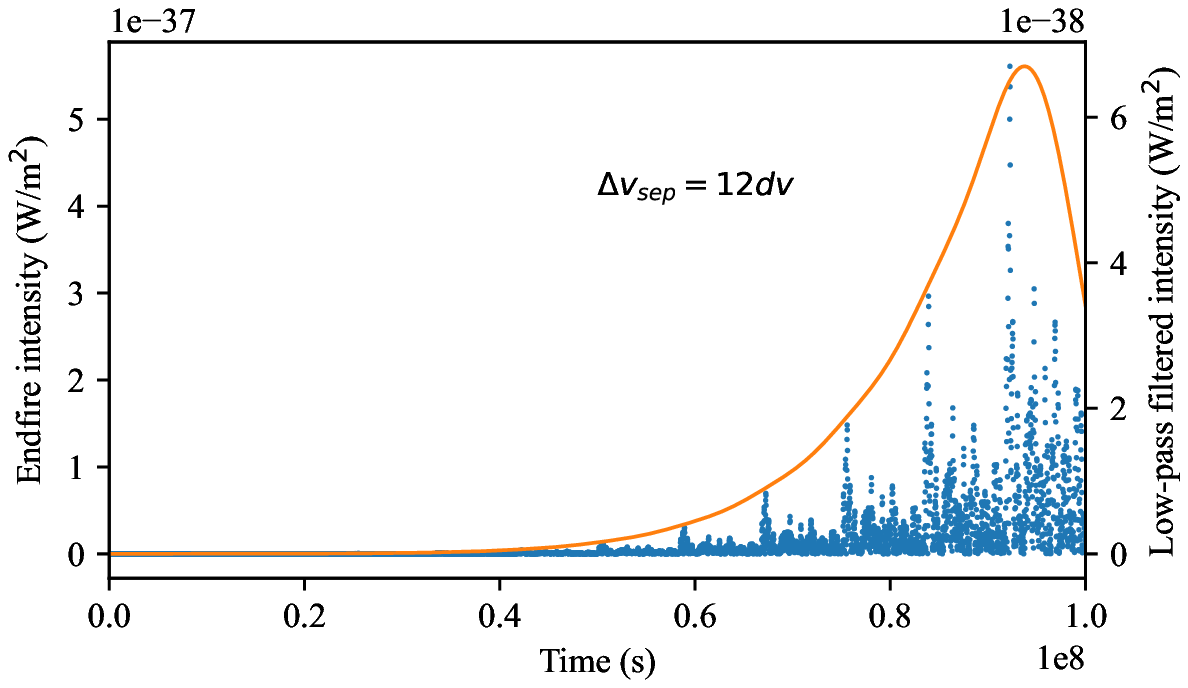}
    \includegraphics[width=1.\columnwidth, trim=.4cm .8cm .5cm .4cm, clip]{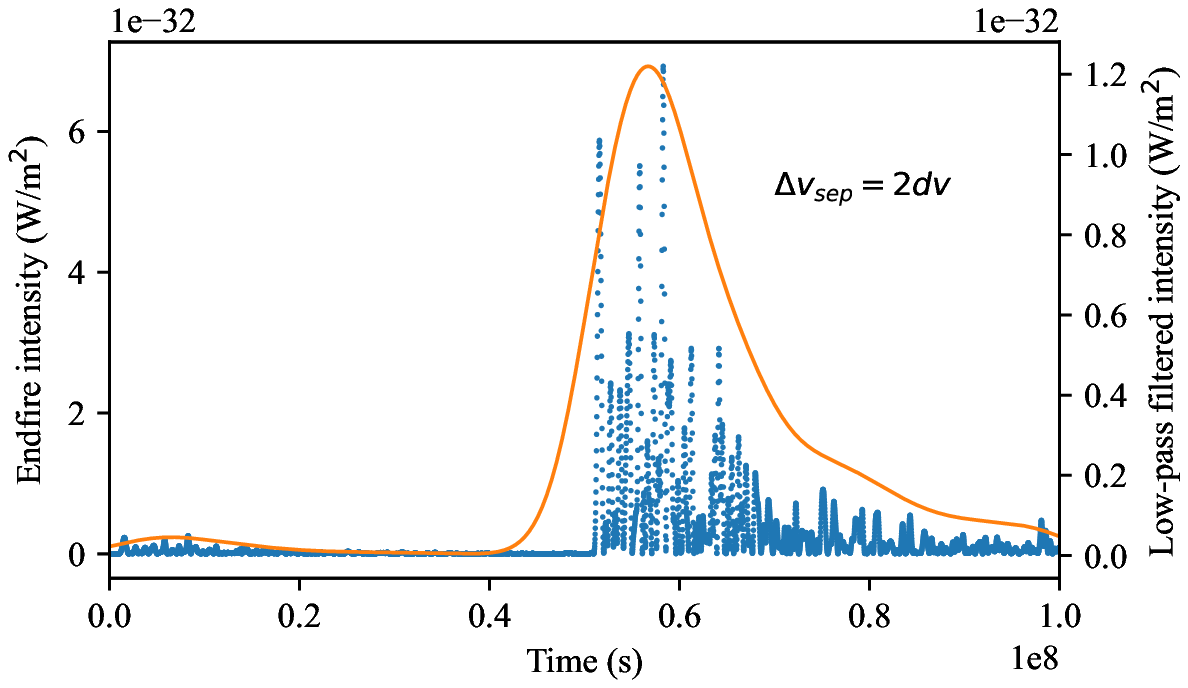}
    \includegraphics[width=1.\columnwidth, trim=.4cm .5cm .5cm .4cm, clip]{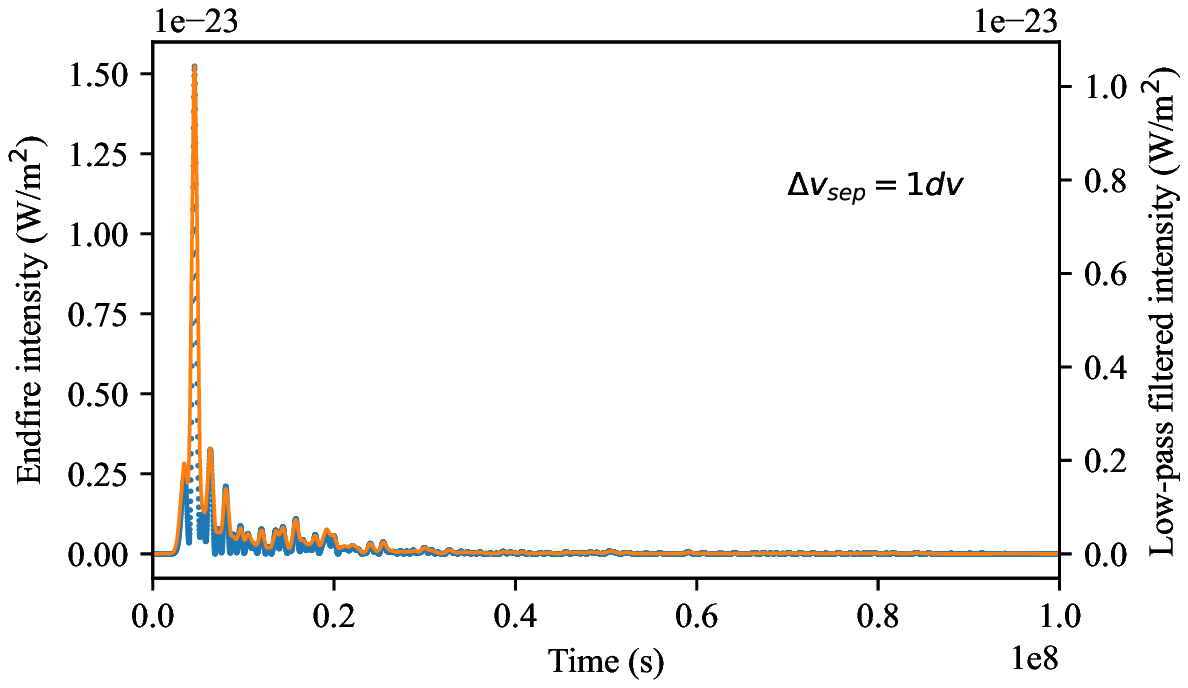}
    \caption{Intensity transients at the endfire of a comb distribution of 161 channels and initial population inversion $N_{0,\textrm{ med}} = 6.7\times {10}^{-8} \textrm{ m}^{-3}$ for velocity separations $\Delta v_\textrm{sep}/dv=12,\:2,\textrm{ and } 1$ (top to bottom). Filters of top two panels equal to those described in text accompanying Figure \ref{fig:transverse} (sliding 20-channel FFT band mask); bottom panel filter modified to 80-channel FFT mask for resolving faster temporal response. Wide spectral correlation is found to develop in the final transition from  $\Delta v_\textrm{sep} = 2 dv$ to $\Delta v_\textrm{sep} = dv$.}
    \label{fig:comb_161channels_medsat_toothsepdecreasing}
\end{figure}

The top panel of Figure \ref{fig:comb_161channels_medsat_toothsepdecreasing} depicts the intensity at the endfire of the sample for $\Delta v_\textrm{sep}=12 dv$, and demonstrates temporal structure of very long duration (in fact, $\tau_\textrm{d}>T$). The temporal structure is sustained through successively finer $\Delta v_\textrm{sep}$; in fact, the retention of finite temporal structure is apparent even at $\Delta v_\textrm{sep}=2dv$, depicted in the second panel of Figure \ref{fig:comb_161channels_medsat_toothsepdecreasing}. Our earlier discussion of Section \ref{subsubsec:swept-comb-hisat} regarding the spurious periodic modulation of the slow transient is relevant here again, where now the channel separation of $2 dv$ explains (see Appendix \ref{app:comb_spurious}) the sharp rise at $T/2$ in the time domain. Ignoring this non-physical artifact of the comb distribution, the case $\Delta v_\textrm{sep}=2dv$ is seen to yield finite temporal duration.

If we cross below the coupling threshold which we configured to occur (via our choice of $N_{0,\textrm{ mod}}=N_{0,\textrm{ high}}/12$) at $\Delta v_\textrm{sep}=1dv$, however, the response changes dramatically. In the bottom panel of Figure \ref{fig:comb_161channels_medsat_toothsepdecreasing}, the reduction to $\Delta v_\textrm{sep}=1dv$ yields extremely wide bandwidth velocity coupling with a collapse of the SR peak intensity duration to a very short duration. Additionally, we observe a loss of population inversion in only the final $10\%$ of this sample's length. Our act of configuring the sample to yield channel coupling precisely when $\Delta v_\textrm{sep} = 1dv$ has, in fact, also selected an initial population inversion which is just above the critical threshold required to initiate a non-linear SR process. This is not a coincidence, but an important result; we demonstrate next in Section \ref{subsubsec:swept-comb-lowsat} that the development of wide velocity spectral coherence only occurs (for a smooth distribution with a swept inversion) when the sample enters the non-linear SR regime.

\subsubsection{Third experiment: initial population inversion below the critical threshold}\label{subsubsec:swept-comb-lowsat}

\begin{figure}
    \centering
    \includegraphics[width=1.\columnwidth, trim=.4cm .8cm .5cm .4cm, clip]{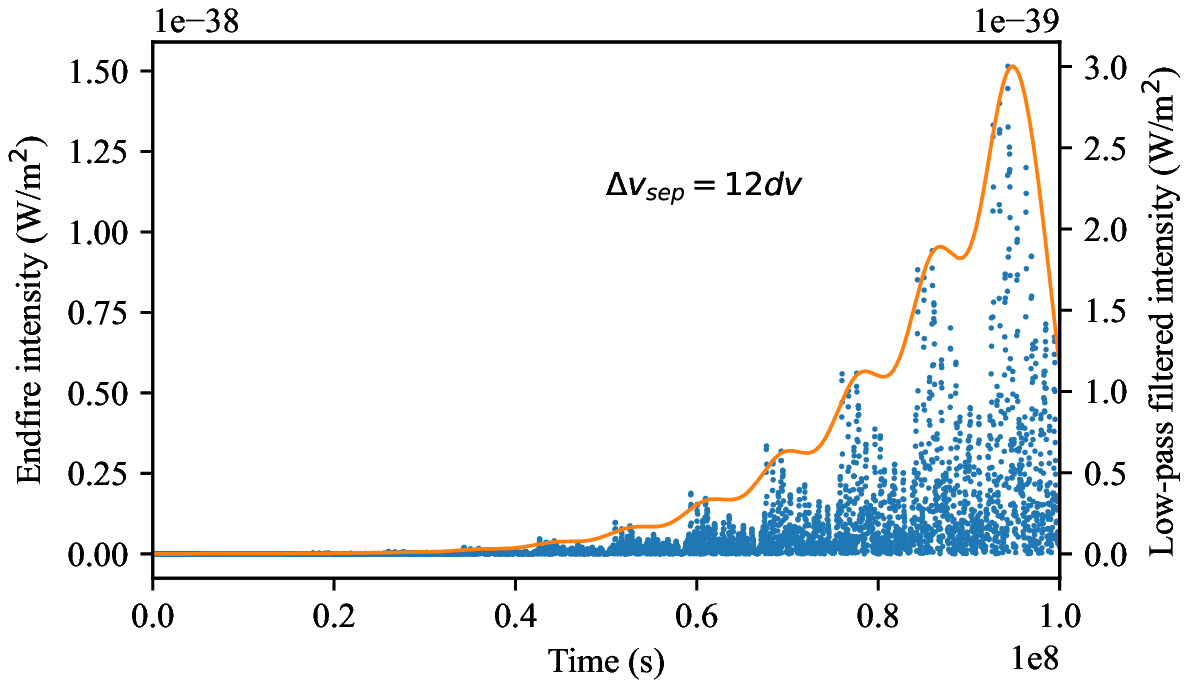}
    \includegraphics[width=1.\columnwidth, trim=.4cm .8cm .5cm .4cm, clip]{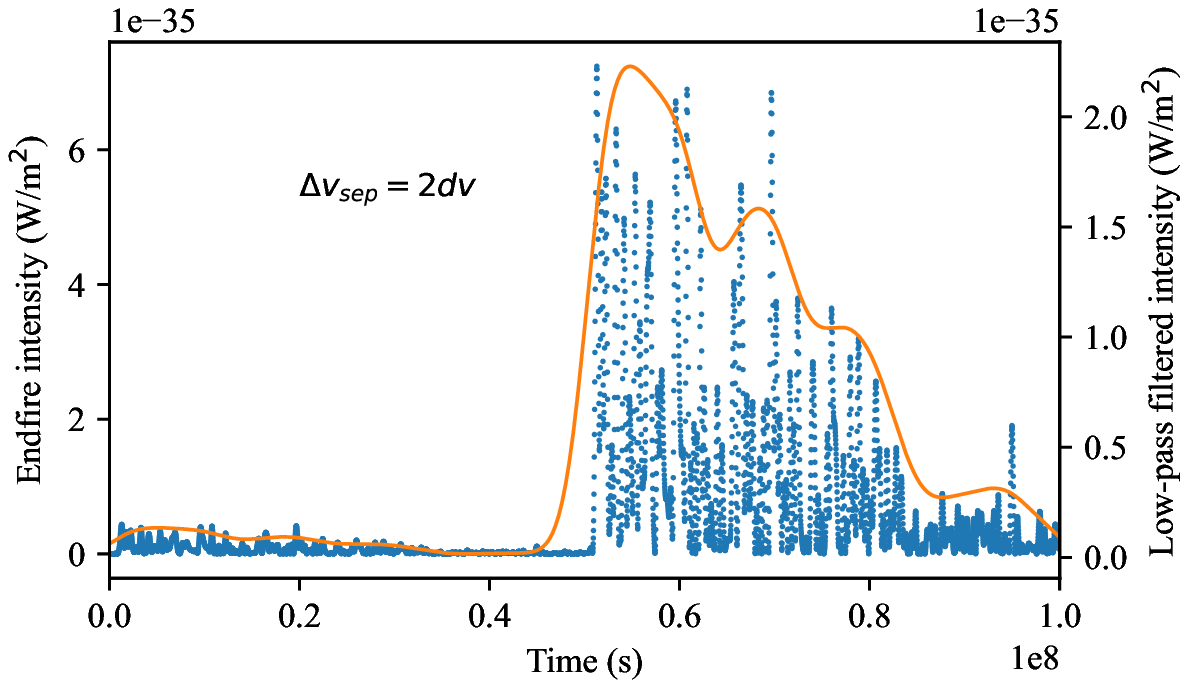}
    \includegraphics[width=1.\columnwidth, trim=.4cm .5cm .3cm .4cm, clip]{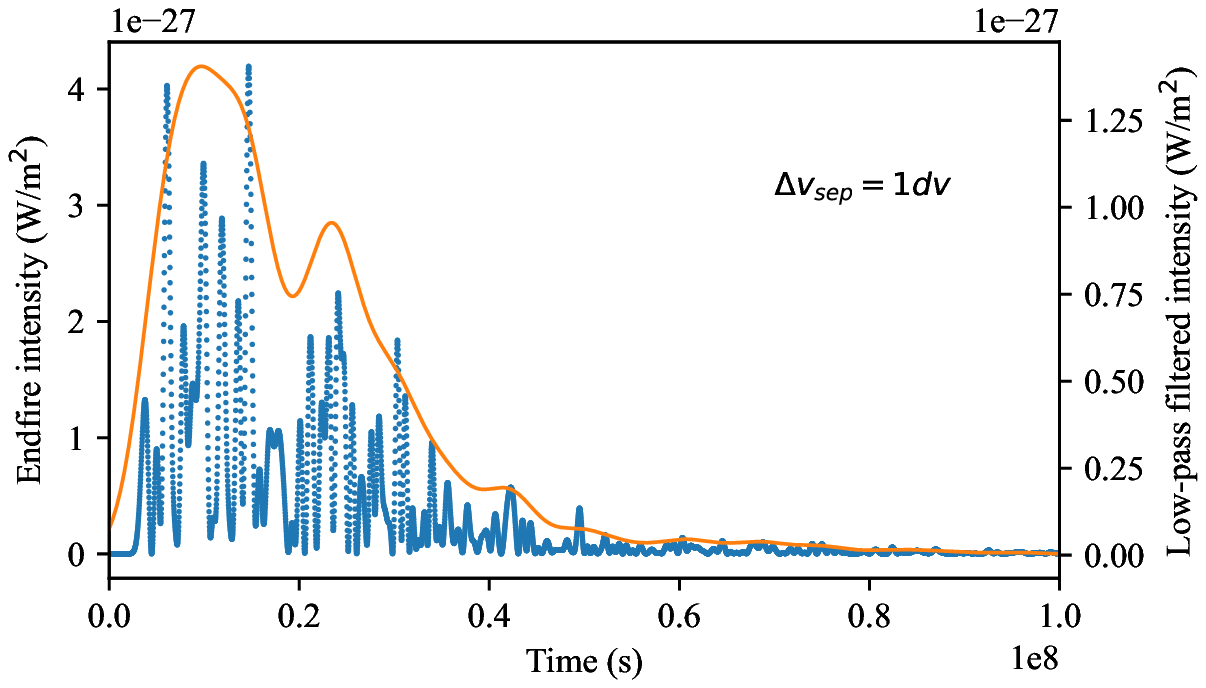}
    \caption{Intensity transients at the endfire of a comb distribution of 161 channels and initial population inversion $N_{0,\textrm{ low}} = 3.3\times {10}^{-8} \textrm{ m}^{-3}$ for velocity separations $\Delta v_\textrm{sep}/dv=12,\:2,\textrm{ and } 1$ (top to bottom). Filters of all panels equal to those described in text accompanying Figure \ref{fig:transverse} (sliding 20-channel FFT band mask). Wide spectral correlation is no longer found to develop in the final transition from  $\Delta v_\textrm{sep} = 2 dv$ to $\Delta v_\textrm{sep} = dv$.}
    \label{fig:comb_161channels_lowsat_toothsepdecreasing}
\end{figure}

We now set our initial population inversion in each velocity channel to $N_{0,\textrm{ low}} = N_{0,\textrm{ med}}/2 = 3.3\times {10}^{-8} \textrm{ m}^{-3}$, which is below the critical threshold required to initiate a non-linear SR process. The resulting endfire intensities generated by comb distributions of $\Delta v_\textrm{sep}/dv = 12,\:2,\textrm{ and }1$ are shown from top to bottom in Figure \ref{fig:comb_161channels_lowsat_toothsepdecreasing}, respectively. In the terminal continuum case, where $\Delta v_\textrm{sep} = 1dv$ (Figure \ref{fig:comb_161channels_lowsat_toothsepdecreasing}, bottom), the transient response possesses a finite temporal duration in its exponential decay with timescale equal to $T_2 / 2$. Additionally, the population inversion never departs from the linear regime ($N_\omega \approx N_{0,\textrm{ low}}$ for all $\omega \textrm{ and } z$ throughout the simulation).

This result suggests that increasing the initial population inversion above that critical threshold which initiates a non-linear SR process induces not only a change in radiative process, but also a ``phase transition'' in the sample's ability to develop wide velocity bandwidth phase correlation. We make this concept more precise and study it further in Section \ref{subsec:cont-polphases}.


\section{Swept Inversion Processes Part II: \\ \hspace{11 pt} Continuum Distributions}\label{sec:swept-cont}

The terminal cases of Sections \ref{subsubsec:swept-comb-hisat}, \ref{subsubsec:swept-comb-medsat}, and \ref{subsubsec:swept-comb-lowsat} with $\Delta v_\textrm{sep} = 1dv$ described continuum distributions, but represented exceedingly small fractions of realistic velocity distributions as discussed in our introductory remarks of Section \ref{sec:intro}. We now turn to simulate much wider velocity distributions which capture the behaviour of a system in its transition to the WDB limit. Our objective is to reach conclusions regarding a sample's peak intensity scaling, its transient profile, and its ability to sustain temporal SR structure in the WDB limit under various configurations.

\subsection{Polarisation phase correlations}\label{subsec:cont-polphases}

\begin{figure}
    \centering
    \includegraphics[width=1.\columnwidth, trim=.4cm .4cm .4cm .3cm, clip]{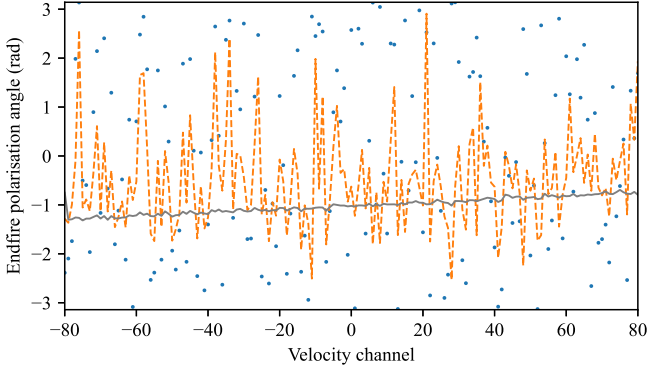}
    \caption{Endfire polarisation phases at three successive times for a sample slightly above the SR critical threshold (Section \ref{subsubsec:swept-comb-medsat}). Scatter: $\tau=0$; dashed: $\tau=T/600$; solid: $\tau=T/300$. Wide polarisation phase correlation quickly develops above the SR critical threshold.}
    \label{fig:comb_161channels_medsat_polphases}
\end{figure}

The quenching of SR temporal duration is related to the bandwidth of polarisation phase correlation which develops across velocity channels during the transient SR process. In Figure \ref{fig:comb_161channels_medsat_polphases} we show the polarisation phases across all velocity channels at three successive moments ($\tau=0$, $\tau=T/600$, and $\tau=T/300$) during the initial stages of the simulation executed in Section \ref{subsubsec:swept-comb-medsat}. The initial uniform random distribution of polarisation phases over $\left[-\pi,\pi\right]$ quickly converges to a highly correlated distribution.

We quantify the development of phase correlation in a simulation as a function of the initial population inversion by first defining a so-called critical coherence time $T_\textrm{cc}$, as follows. Suppose that a distribution is simulated which spans some total velocity extent $\Delta v_\textrm{tot}$ that is a multiple $n_\textrm{ch}$ of the fundamental velocity interval $dv = 1/T$ (for example, the experiments of Sections \ref{subsubsec:swept-comb-hisat}, \ref{subsubsec:swept-comb-medsat}, and \ref{subsubsec:swept-comb-lowsat} each simulated distributions with $n_\textrm{ch} = 160$). If all channel polarisations are initiated in phase with each other, and if all channels evolve independently, then channel polarisations will interfere constructively until approximately $\tau = T/\left(2n_\textrm{ch}\right)$. We therefore define a critical coherence time for a simulation of duration $T$ and distribution extent $\Delta v_\textrm{tot} = n_\textrm{ch} dv$ as $T_\textrm{cc} = T/\left(2n_\textrm{ch}\right)$.

\begin{figure}
    \centering
    \includegraphics[width=1.\columnwidth, trim=.4cm .4cm .4cm .3cm, clip]{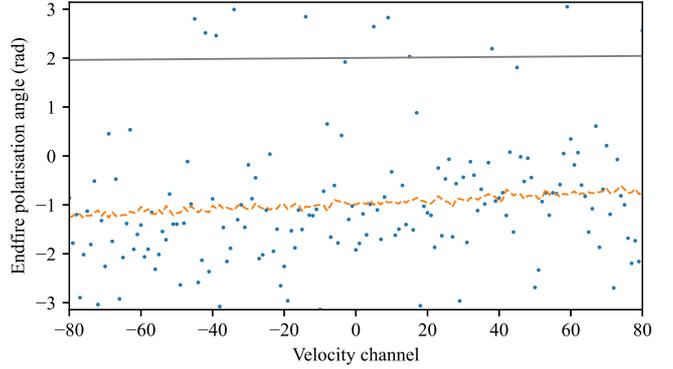}
    \caption{Endfire polarisation phases at the critical coherence time $T_\textrm{cc} = 3.1 \times 10^5 \textrm{ s}$ for initial population inversion density below the SR critical threshold (scatter, see Section \ref{subsubsec:swept-comb-lowsat}), slightly above the SR critical threshold (dashed, see Section \ref{subsubsec:swept-comb-medsat}), and well above the SR critical threshold (solid, see Section \ref{subsubsec:swept-comb-hisat}). Below the SR critical threshold polarisation phase correlation is nearly absent, while above the SR critical threshold global polarisation phase correlation emerges.}
    \label{fig:comb_161channels_allsats_polphases}
\end{figure}

Although neither of the aforementioned conditions of in-phase polarisation initiation or channel independence are true of our simulations, $T_\textrm{cc}$ defines a meaningful timescale over which to assess the development of polarisation phase correlation. If wide polarisation phase correlation develops within a sample at or before $T_\textrm{cc}$, then we expect the system dynamics to be dominated by a strong, coherent electric field. Figure \ref{fig:comb_161channels_allsats_polphases} shows the polarisation phases at $T_\textrm{cc}$ for each of the three experiments of Sections \ref{subsubsec:swept-comb-hisat}, \ref{subsubsec:swept-comb-medsat}, and \ref{subsubsec:swept-comb-lowsat}. It is clearly apparent that the polarisation phases are widely correlated at $T_\textrm{cc}$ only for those experiments above the critical threshold (Sections \ref{subsubsec:swept-comb-hisat} and \ref{subsubsec:swept-comb-medsat}).

\begin{figure}
    \centering
    \includegraphics[width=1.\columnwidth, trim=.4cm .8cm .4cm .4cm, clip]{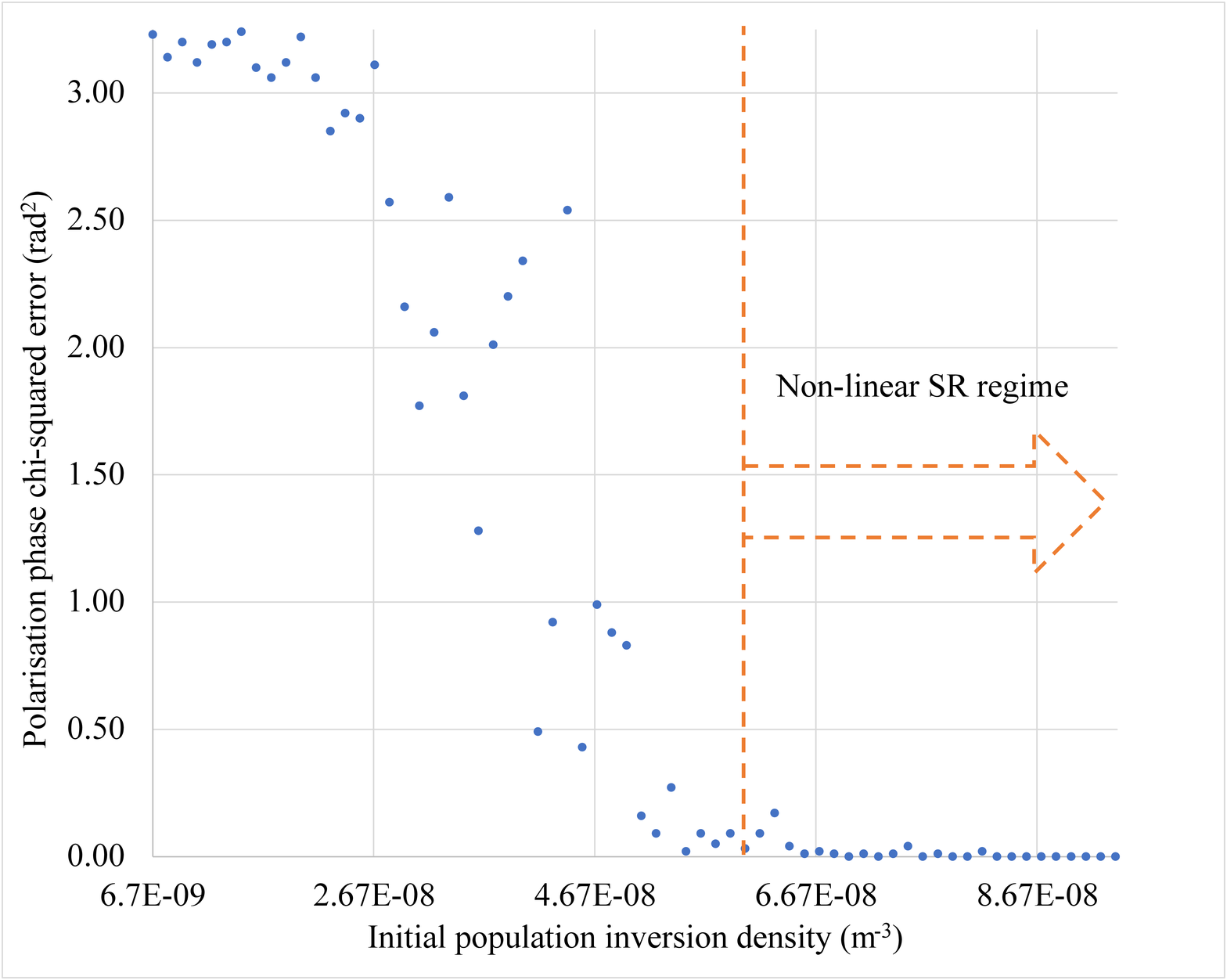}
    \caption{Chi-squared error in a linear fit to the polarisation angles across velocity channels as a function of initial population inversion density. A clear transition to wide polarisation phase correlation occurs as a sample's initial population inversion is made to cross the SR critical threshold.}
    \label{fig:comb_1281channels_polphase_chisq}
\end{figure}

The introduction of $T_\textrm{cc}$ helps us visualise a transition in the polarisation phase correlation as the initial population inversion density is increased. Referring to Figure \ref{fig:comb_161channels_allsats_polphases}, we identify the development of wide polarisation phase correlation with a low chi-squared error in a linear fit to the polarisation phases across velocity channels. In Figure \ref{fig:comb_1281channels_polphase_chisq} we therefore plot, as a function of initial population inversion density, the mean chi-squared error in a linear fit to the polarisation phases at $T_\textrm{cc}$ across all velocity channels. Figure \ref{fig:comb_1281channels_polphase_chisq} is in fact generated from simulations of distributions possessing 1281 velocity channels; remarkably, polarisation phase correlation develops across the entire velocity extent as long as the initial population inversion density exceeds the non-linear SR critical threshold of approximately $6\times 10^{-8} \textrm{ m}^{-3}$.

\subsection{Candidate stochastic distributions for sustaining temporal structure}\label{subsec:cont-candidates}

It is important to recognise that the transition preceding the onset of the non-linear SR regime when moving from left to right in figure \ref{fig:comb_1281channels_polphase_chisq} describes a global correlation emerging from a system with only local physical interactions. In Section \ref{subsec:cont-polphases} we demonstrated the emergence of global polarisation phase correlation when $N_0$ exceeded $N_{0,\textrm{ crit}} \approx 6 \times 10^{-8} \textrm{ m}^{-3}$ in each velocity channel, and in Section \ref{subsec:swept-two} we demonstrated that the Doppler-broadened MB equations do not explicitly couple distant velocity channels. The global phase correlation in Figure \ref{fig:comb_1281channels_polphase_chisq} must therefore be a consequence of correlations propagating transitively through the connectivity of the full velocity distribution.

The picture is somewhat analogous to a phase transition in a crystal lattice, although such an analogy is not strictly true due to the finite length over which the transition takes place. Still, the analogy of defects in a crystal lattice suggests potential WDB configurations which may sustain SR temporal structure: because the global phase correlation is a transitive phenomenon, it may be eliminated by ``breaking'' the connectivity of the velocity distribution. We therefore seek to establish discontinuities in our velocity distribution analogous to defects in a crystal lattice; such defects facilitate the development of numerous correlated sub-regions in a lattice which are globally decoherent; similarly, we expect ``defects'' in a WDB velocity distribution to limit the correlation between velocity channels to finite neighbourhoods.

We introduce discontinuities into our WDB distribution by considering a stochastic velocity distribution; however, $F_\omega$ must possess certain statistical properties if it is to inhibit the connectivity of the distribution and thereby sustain finite SR temporal structure. If typical variations in $F_\omega$ occur over too short of velocity bandwidths, then coupling between velocity channels will simply extend over the variations and the distribution will continue to develop global correlations; on the other hand, if typical variations in $F_\omega$ occur over very wide velocity bandwidths, then sub-correlated regions which develop within a typical variation width will be of such wide bandwidths that their temporal structures will each individually collapse. Finally, standard deviations in $F_\omega$ must be of sufficient amplitude to facilitate SR in some regions while eliminating it in others; that is, they must be on the order of the phase transition length such as, for example, the region from $2.5\times 10^{-8} \textrm{ m}^{-3}$ to $5.0\times 10^{-8} \textrm{ m}^{-3}$ on the horizontal axis of Figure \ref{fig:comb_1281channels_polphase_chisq}.

These constraints are admittedly fine-tuned, and it is not our present objective to investigate the feasibility of such configurations occurring in nature; however, three comments may be made in this regard. First, turbulence in the ISM may provide a physical mechanism for developing a stochastic $F_\omega$, although the bandwidth and amplitude of velocity distribution variations induced by the turbulent property of intermittency requires further research. Second, although the extremely long time scales of the examples provided in this paper imply a demand for very small bandwidth velocity variations, they also provide very wide velocity extents over which such variations could somewhere occur. That is, although the vast majority of a WDB velocity distribution might not possess the specific statistical characteristics outlined above, we pointed out in our opening remarks of Section \ref{sec:intro}) that simulations of this paper describe only a negligible fraction of a full distribution; thus, even if the majority of a real distribution is incapable of supporting SR transient processes, it is not (as of yet) unreasonable to suppose that the fine-tuned statistical requirements might be realised at least somewhere within the distribution. Third, in the event that further research into the particular long time scale examples of this paper should find these fine-tuned constraints practically unfeasible, the following results are applicable to other WDB systems which may demonstrate SR processes over shorter time scales and thereby demand less finely tuned noise statistics.

\subsection{Stochastic velocity distribution simulations}\label{subsec:cont-noisysims}

\begin{figure}
    \centering
    \includegraphics[width=1.\columnwidth, trim=.4cm .2cm .4cm .4cm, clip]{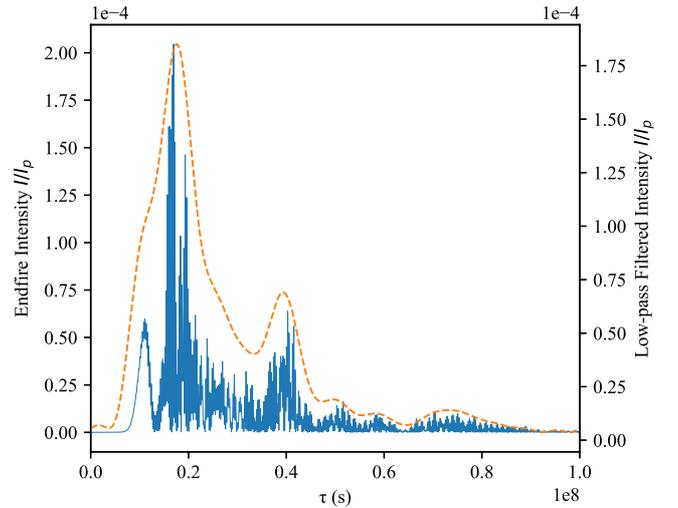}
    \caption{Endfire intensity transient generated from a CTD simulation of a stochastic distribution $1{\small,}023$ channels wide. Normalised to $I_\textrm{p}$. Filter equal to that described in text accompanying Figure \ref{fig:transverse} (sliding 20-channel FFT band mask). The stochastic distribution is found to sustain temporal SR structure in the WDB limit even in response to a swept inversion.}
    \label{fig:ctd_nsch511_intensity}
\end{figure}

Motivated by the discussion of Section \ref{subsec:cont-candidates}, we investigate now the response of a noisy WDB velocity distribution. We suspect that a distribution possessing appropriate statistical characteristics will inhibit the development of global phase correlations and thereby justify the application of the SF algorithm, but we must verify this result with a CTD simulation first. We therefore begin with a CTD simulation of a noisy distribution of moderate velocity extent ($1{\small,}023$ channels) constructed to satisfy the statistical requirements of Section \ref{subsec:cont-candidates} as follows. First, the distribution $F_\omega$ is a normal distribution of mean $N_{0\textrm{ crit}}$ and standard deviation $(2/3)N_{0\textrm{ crit}}$; second, the bandwidth of the noise is reduced via convolution with a Gaussian of width $\sigma$ equal to $20$ velocity channels. The resulting intensity shown in Figure \ref{fig:ctd_nsch511_intensity} indeed sustains finite temporal structure as expected.

\begin{figure}
    \centering
    \includegraphics[width=1.\columnwidth, trim=.4cm .5cm .4cm .3cm, clip]{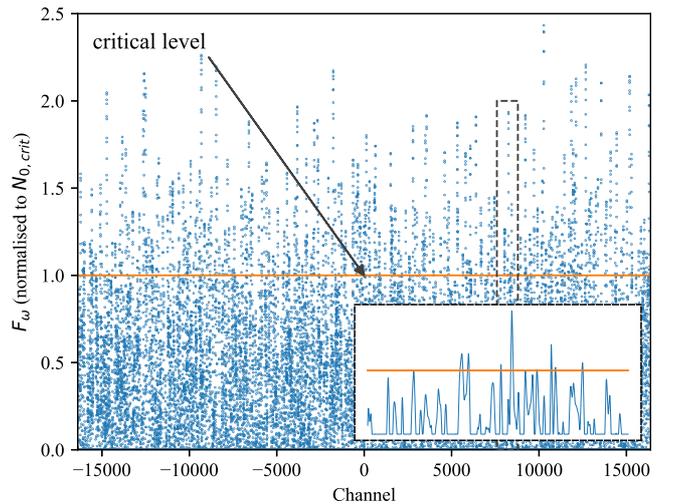}
    \caption{Wide distribution of $32{\small,}767$ total velocity channels. Inset: zoom into the portion of the stochastic distribution highlighted, showing the finite bandwidth structure to its statistics.}
    \label{fig:sfa_fv}
\end{figure}

The distribution generating Figure \ref{fig:ctd_nsch511_intensity} does not represent a statistically significant total velocity extent. The reduction of noise bandwidth in $F_\omega$ by $\sigma=20$ effectively modifies the distribution to represent on the order of $n_\textrm{total channels} / \sigma \approx 50$ sub-groups of varying population sizes. Although the sustained temporal structure demonstrated in Figure \ref{fig:ctd_nsch511_intensity} implies the independence of these sub-groups, accurate statistical statements regarding intensity transients in the WDB limit cannot be made without simulating a much wider distribution. We therefore now widen our total velocity extent by a factor of $32$ to the distribution $F_\omega$ of $32{\small,}767$ total velocity channels shown in Figure \ref{fig:sfa_fv}. The zoomed inset panel shows the limited bandwidth structure to the statistics in $F_\omega$.

\begin{figure}
    \centering
    \includegraphics[width=1.\columnwidth, trim=.4cm .2cm .4cm .4cm, clip]{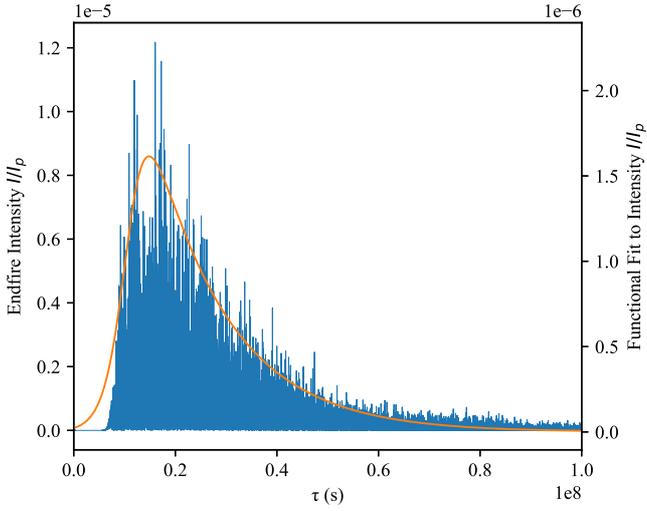}
    \caption{Endfire intensity transient generated from the distribution shown in Figure \ref{fig:sfa_fv}. The FFT-based SF algorithm was used with a rectangular kernel of width $160 dv$. Normalised to $I_\textrm{p}$, smooth curve corresponds to functional fit (see text). A pulse-like temporal structure is sustained in the WDB limit.}
    \label{fig:sfa_nsch16383_intensity}
\end{figure}

The CTD simulation of Figure \ref{fig:ctd_nsch511_intensity} was generated in approximately 16 minutes on a 2018 CPU, while the same distribution would take only $11$ minutes to simulate with the SF algorithm. This performance gap widens dramatically as we expand to $32{\small,}767$ channels: the CTD algorithm would take more than $11$ days to simulate such a distribution, while the SF algorithm simulates it in under $6$ hours. The resulting intensity transient is shown in Figure \ref{fig:sfa_nsch16383_intensity}.

In order to investigate the behaviour of temporal structure and peak intensity in the WDB limit, we define a simple functional form with three free parameters to fit against the intensity transients generated. We expect the intensity to decay at large times with time constant $T_2/2$, and therefore choose the envelope factorisation
\begin{equation}
    I_\textrm{fit} = I_\textrm{amp} S_\textrm{w}\left(\tau-\tau_0\right) e^{-2\tau/T_2},
\end{equation}
where $I_\textrm{amp}$, $w$, and $\tau_0$ are fitting parameters. The envelope $S_\textrm{w}$ should smoothly merge into a constant value after some time and should tend to zero as $\tau \rightarrow -\infty$. A simple choice of $S_\textrm{w}$ is the $S$-shaped curve generated by a hyperbolic tangent of varying offset $\tau_0$ and characteristic width $w$; specifically,
\begin{equation}
    S_\textrm{w}\left(\tau-\tau_0\right) = 1+\tanh\left(\frac{\tau-\tau_0}{w}\right).
\end{equation}
A fit of this form is superimposed on the transient of Figure \ref{fig:sfa_nsch16383_intensity} and provides a peak intensity estimate of $1.7\times{10}^{-6} I_\textrm{p}$ at a time of $1.5\times{10}^{7} \textrm{ s}$ for the simulation of $32{\small,}767$ total velocity channels.

The finite delay to peak intensity is an important statistical result. The number of velocity channels simulated may be very large as a number of fundamental interacting velocity slices, but remains small on the scale of observation bandwidth. For the present system, $32{\small,}767$ velocity channels corresponds to a measurement bandwidth of only $32{\small,}767 / T \approx 3\times{10}^{-4} \textrm{ Hz}$ (recall that the velocity slices, indexed by their Doppler-shifted natural angular frequencies, are discretised by the fundamental angular frequency differential $d\omega = 2\pi/T$). An observer averaging the power across some finite bandwidth about some central observation frequency over some observation time would measure a pulse emerging after a delay on the order of $1.5\times{10}^{7} \textrm{ s}$. This delay is the statistical outcome of many decoherently interfering SR processes developing in small, independent sub-groups of velocity channels across the bandwidth observed.

The large number statistics of the ensemble of fine bandwidth SR processes of varying amplitudes and delays ultimately defines the total transient of Figure \ref{fig:sfa_nsch16383_intensity}; however, due to the finite bandwidth of the measuring apparatus these sub-processes would be unidentifiable to an observer (at least for the present system). It is nonetheless informative to inspect the transients of narrow bandwidth windows at various locations across the velocity distribution for two reasons. First, observation of SR sub-processes within the distribution confirms the physical intuition we have built regarding independent sub-samples decoupled by the discontinuity of the distribution. Second, these results would be applicable to other WDB SR systems of much shorter time scales, wherein the observation bandwidth required to observe the SR sub-processes may widen to practically achievable levels. For example, the fast radio bursts modelled in \citet{Houde2018a, Houde2019} are of durations on the order of milliseconds.

\begin{figure}
    \centering
    \includegraphics[width=1.\columnwidth, trim=.4cm .4cm .4cm .4cm, clip]{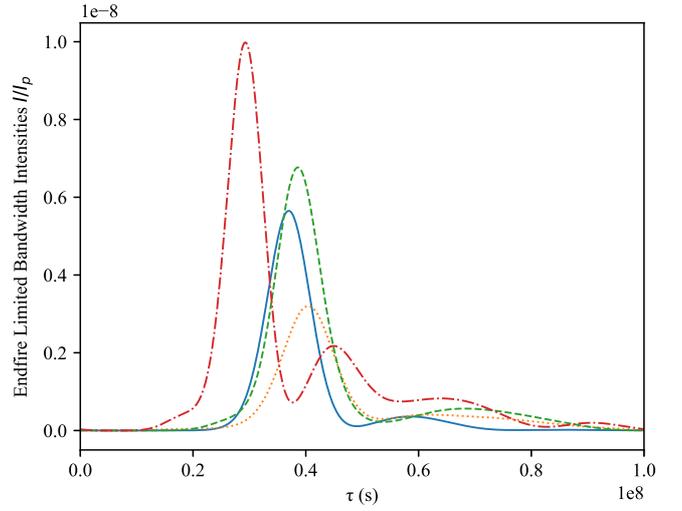}
    \caption{Specific endfire intensities from finite bandwidths centred upon four arbitrary frequencies across the total spectrum. Extracted from the simulation of Figure \ref{fig:sfa_nsch16383_intensity}. Linetypes vary for visual differentiation and not for identification. Transient SR processes of varying saturation are observed across the spectrum, and the large-number statistics of many such underlying processes is responsible for the emergent transient shown previously in Figure \ref{fig:sfa_nsch16383_intensity}.}
    \label{fig:sfa_temporal_struc}
\end{figure}

We choose four central channel indices $-9,000$, $-3,000$, $3,000$, and $9,000$ (well separated across the distribution) and truncate the FFT of the electric field to a neighbourhood of angular frequencies $\pm 10 d\omega$ about the natural frequency of each central channel. We plot the resulting four intensity transients in Figure \ref{fig:sfa_temporal_struc}. The observed transients are clearly SR processes, each demonstrating the salient SR features of a delay to peak intensity, a peak intensity duration proportional to the particular delay realised, and an ensuing ringing thereafter.

\begin{figure}
    \centering
    \includegraphics[width=1.\columnwidth, trim=.4cm .2cm .4cm .4cm, clip]{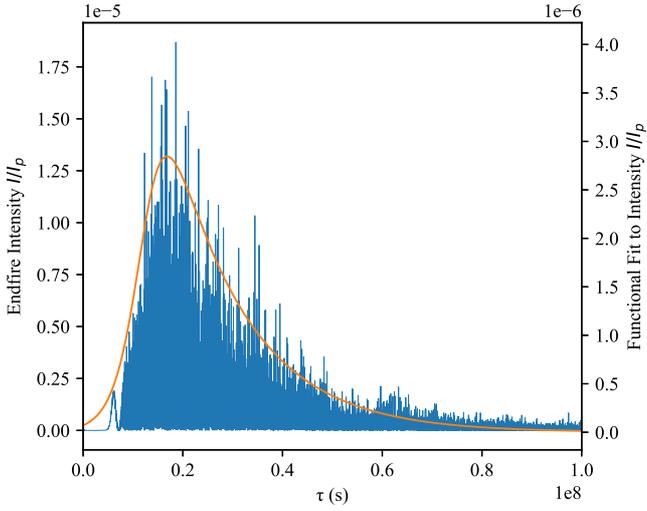}
    \caption{Endfire intensity transient generated from a distribution of $16{\small,}383$ channels. The FFT-based SF algorithm was used with a rectangular kernel of width $160 dv$.  Normalised to $I_\textrm{p}$, smooth curve corresponds to functional fit (see text). The emergent amplitude is half that of Figure \ref{fig:sfa_nsch16383_intensity} (where in drawing such a comparison we have accounted for the fact that $I_\textrm{p}$ varies quadratically with the total population).}
    \label{fig:sfa_nsch8191_intensity}
\end{figure}

Although the pulse of Figure \ref{fig:sfa_nsch16383_intensity} does not demonstrate ringing, we can investigate other scaling features against those of a resonant (single velocity) SR system. First, we expect the system to be decoherent in the total velocity width simulated; we therefore reduce the distribution width by a factor of two while leaving the number of molecules per velocity differential of the distribution constant. The resulting intensity transient is shown in Figure \ref{fig:sfa_nsch8191_intensity} and possesses a peak intensity that is indeed half that of Figure \ref{fig:sfa_nsch16383_intensity}. This is a trivial result in light of our construction of the distribution, which intentionally eliminated wide correlations between velocity channels; still, it provides a good consistency check.

\begin{figure}
    \centering
    \includegraphics[width=1.\columnwidth, trim=.4cm .2cm .5cm .4cm, clip]{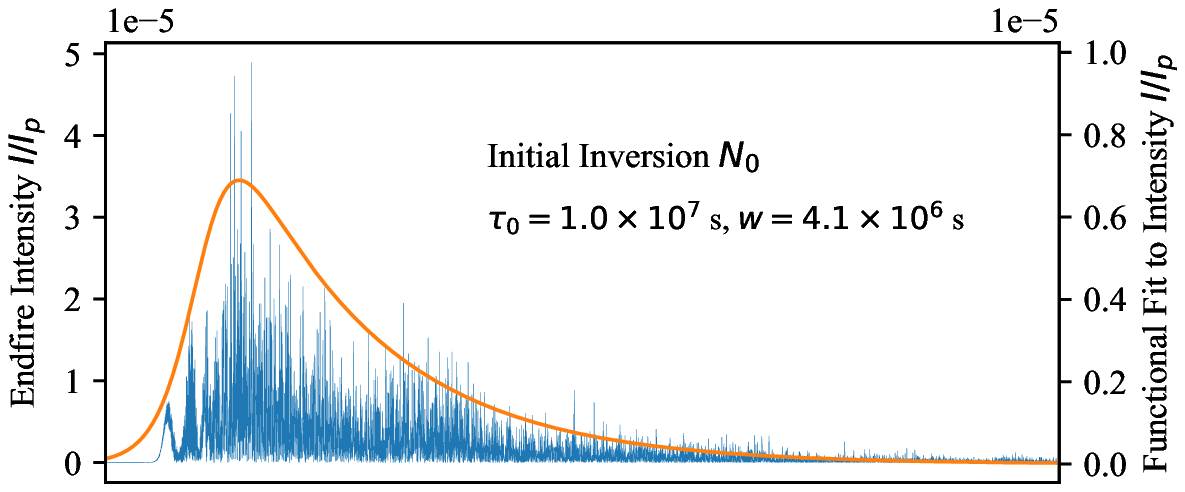}
    \includegraphics[width=1.\columnwidth, trim=.4cm .2cm .5cm .2cm, clip]{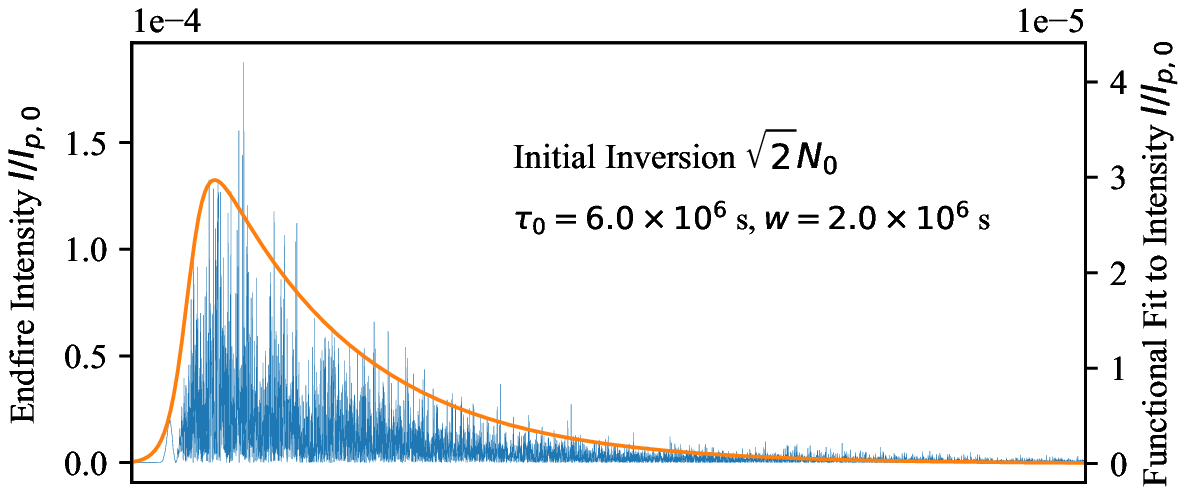}
    \includegraphics[width=1.\columnwidth, trim=.4cm .2cm .5cm .2cm, clip]{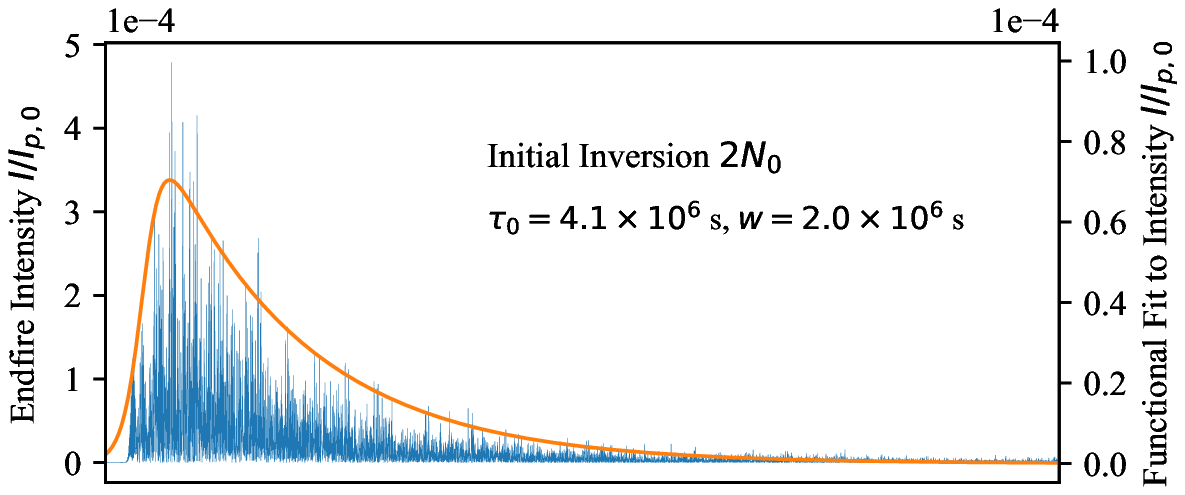}
    \includegraphics[width=1.\columnwidth, trim=.5cm .2cm .4cm 0cm, clip]{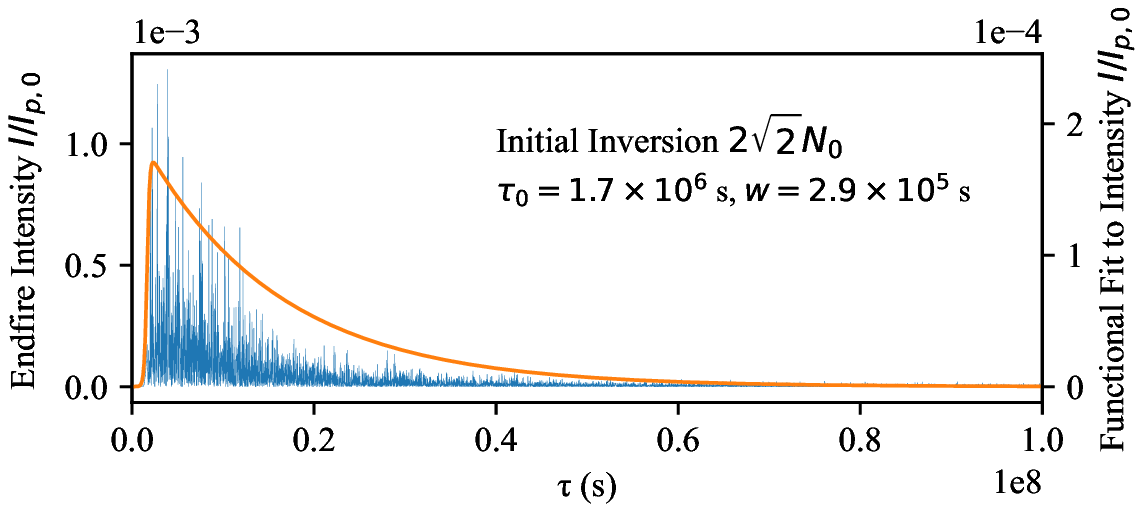}
    \caption{Endfire intensity transients generated from four simulations of $8,191$ channels. Initial population inversion per velocity differential increases by a factor of $\sqrt{2}$ in each successive panel, from top to bottom. The FFT-based SF algorithm was used with a rectangular kernel $s_\delta$ of width $\delta = 160 dv$ in the top three panels, and increased to $240dv$ in the bottom panel (see discussion).  Normalised to $I_\textrm{p,0}=I_\textrm{p} \textrm{ of the top panel}$, smooth curves correspond to functional fit (see text).}
    \label{fig:sfa_nsch4095_progression}
\end{figure}

We next investigate the dependency of the intensity transient upon increasing initial population inversion with fixed total velocity extent. We simulate four successive systems of $8,191$ channels, starting from a distribution possessing the same density of population inversion per velocity differential and statistical characteristics as simulated previously (a cropped portion of Figure \ref{fig:sfa_fv}). We define $I_\textrm{p,0}$ as that $I_\textrm{p}$ corresponding to the first simulation, shown in the top panel of Figure \ref{fig:sfa_nsch4095_progression}. We proceed to increase the initial population inversion by successive factors of $\sqrt{2}$, plotting the resulting intensity transients in units of $I_\textrm{p,0}$. The first three experiments, depicted in the top three panels of Figure \ref{fig:sfa_nsch4095_progression}, in fact demonstrate greater than quadratic scaling in their peak intensities. This is due to the fact that the starting distribution $F_\omega$ only slightly exceeds the critical threshold in rare locations; increasing the initial population inversion thus increases not only the saturation, but also the number of sub-groups within the SR regime. The fourth transient of Figure \ref{fig:sfa_nsch4095_progression}, where the majority of sub-groups within the distribution are now in the SR regime, indeed demonstrates quadratic scaling in its peak intensity equal to twice that of the prior transient above it.

The progression of intensity transients in Figure \ref{fig:sfa_nsch4095_progression} demonstrates two important temporal structure features. First, the delay to peak intensity is inversely proportional to the initial population inversion, just as it is in a resonant SR sample. Second, there appears to be an abrupt transition in the profile shape of the fourth plot: whereas the top three transients are relatively slowly and smoothly increasing pulses (ignoring stochastic noise), the fourth transient shows a delay followed by an abrupt onset of intensity which thereafter decays exponentially with time constant $T_2 / 2$. These features are most clearly visible in the functional fits to the noisy transients, which possess sufficient degrees of fitting freedom to capture the change in profile shape and are indeed seen to abruptly change between the third and fourth plots; notice specifically the sudden transition in the fit parameter $w$.

The abrupt transition in Figure \ref{fig:sfa_nsch4095_progression} may be understood in the context of the statistical properties of $F_\omega$. The stochastic velocity distribution of Figure \ref{fig:sfa_fv} was constructed via convolution with a Gaussian of width $\sigma$ on the order of the bandwidth of an intensity transient generated by a sample just above the non-linear SR threshold (see, for example, Figure \ref{fig:two_distrns_vsep64}). In the phase transition discussion closing Section \ref{subsec:cont-candidates}, we proposed that ``defects'' in the continuity of $F_\omega$ should occur over the spectral interaction distance (see Section \ref{subsec:swept-two}) if temporal structure is to be retained in the WDB limit. As $N_0$ is raised from the third to the fourth panel of Figure \ref{fig:sfa_nsch4095_progression}, however, the higher saturation extends the bandwidth of typical SR transients developing within sub-groups of the distribution to become larger than the defect scale $\sigma$. In other words, the distribution is nearly smooth on the scale relevant to the higher saturation SR processes of the fourth simulation. Note that the SF algorithm required a wider interaction kernel and finer temporal step size in order to converge to a solution in the fourth transient of Figure \ref{fig:sfa_nsch4095_progression} (see caption), which is consistent with this concept of wider spectral interaction distance occurring within the system.


\section{Summary and Future Work}\label{sec:summary}

We demonstrated that a sample inverted by a transverse pump will sustain a finite intensity duration in the WDB limit whenever its length is comparable to the corresponding cooperation length of the non-linear SR emission process. At such a length scale and under such a pumping process, the emission duration was found to be proportional to the length of the sample.

We demonstrated in Sections \ref{sec:swept-disc} and \ref{sec:swept-cont} that a velocity distribution inverted above the critical threshold $N_{0,\textrm{ crit}}$ yielding non-linear dynamics in the MB equations cannot generically sustain temporal structure in the WDB limit. A smooth distribution inverted above $N_{0,\textrm{ crit}}$ by a swept pumping mechanism will develop a global polarisation phase correlation, which has the effect of quenching temporal structure. During a realistic finite duration swept pumping event the endfire intensity generated by non-linear SR emission within a WDB sample will follow the pump's transient profile.

Having recognised in Section \ref{subsec:swept-two} that the MB equations indeed capture the independence of SR samples separated by at least the bandwidth of the transient process, we attributed the formation of global polarisation correlation in a WDB distribution to transitive coupling across its velocity channels. This interpretation suggested certain statistical characteristics of a velocity distribution which might inhibit the transitive formation of correlation; namely, that variations in the distribution exist over the bandwidth of the transient SR pulse and that the amplitude of said variations be on the order of the saturation phase transition distance. We demonstrated in Section \ref{subsec:cont-noisysims} that such distributions do indeed sustain a finite time delay and duration in their peak intensity response in the WDB limit. Such constructions were admittedly fine-tuned for the long timescale processes described in this paper, but these results also hold for shorter timescale processes which could reasonably realise the necessary statistical characteristics. A turbulent medium demonstrating intermittency in its velocity distribution, or a population not in thermal equilibrium, for example, could both demonstrate finite temporal structure in the WDB limit under a swept inversion above the non-linear SR threshold.

The order of magnitude of an astrophysical SR process timescale strongly affects the application of this work. Even under modest thermal broadening, very long timescale SR processes (such as those modelled in Sections \ref{sec:transverse}, \ref{sec:swept-disc}, and \ref{sec:swept-cont}) must be classified as WDB. Consequently, the temporal duration of such processes is determined by the swept or transverse character of the pumping mechanism.\footnote{We assume for this discussion a smooth velocity distribution.} In such environments the temporal duration of SR emission from a fixed population inversion can vary dramatically, depending upon the orientation of the triggering mechanism; i.e., both short duration radiation flares and long duration emission events can be generated in the same region via different SR triggering mechanisms. Astrophysical processes possessing shorter timescales (not modelled in this paper but existing in the literature as described in Section \ref{sec:intro}), on the other hand, could permit the observation of temporal SR structure in response to even a swept inversion mechanism, as the bandwidth of SR features may extend over a bandwidth resolvable by the measuring apparatus. Finally, in the most trivial case, an SR process may be of such short temporal duration that its bandwidth becomes broader than its thermal broadening extent; such a sample is accurately modelled by the resonant case of the MB equations and thereby renders the methods of this work unnecessary.


\section*{Acknowledgements}
C.W. is supported by the Natural Sciences and Engineering Research Council of Canada (NSERC) through the doctoral postgraduate scholarship (PGS D). F.R.’s research at Perimeter Institute is supported in part by the Government of Canada through the Department of Innovation, Science and Economic Development Canada and by the Province of Ontario through the Ministry of Economic Development, Job Creation and Trade. M.H.'s research is funded through the Natural Sciences and Engineering Research Council of Canada Discovery Grant RGPIN-2016-04460 and the Western Strategic Support for NSERC Success Accelerator program.

\section*{Data Availability Statement}
The data pipeline is made available at: \url{https://github.com/cwyenberg/MandL-to-Superradiance} and maintained by C.M.W. The figures in this paper were prepared using the {\tt matplotlib} package \citep{Hunter2007}.




\bibliographystyle{mnras}
\bibliography{scibib}

\begin{thebibliography}{}
\makeatletter
\relax
\def\mn@urlcharsother{\let\do\@makeother \do\$\do\&\do\#\do\^\do\_\do\%\do\~}
\def\mn@doi{\begingroup\mn@urlcharsother \@ifnextchar [ {\mn@doi@}
  {\mn@doi@[]}}
\def\mn@doi@[#1]#2{\def\@tempa{#1}\ifx\@tempa\@empty \href
  {http://dx.doi.org/#2} {doi:#2}\else \href {http://dx.doi.org/#2} {#1}\fi
  \endgroup}
\def\mn@eprint#1#2{\mn@eprint@#1:#2::\@nil}
\def\mn@eprint@arXiv#1{\href {http://arxiv.org/abs/#1} {{\tt arXiv:#1}}}
\def\mn@eprint@dblp#1{\href {http://dblp.uni-trier.de/rec/bibtex/#1.xml}
  {dblp:#1}}
\def\mn@eprint@#1:#2:#3:#4\@nil{\def\@tempa {#1}\def\@tempb {#2}\def\@tempc
  {#3}\ifx \@tempc \@empty \let \@tempc \@tempb \let \@tempb \@tempa \fi \ifx
  \@tempb \@empty \def\@tempb {arXiv}\fi \@ifundefined
  {mn@eprint@\@tempb}{\@tempb:\@tempc}{\expandafter \expandafter \csname
  mn@eprint@\@tempb\endcsname \expandafter{\@tempc}}}

\bibitem[\protect\citeauthoryear{{Arecchi} \& {Courtens}}{{Arecchi} \&
  {Courtens}}{1970}]{Arecchi1970}
{Arecchi} F.~T.,  {Courtens} E.,  1970, \mn@doi [\pra]
  {10.1103/PhysRevA.2.1730}, \href
  {http://adsabs.harvard.edu/abs/1970PhRvA...2.1730A} {2, 1730}

\bibitem[\protect\citeauthoryear{Benedict et~al.}{Benedict
  et~al.}{1996}]{Benedict1996}
Benedict M.~G.,  et~al., 1996, Super-radiance: Multiatomic Coherent Emission.
IOP Publishing Ltd

\bibitem[\protect\citeauthoryear{Bracewell}{Bracewell}{1978}]{Bracewell1978}
Bracewell R.,  1978, The Fourier Transform and its Applications, second edn.
McGraw-Hill Kogakusha, Ltd., Tokyo

\bibitem[\protect\citeauthoryear{Dicke}{Dicke}{1954}]{Dicke1954}
Dicke R.~H.,  1954, \mn@doi [Phys. Rev.] {10.1103/PhysRev.93.99}, \href
  {http://adsabs.harvard.edu/abs/1954PhRv...93...99D} {93, 99}

\bibitem[\protect\citeauthoryear{Feld \& MacGillivray}{Feld \&
  MacGillivray}{1980}]{Feld1980}
Feld M.,  MacGillivray J.,  1980, in , Coherent Nonlinear Optics.
Springer, pp 7--57

\bibitem[\protect\citeauthoryear{Gross \& Haroche}{Gross \&
  Haroche}{1982}]{Gross1982}
Gross M.,  Haroche S.,  1982, \physrep, 93, 301

\bibitem[\protect\citeauthoryear{Houde, Mathews  \& Rajabi}{Houde
  et~al.}{2018}]{Houde2018a}
Houde M.,  Mathews A.,   Rajabi F.,  2018, \mn@doi [\mnras]
  {10.1093/mnras/stx3205}, \href
  {http://adsabs.harvard.edu/abs/2018MNRAS.475..514H} {475, 514}

\bibitem[\protect\citeauthoryear{{Houde}, {Rajabi}, {Gaensler}, {Mathews}  \&
  {Tranchant}}{{Houde} et~al.}{2019}]{Houde2019}
{Houde} M.,  {Rajabi} F.,  {Gaensler} B.~M.,  {Mathews} A.,   {Tranchant} V.,
  2019, \mn@doi [\mnras] {10.1093/mnras/sty3046}, \href
  {http://adsabs.harvard.edu/abs/2019MNRAS.482.5492H} {482, 5492}

\bibitem[\protect\citeauthoryear{Hunter}{Hunter}{2007}]{Hunter2007}
Hunter J.~D.,  2007, \mn@doi [Computing in Science \& Engineering]
  {10.1109/MCSE.2007.55}, 9, 90

\bibitem[\protect\citeauthoryear{{MacGillivray} \& {Feld}}{{MacGillivray} \&
  {Feld}}{1976}]{MacGillivray1976}
{MacGillivray} J.~C.,  {Feld} M.~S.,  1976, \mn@doi [\pra]
  {10.1103/PhysRevA.14.1169}, \href
  {https://ui.adsabs.harvard.edu/abs/1976PhRvA..14.1169M} {14, 1169}

\bibitem[\protect\citeauthoryear{Menegozzi \& Lamb}{Menegozzi \&
  Lamb}{1978}]{Menegozzi1978}
Menegozzi L.~N.,  Lamb W.~E.,  1978, \mn@doi [Phys. Rev. A]
  {10.1103/PhysRevA.17.701}, 17, 701

\bibitem[\protect\citeauthoryear{Rajabi \& Houde}{Rajabi \&
  Houde}{2016}]{Rajabi2016b}
Rajabi F.,  Houde M.,  2016, \apj, 828, 57

\bibitem[\protect\citeauthoryear{{Rajabi} \& {Houde}}{{Rajabi} \&
  {Houde}}{2017}]{Rajabi2017}
{Rajabi} F.,  {Houde} M.,  2017, \mn@doi [Sci. Adv.] {10.1126/sciadv.1601858},
  \href {http://adsabs.harvard.edu/abs/2017SciA....3E1858R} {3, e1601858}

\bibitem[\protect\citeauthoryear{{Rajabi} \& {Houde}}{{Rajabi} \&
  {Houde}}{2020}]{Rajabi2020}
{Rajabi} F.,  {Houde} M.,  2020, \mn@doi [\mnras] {10.1093/mnras/staa1067},
  \href {https://ui.adsabs.harvard.edu/abs/2020MNRAS.494.5194R} {494, 5194}

\bibitem[\protect\citeauthoryear{Rajabi \& Houde}{Rajabi \&
  Houde}{016a}]{Rajabi2016a}
Rajabi F.,  Houde M.,  2016a, \apj, 826, 216

\bibitem[\protect\citeauthoryear{{Rajabi}, {Houde}, {Bartkiewicz}, {Olech},
  {Szymczak}  \& {Wolak}}{{Rajabi} et~al.}{2019}]{Rajabi2019}
{Rajabi} F.,  {Houde} M.,  {Bartkiewicz} A.,  {Olech} M.,  {Szymczak} M.,
  {Wolak} P.,  2019, \mn@doi [\mnras] {10.1093/mnras/stz074}, \href
  {https://ui.adsabs.harvard.edu/abs/2019MNRAS.484.1590R} {484, 1590}

\bibitem[\protect\citeauthoryear{{Rajabi}, {Chamma}, {Wyenberg}, {Mathews}  \&
  {Houde}}{{Rajabi} et~al.}{2020}]{Rajabi2020b}
{Rajabi} F.,  {Chamma} M.~A.,  {Wyenberg} C.~M.,  {Mathews} A.,   {Houde} M.,
  2020, \mn@doi [\mnras] {10.1093/mnras/staa2723}, \href
  {https://ui.adsabs.harvard.edu/abs/2020MNRAS.498.4936R} {498, 4936}

\bibitem[\protect\citeauthoryear{Skribanowitz, Herman, MacGillivray  \&
  Feld}{Skribanowitz et~al.}{1973}]{Skribanowitz1973}
Skribanowitz N.,  Herman I.~P.,  MacGillivray J.~C.,   Feld M.~S.,  1973,
  \mn@doi [Phys. Rev. Lett.] {10.1103/PhysRevLett.30.309}, 30, 309

\bibitem[\protect\citeauthoryear{Wyenberg, Lankhaar, Rajabi, Chamma  \&
  Houde}{Wyenberg et~al.}{2021}]{Wyenberg2021}
Wyenberg C.~M.,  Lankhaar B.,  Rajabi F.,  Chamma M.~A.,   Houde M.,  2021,
  \mn@doi [MNRAS] {10.1093/mnras/stab2222}, 507, 4464

\makeatother
\end{thebibliography}



\appendix


\section{List of Abbreviations}\label{app:abbr}

\begin{description}
    \item[CTD:] Conventional Time Domain
    \item[IF:] Integral Fourier
    \item[$\textrm{LMI}^\textrm{IF}$:] Local Mode Interaction (in the IF algorithm)
    \item[$\textrm{LMI}^\textrm{SF}$:] Local Mode Interaction (in the SF algorithm)
    \item[MB:] Maxwell-Bloch
    \item[SF:] Supplementary Fields
    \item[SM:] Statistical Mechanic(al/s)
    \item[SR:] Superradian(ce/t)
    \item[SVEA:] Slowly-Varying Envelope Approximation
    \item[WDB:] Widely Doppler Broadened

\end{description}

\section{The Method of Supplementary Fields}\label{app:sf_algorithm}

We provide in this appendix the derivation and a brief study of the SF method. To motivate this new approach, notice that the $\mathcal{O}\left(n^2\right)$ complexity of equations \eqref{eq:MB_TD_norm-1}---\eqref{eq:MB_TD_norm-3} is rooted in the channel asymmetry introduced by the choice of envelope factorisation $\omega_0$; that is, channels of natural frequencies far removed from the central envelope introduce fast oscillating exponentials which demand fine time stepping to avoid their aliasing.

Motivated by a desire to symmetrise equations \eqref{eq:MB_TD_norm-1}---\eqref{eq:MB_TD_norm-3} across velocity channels, let us introduce supplementary electric field envelopes: one electric field factorisation for each natural frequency available across velocity channels. Explicitly, we define the array of electric field envelopes
\begin{equation}
    \bar{\mathcal{E}}^\pm_\omega \equiv \mathcal{E}^\pm e^{\pm i \omega \tau}
\end{equation}
such that equations \eqref{eq:MB_TD_norm-1}---\eqref{eq:MB_TD_norm-3} now read
\begin{align}
    \frac{\partial Z_\omega}{\partial\tau} &= \frac{i}{T_\textrm{R}} \left(\bar{\mathcal{P}}^+_\omega \bar{\mathcal{E}}^+_\omega - \bar{\mathcal{P}}^-_\omega\bar{\mathcal{E}}^-_\omega\right) - \frac{Z_\omega - 1}{T_1} + \Lambda^{(N)} \label{eq:MB_TD-1_lia} \\
    \frac{\partial\bar{\mathcal{P}}^+_\omega}{\partial\tau} &= \frac{2i}{T_\textrm{R}} \bar{\mathcal{E}}^-_\omega Z_\omega - \frac{\bar{\mathcal{P}}^+_\omega}{T_2} + \Lambda^{(P)} \label{eq:MB_TD-2_lia} \\
    \frac{\partial\bar{\mathcal{E}}^+_\omega}{\partial z} &= \frac{i}{2 L} e^{i\omega\tau} \int \mathrm{d} \omega' F_{\omega'} \mathcal{\bar{P}}^-_{\omega'} e^{-i\omega'\tau}. \label{eq:MB_TD-3_lia}
\end{align}

We have thus far only added to the complexity of the problem by simulating an array of supplementary fields described by the many instances of equation \eqref{eq:MB_TD-3_lia} which are, at present, not independent: they are algebraically related by a simple exponential multiplication factor. We will shortly break this algebraic relation, however, and evolve each $\bar{\mathcal{E}}^+_\omega$ independently; doing so will ultimately lead to a final algorithm of reduced complexity in many sample configurations.

Equation \eqref{eq:MB_TD-3_lia} is not yet manifestly symmetric across velocity channels, as higher Doppler shifted velocity channels are driven by faster oscillating exponentials appearing in the term multiplying the integral. We now make a change of integration variable $\omega'\rightarrow\omega'+\omega$ in equation \eqref{eq:MB_TD-3_lia} in order to cast the entire problem into a channel-symmetric form, viz.,
\begin{equation}
    \frac{\partial\bar{\mathcal{E}}^+_\omega}{\partial z} = \frac{i}{2 L} \int \mathrm{d} \omega' F_{\omega - \omega'} \mathcal{\bar{P}}_{\omega - \omega'}^- e^{i\omega'\tau}. \label{eq:MB_TD-3_symmetric}
\end{equation}
Equations \eqref{eq:MB_TD-1_lia}, \eqref{eq:MB_TD-2_lia}, and \eqref{eq:MB_TD-3_symmetric} are manifestly symmetric across velocity channels, as equation \eqref{eq:MB_TD-3_symmetric} couples all field modes to the medium in identical form. For any given instance $\omega$ of equation \eqref{eq:MB_TD-3_symmetric}, a common kernel $e^{i\omega'\tau}$ is integrated against medium properties of that velocity neighbourhood possessing central natural frequency matching the envelope factorisation frequency of $\bar{\mathcal{E}}^+_\omega$. We herein refer to this system as the supplementary fields (SF) representation, and any of the solution methods listed to follow as SF algorithms.

\subsection{Local mode interaction kernels in the SF representation}

The channel-symmetric form of equations \eqref{eq:MB_TD-1_lia}, \eqref{eq:MB_TD-2_lia}, and \eqref{eq:MB_TD-3_symmetric} allow us to apply a natural, physically motivated approximation which achieves $\mathcal{O}\left(n\right)$ numerical complexity whenever the velocity coherence developing within an SR process is of a moderate bandwidth.

As a WDB distribution is constructed via addition of velocity channels to the wings of an initially narrow distribution, the extent to which polarisation phase correlation (and an associated reduction in transient timescales) will develop is not a priori known. Indeed, it is the task of the following sections to determine just how temporal SR features collapse as various distributions are constructed.

In many cases, however, phase correlation across velocity channels will be limited to some bandwidth $\delta$; that is, the continued addition of velocity channels to the wings of the velocity distribution will introduce independent SR processes within finite groups of coherent velocity channels, each group being randomly phased relative to another. In the WDB limit, the total electric field at the endfire of such a sample will become the random interference of multiple coherent signals. Such an electric field transient will appear, on coarser timescales, to be a slow envelope with a timescale of variation on the order of the SR transient timescales characteristic of the separate coherent velocity groups.\footnote{Looking ahead, for example, one may refer to Figure \ref{fig:sfa_nsch16383_intensity}.}

In such systems the polarisation of a velocity channel of angular frequency $\omega$ is correlated only to fields of angular frequency $\omega'$ within $|\omega' - \omega| \lessapprox \delta$; i.e., we do not expect $\bar{\mathcal{E}}^\pm_\omega$ of equation \eqref{eq:MB_TD-3_symmetric} to be correlated to the integrand $\bar{\mathcal{P}}^-_{\omega-\omega'}$ for $|\omega'|>\delta$. Let us introduce a kernel $s_\delta \left(\omega'\right)$ of width $\delta$ to the integrand such that equation \eqref{eq:MB_TD-3_symmetric} becomes
\begin{equation}
    \frac{\partial\bar{\mathcal{E}}^+_\omega}{\partial z} = \frac{i}{2 L} \int \mathrm{d} \omega' F_{\omega - \omega'} \mathcal{\bar{P}}_{\omega - \omega'}^- e^{i\omega'\tau} s_\delta\left(\omega'\right). \label{eq:MB_TD-3_kernel}
\end{equation}
Note that the kernel introduction to equation \eqref{eq:MB_TD-3_kernel} has the effect of breaking the algebraic relationship between differing $\bar{\mathcal{E}}^+_\omega$; i.e., it is no longer true that $\bar{\mathcal{E}}^+_\omega = \mathcal{E}^+ e^{i \omega \tau}$. Instead, $\bar{\mathcal{E}}^+_\omega$ is $\mathcal{E}^+$ shifted in the Fourier domain by an amount $\omega$, and then truncated to some finite bandwidth neighbourhood of width $\delta$.

Equations \eqref{eq:MB_TD-1_lia}, \eqref{eq:MB_TD-2_lia}, and \eqref{eq:MB_TD-3_kernel} together form an $\mathcal{O}\left(n\right)$ complex algorithm. The problematic aliasing of fast oscillating exponentials has been removed by effectively truncating integration ranges over a finite velocity bandwidth due to the finite width of $s_\delta$, such that additional velocity channels can be indefinitely added to the system without concurrently reducing the temporal step size in a Runge-Kutta temporal propagation. Although equation \eqref{eq:MB_TD-3_kernel} has a convolution structure (which suggests greater than $\mathcal{O}\left(n\right)$ complexity in the number of velocity channels simulated), the convolution is in fact $\mathcal{O}\left(n\right)$ since the kernel $e^{i\omega'\tau} s_\delta \left(\omega'\right)$ does not increase in width with increasing channel count. In fact, we demonstrate in Appendix \ref{appsub:rect_kernel} that a rectangular kernel has the special status of circumventing the convolution calculation altogether; however, the FFT method detailed next in Section \ref{subsubsec:FFT_accel} performs optimally in our regimes of interest and is therefore used wherever the SF algorithm is applied in this paper.

The functional form $s_\delta(\omega')$ is at this point arbitrary; in fact, it can be any kernel with a central value of $1$ extending over a width of at least $\delta$. Its purpose is not to describe the lineshape of the SR emission process, which is determined by simulation; instead, it is a tool to reduce the numerical complexity of the problem. Any choice of kernel shape may be arbitrarily widened\footnote{At the cost of numerical complexity, see discussion below.} to achieve accurate results, and its width should be tuned against a full fidelity CTD simulation at a moderate velocity distribution width prior to applying it to the simulation of a WDB distribution.

\subsection{Fourier domain acceleration}\label{subsubsec:FFT_accel}

The FFT provides a further speedup over computing the raw convolution of equation \eqref{eq:MB_TD-3_kernel}, but only gains supremacy when the polarisation correlation width $\delta$ exceeds some threshold. If $\delta$ covers a small number of velocity channels, a raw convolution method will be faster than the FFT method; however, if $\delta$ covers a large number of velocity channels, the FFT method may overtake the raw convolution. Additionally, the introduction of the kernel $s_\delta$ further accelerates the computation of the FFT, as described momentarily.

Let $\underset{\omega \leftrightarrow \xi}{\mathcal{F}}$ denote a Fourier transform (or FFT) operation which transforms the space indexed by $\omega$ to a space indexed by $\xi$, and $\underset{\omega \leftrightarrow \xi}{\mathcal{F}^{\:-1}}$ its inverse operation. We denote the transform of a quantity with a circumflex; for example,
\begin{align}
    \widehat{F\bar{\mathcal{P}}}_\xi = \underset{\omega \leftrightarrow \xi}{\mathcal{F}} \left\{F\bar{\mathcal{P}}_\omega\right\}
\end{align}
(which, importantly, is not the same as $\widehat{F}\widehat{\bar{\mathcal{P}}}$). It is essential to note that the transform is performed with respect to the indices over which the convolution of equation \eqref{eq:MB_TD-3_kernel} occurs, and not with respect to either $\tau$ or $z$. Furthermore, the original space's index $\omega$ is an angular frequency index, making the transformed space's index $\xi$ a time-like index; this is opposite to traditional usages of a Fourier transform to move from the time domain to the frequency domain, and so we will not use terms such as ``bandwidth'' or ``duration'' in references to the ranges of either index (in order to avoid confusion).

We now write the convolution of equation \eqref{eq:MB_TD-3_kernel} as
\begin{align}
    \left(F \mathcal{\bar{P}}\right) \star \left( e^{i \omega \tau} s_\delta\right) &= \underset{\omega \leftrightarrow \xi}{\mathcal{F}^{\:-1}} \left\{ \widehat{F\bar{\mathcal{P}}} \left[ \delta\left(\xi - \tau\right) \star \widehat{s}_\delta\left(\xi\right)\right]\right\} \\
    &= \underset{\omega \leftrightarrow \xi}{\mathcal{F}^{\:-1}} \left\{ \widehat{F\bar{\mathcal{P}}} \widehat{s}_\delta\left(\xi-\tau\right)\right\}\label{eq:conv_simp}.
\end{align}
Expression \eqref{eq:conv_simp} demonstrates further acceleration over a conventional FFT-based approach (which itself is already $\mathcal{O}\left(n\right)$ complex) to the convolution of equation \eqref{eq:MB_TD-3_kernel}. The finite width of $\widehat{s}_\delta$ in $\xi$-space (proportional to $1/\delta$) means that its multiplication against $\widehat{F\bar{\mathcal{P}}}$ need only be performed over a finite width centred upon $\tau$, which offers the advantage of calculating a band-limited FFT of $F\bar{\mathcal{P}}$ (a sub-region of the full FFT). A band-limited FFT may be computed by multiplying $F\bar{\mathcal{P}}$ by $e^{i\omega\tau}$, downsampling, and performing an FFT on the resulting smaller array; the steps are reversed for a band-limited inverse FFT (IFFT). Numerical libraries also exist for computing limited bandwidth FFTs and IFFTs. In \texttt{Python3}, for example, the routine \verb|scipy.signal.zoom_fft| performs a band-limited FFT based upon the chirp z-transform; note, however, that a full bandwidth \verb|numpy.fft.fft| call remains faster below some critical size of $F \bar{\mathcal{P}}$.

A full simulation may now proceed according to the following steps:
\begin{enumerate}
    \item \noindent \label{first_sf_conv_step} At the start of the simulation, compute the FFT $\widehat{s}_\delta$ for use in each looping time step.
    \item \noindent \label{sf_conv_loop_step} Propagate all population inversion densities $Z_\omega$ and polarisations $\bar{\mathcal{P}}_\omega$ forward in time from the electric fields $\bar{\mathcal{E}}_\omega$ via equations \eqref{eq:MB_TD-1_lia} and \eqref{eq:MB_TD-2_lia}.
    \item \noindent \label{third_sf_conv_step} Compute the band-limited FFT $\widehat{F\bar{\mathcal{P}}}$ centred upon $\xi$-space index $\tau$ (i.e., a different region of $\xi$-space is required at each time step, and slides forward throughout the simulation),
    \item \noindent \label{fourth_sf_conv_step} Multiply the result by the pre-computed $\widehat{s}_\delta$,
    \item \noindent \label{fifth_sf_conv_step} Perform the band-limited IFFT, and
    \item \noindent Use this result to propagate the electric fields $\bar{\mathcal{E}}_\omega$ down the length of the sample via equation \eqref{eq:MB_TD-3_kernel}.
    \item \noindent \label{last_sf_conv_step} Repeat at step \ref{sf_conv_loop_step}.
\end{enumerate}

Steps \ref{first_sf_conv_step}---\ref{last_sf_conv_step} make more precise our earlier remarks concerning the supremacy of an FFT-based convolution over a raw convolution. As $\delta$ is increased, the smaller width (proportional to $\delta^{-1}$) of $\widehat{s}_\delta$ in $\xi$-space narrows the band-limited FFT and IFFT employed in steps \ref{third_sf_conv_step} and \ref{fifth_sf_conv_step}, and increases the complexity of a raw convolution of equation \eqref{eq:MB_TD-3_kernel}.

As a final remark on the FFT-based SF algorithm we point out that a kernel of the form $s_\delta\left(\omega\right) = \textrm{sinc}\left(\omega\right)$ is particularly efficient, as its FFT is a rectangular function and therefore eliminates step \ref{fourth_sf_conv_step} above. For such a $\textrm{sinc}$ kernel, the convolution of equation \eqref{eq:MB_TD-3_kernel} is realised by the operations of applying a band-limited FFT (centred on $\tau$) followed by a band-limited IFFT. For this reason, the $\textrm{sinc}$ kernel is used wherever the SF algorithm is applied in this paper.

\subsection{High temporal fidelity reconstruction}\label{subsubsec:theory-afalg-hifi}

The $\textrm{LMI}^\textrm{SF}$ approximation established by the finite kernel introduced in equation \eqref{eq:MB_TD-3_kernel} essentially means that each transient $\bar{\mathcal{E}}^+_\omega(z,\tau)$ is a bandwidth-limited window into the full transient $\mathcal{E}^+(z,\tau)$, centred upon angular frequency $\omega$. In many cases this is advantageous: by generating transients of every $\bar{\mathcal{E}}^+_\omega(z,\tau)$ during the SF algorithm execution, we have immediate access to a bandwidth-limited transient response of the electric field in any desired region of the spectrum. In the original CTD algorithm, such responses would have to be filtered out by band masking and frequency shifting the Fourier transform of the electric field transient.

On the other hand, the SF algorithm does not (immediately) provide a high fidelity, wide-bandwidth transient of the full electric field $\mathcal{E}^+(z,\tau)$. Suppose, for instance, that we were to plot as a function of $\tau$ the central supplementary field $\bar{\mathcal{E}}^+_{\omega=0}(z',\tau)$ at some position $z'$. The fine detail of the transient so generated would be limited to time steps defined by the kernel width $\delta$. If $\delta$ is $m$ multiples of the fundamental angular frequency differential $d\omega$, then the present transient will only resolve features down to temporal width $T/m$.

Higher temporal resolution is, however, available from the results of the SF algorithm by assembling together all $\bar{\mathcal{E}}^+_\omega$ responses. We begin the reconstruction in the Fourier domain by recognising that
\begin{equation}
    \mathcal{F}\left\{\mathcal{E}^+\right\}\left(\omega\right) = \int \textrm{d}\tau e^{-i \omega \tau} \mathcal{E}^+(\tau) = \int \textrm{d}\tau \bar{\mathcal{E}}^+_{-\omega}\left(\tau\right),
\end{equation}
where $\mathcal{F}$ denotes the Fourier transform with respect to $\tau$ and we have suppressed all $z$ dependence. The $\omega$ mode of the Fourier transform of $\mathcal{E}^+$ is, therefore, the integral over all time of the $\bar{\mathcal{E}}^+_{-\omega}$ transient.

At present we have in our possession the Fourier transform of $\mathcal{E}^+$ between $\omega_\textrm{min} = \omega_0 v_\textrm{min}/c$ and $\omega_\textrm{max} = \omega_0 v_\textrm{max}/c$ for a distribution $F_\omega$ defined on $\omega_\textrm{min} \leq \omega \leq \omega_\textrm{max}$; however, we can also extend beyond the trailing and leading ends of the current spectrum by an amount $\delta / 2$ by considering the Fourier transforms of $\bar{\mathcal{E}}^+_{\omega_\textrm{max}}$ and $\bar{\mathcal{E}}^+_{\omega_\textrm{min}}$, respectively. For example, for a frequency a distance $\eta \leq \delta / 2$ beyond the leading end of the current spectrum, we compute
\begin{align}
    \mathcal{F}\left\{\mathcal{E}^+\right\}\left(\omega_\textrm{max} + \eta \right) &= \int \textrm{d} \tau \bar{\mathcal{E}}^+_{\omega_\textrm{min}}(\tau) e^{-i \eta \tau} \\
    &= \mathcal{F}\left\{\bar{\mathcal{E}}^+_{\omega_\textrm{min}}\right\}\left(\eta \right),
\end{align}
which is the $\eta$ mode of the Fourier transform of the minimum velocity supplementary field.\footnote{This assumes that $F_\omega$ is equally distributed about $\omega=0$; i.e., that $\omega_\textrm{min} = -\omega_\textrm{max}$.} A similar result holds at the trailing end of the spectrum.

\begin{figure}
    \centering
    \includegraphics[width=1.\columnwidth, trim=.5cm .4cm .3cm .2cm, clip]{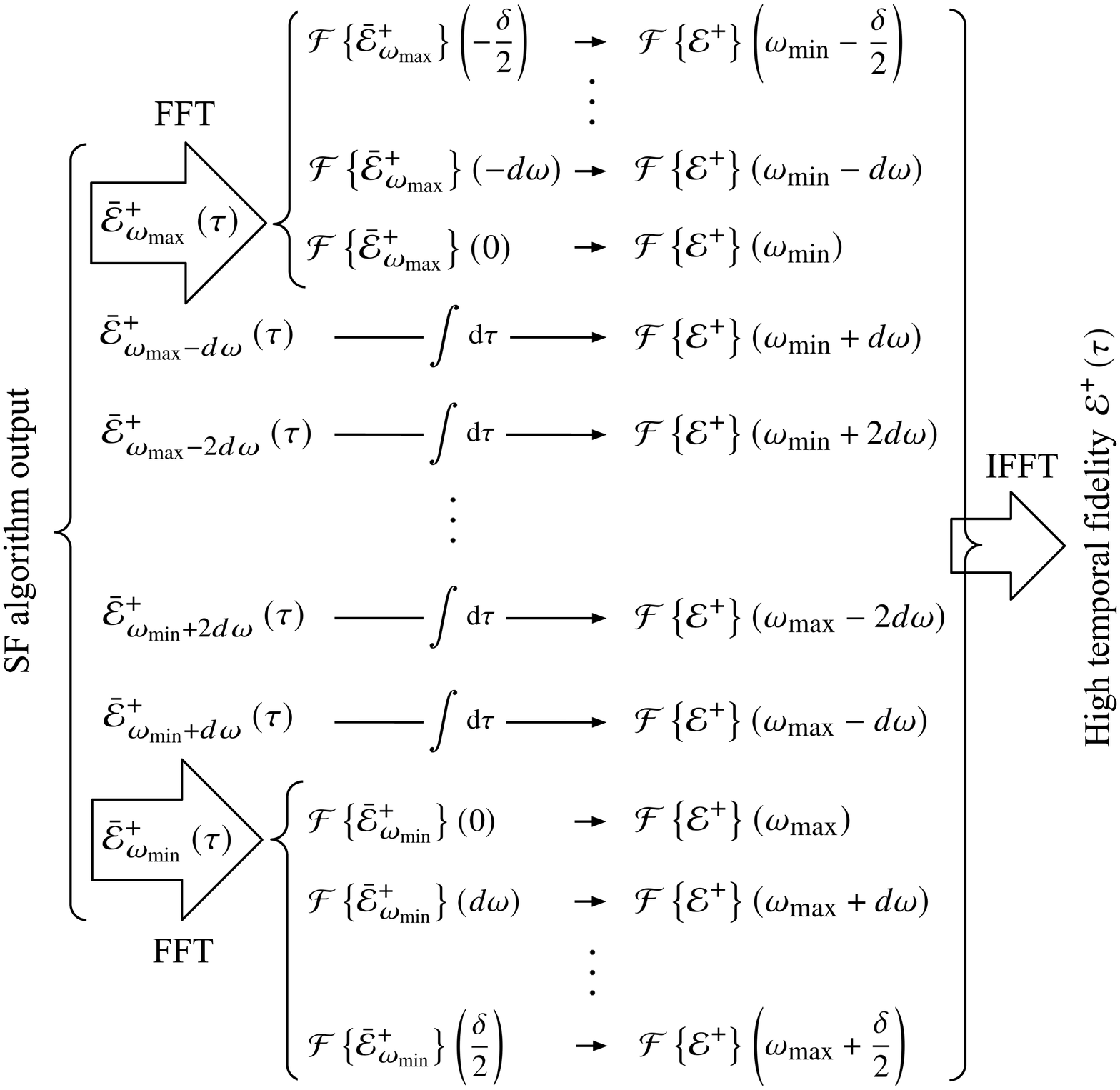}
    \caption{Schematic of the high temporal fidelity reconstruction procedure from the output of the SF algorithm.}
    \label{fig:sf_temporal_reconstruction}
\end{figure}

The previous discussion is admittedly high-level, and should be made precise by (i) discretising all quantities over $\omega$ into arrays of velocity separation $d\omega$ and (ii) re-expressing all (continuous) Fourier transforms as discrete Fourier transforms (DFTs) of mode separation $d\omega = 2\pi/T$. We omit these trivial details, but point out that an FFT is used to compute the DFTs of $\bar{\mathcal{E}}^+_{\omega_\textrm{min}}$ and $\bar{\mathcal{E}}^+_{\omega_\textrm{max}}$ for extending the spectrum, and that an IFFT is used to finally recover the full high fidelity electric field transient. The entire reconstruction procedure is illustrated in Figure \ref{fig:sf_temporal_reconstruction}.

\subsection{Special status of the rectangular kernel}\label{appsub:rect_kernel}

A rectangular kernel
\begin{equation}
    s_\delta\left(\omega\right) = \begin{cases} 1 &|\omega|<\frac{\delta}{2} \\
    0 &\textrm{otherwise}\end{cases} \label{eq:sdelta_Pi}
\end{equation}
leads to an especially efficient algorithm for computing the convolution of equation \ref{eq:MB_TD-3_kernel}. Inserting expression \eqref{eq:sdelta_Pi} for $s_\delta$ into equation \ref{eq:MB_TD-3_kernel} and changing the variable of integration, we return to the asymmetric formulation (but now with a finite range of integration)
\begin{equation}
    \frac{\partial\bar{\mathcal{E}}^+_\omega}{\partial z} = \frac{i}{2 L} e^{i \omega \tau} \int_{\omega-\delta/2}^{\omega+\delta/2} \mathrm{d} \omega' F_{\omega'} \mathcal{\bar{P}}^-_{\omega'} e^{-i\omega'\tau}. \label{eq:MB_TD-3_kernel_as}
\end{equation}
Equations \eqref{eq:MB_TD-3_kernel_as} contain what appear to be fast oscillating exponentials; however, said form being mathematically identical to equation \eqref{eq:MB_TD-3_kernel}, it is nonetheless physically correct for large temporal step sizes (even if those step sizes alias the terms explicit in this formulation).

Equation \eqref{eq:MB_TD-3_kernel_as} is much more computationally efficient than equation \eqref{eq:MB_TD-3_kernel}, if only a numerical solution implements the following order of operations:
\begin{enumerate}
    \item \noindent \label{first_af_step} Compute an array $I_{\omega'}$ representing the integrand of equation \eqref{eq:MB_TD-3_kernel_as}; i.e.,
    \begin{equation}
        I_{\omega'} = F_{\omega'} \mathcal{\bar{P}}^-_{\omega'} e^{-i \omega' \tau}.
    \end{equation}
    \item \noindent Defining the following array pending computation,
    \begin{equation}
        J_\omega = \int_{\omega-\delta/2}^{\omega+\delta/2} \textrm{d} \omega' I_{\omega'},
    \end{equation}
    compute its first element $J_{\omega,\textrm{ min}}$ by summing over the appropriate number of starting elements of $I_{\omega'}$,
    \item \noindent Proceed by induction to compute $J_{\omega+d\omega}$ from $J_\omega$ by simply adding to $J_\omega$ that element of $I_{\omega'}$ which is introduced by advancing the integration window forward one step, and subtracting from it that element of $I_{\omega'}$ which is lost by advancing the integration window forward one step.
    \item \noindent \label{last_af_step} Compute $\partial \bar{\mathcal{E}}^+_\omega / \partial z$ by element-wise multiplication between the array preceding the integral of equation \eqref{eq:MB_TD-3_kernel_as} and the array $J_\omega$.
\end{enumerate}

\section{The integral Fourier representation of the Maxwell-Bloch equations}\label{app:if_representation}

The integral Fourier (IF) representation of a system of partial differential equations provides a system of mode relations which, when solved, yield a convergent Fourier series representation of a system's transient behaviour. Such a representation is importantly distinct from the typical Fourier mode relations of, for example, \citet{Menegozzi1978}, which describe only the steady-state regime of a dynamical system. The derivation of the IF representation can be found in \citet{Wyenberg2021}.

The inversion and polarisation of each velocity channel, as well as the electric field, are expanded in the Fourier series
\begin{align}
    N_{p} &= \sum_{m} \mathbb{N}_{p,m} \left(z\right) e^{i m d\!\omega\tau} \\
    \mathcal{\bar{P}}_{p}^{\pm} &= \sum_{m} \bar{\mathbb{P}}_{p,m}^{\pm} \left(z\right) e^{\pm i m d\!\omega\tau} \\
    \mathcal{E}^{\pm} &=\sum_{m} \mathbb{E}_{m}^{\pm} \left(z\right) e^{\mp i m d\!\omega\tau} \\
    \Lambda^{\left(N/P\right)} &= \sum_{m} \mathbb{L}^{\left(N/P\right)}_m e^{i m d\!\omega\tau}.
\end{align}

Under such a Fourier series expansion the IF representation of the MB equations yields a set of mode relations which is fundamentally distinct from the typical steady state Fourier mode relations. Specifically, the IF relations read
\begin{align}
    \begin{split}
        \mathbb{N}_{p,0} &= N_{p}\left(0\right) + \frac{i}{\hbar}\sum_{\bar{m}}\left(\Xi_{\bar{m}+p}^{+}\bar{\mathbb{P}}_{p,\bar{m}}^{+}-\Xi_{\bar{m}+p}^{-}\bar{\mathbb{P}}_{p,\bar{m}}^{-}\right) \\
        &\quad +\sum_{m}\mathbb{T}_{m}\left(\mathbb{L}_{m}^{(N)}-\frac{\mathbb{N}_{p,m}}{T_{1}}\right)\label{eq:IF_1}
    \end{split} \\
    \begin{split}
        \mathbb{N}_{p,m\neq0} &= \mathbb{T}_{m} \biggr\{\frac{i}{\hbar} \sum_{\bar{m}} \bigr[\bar{\mathbb{P}}_{p,\bar{m}}^{+} \left(\mathbb{E}_{\bar{m}+p}^{+} - \mathbb{E}_{\bar{m}-m+p}^{+}\right) \\
        &\quad -\bar{\mathbb{P}}_{p,\bar{m}}^{-} \left(\mathbb{E}_{\bar{m}+p}^{-} - \mathbb{E}_{\bar{m}+m+p}^{-}\right) \bigr] \\
        &\quad +\frac{1}{T_{1}} \left(\mathbb{N}_{p,m} - \mathbb{N}_{p,0}\right) + \left(\mathbb{L}_{0}^{(N)} - \mathbb{L}_{m}^{(N)}\right)\biggr\} \label{eq:MB_IF_2}
    \end{split}\\
    \begin{split}
        \bar{\mathbb{P}}_{p,0}^{+} &= \bar{\mathcal{P}}_{p}^{+}\left(0\right) + \sum_{m}\mathbb{T}_{m} \biggr[\frac{2id^{2}}{\hbar}\sum_{\bar{m}}\left(\mathbb{N}_{p,\bar{m}}\mathbb{E}_{m-\bar{m}+p}^{-}\right) \\
        &\quad -\frac{\bar{\mathbb{P}}_{p,m}^{+}}{T_{2}} + \mathbb{L}_{m}^{(P)}\biggr] \label{eq:MB_IF_3}
    \end{split}\\
    \begin{split}
        \bar{\mathbb{P}}_{p,0}^{+} &= \bar{\mathcal{P}}_{p}^{+}\left(0\right)+\frac{2id^{2}}{\hbar}\sum_{\bar{m}}\Xi_{p-\bar{m}}^{-}\mathbb{N}_{p,\bar{m}} \\
        &\quad +\sum_{m}\mathbb{T}_{m}\left(\mathbb{L}_{m}^{(P)}-\frac{\bar{\mathbb{P}}_{p,m}^{+}}{T_{2}}\right),\label{eq:IF_4}
    \end{split}
\end{align}
where the arrays $\mathbb{T}_{m}$ and $\Xi^{\pm}_{a}$ are defined as
\begin{align}
    \mathbb{T}_{m} &= \begin{cases}
        \frac{\pi}{d\omega} & m = 0\\
        \frac{i}{m d\omega} & m \neq 0
    \end{cases} \quad \textrm{and} \\
    \Xi_{a}^{\pm} &= \sum_{m}\mathbb{T}_{m}\mathbb{E}_{a\mp m}^{\pm}.\label{eq:Gamma_Def}
\end{align}

\section{Algorithm comparisons}\label{app:alg_comparisons}

\subsection{Approximation distinctions and domains of validity}\label{subsec:theory-algcomp-approx}

The $\textrm{LMI}^\textrm{IF}$ and $\textrm{LMI}^\textrm{SF}$ approximations, though motivated by similar physical arguments, are not formally identical mathematical operations. The $\textrm{LMI}^\textrm{IF}$ approximation asserts, in the Fourier domain, that a limited bandwidth of the electric field be considered in the evolution of each of the material quantities; conversely, the $\textrm{LMI}^\textrm{SF}$ approximation asserts that a limited velocity width of all available polarisation channels be used to generate local regions of the electric field spectrum. Both LMI approximations are valid in that they generate signals matching those of the full fidelity time domain MB equations. Successful simulation of transient SR processes by the IF algorithm under the $\textrm{LMI}^\textrm{IF}$ approximation is demonstrated in \citet{Wyenberg2021} and the successful simulation of transient SR processes by the SF algorithm under the $\textrm{LMI}^\textrm{SF}$ are demonstrated in Section \ref{sec:swept-cont}.

The two algorithms are found, however, to converge differently depending upon the degree of spectral correlation developing across the velocity distribution. Under a set of simulation parameters yielding polarisation phase correlation across an extremely wide velocity bandwidth, the IF algorithm converges with a moderate fidelity $\textrm{LMI}^\textrm{IF}$ approximation \citep{Wyenberg2021}. In such cases the SF algorithm, on the other hand, requires a higher fidelity $\textrm{LMI}^\textrm{SF}$ approximation and thus loses its computational complexity advantage in such systems. Extremely wide distributions demonstrating wide bandwidth correlations should therefore be simulated with the IF algorithm under the $\textrm{LMI}^\textrm{IF}$ approximation; however, such regimes are not so deeply explored in this paper as to warrant implementation of the IF algorithm in any of the following sections.

\subsection{Execution speeds}\label{subsubsec:theory-algcomp-speed}

\begin{figure}
    \centering
    \includegraphics[width=1.\columnwidth, trim=0cm 1cm 2cm 3cm, clip]{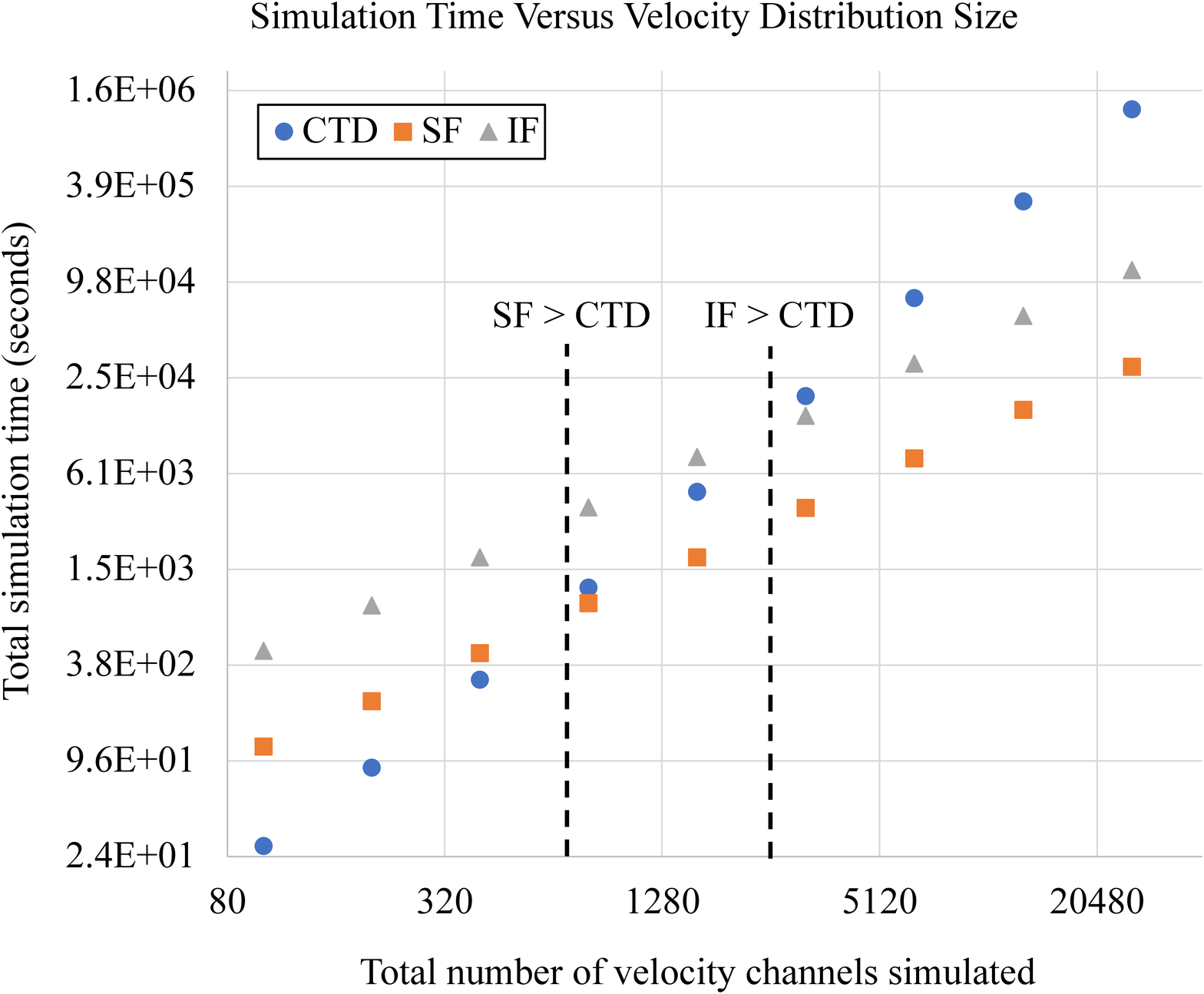}
    \caption{Log-log plot of simulation execution times for the three algorithms as a function of velocity distribution size. In these particular samples wide spectral correlation does not develop. Note that only the first $\sim\!60$ seconds of each simulation were executed, from which total execution time was extrapolated. SF > CTD denotes, for example, that the SF algorithm is \textit{superior} to (shorter execution time than) the CTD algorithm. Note the logarithmic axes.}
    \label{fig:speed_comparisons}
\end{figure}

Figure \ref{fig:speed_comparisons} is a log-log plot of total execution times for the full fidelity CTD algorithm, the SF algorithm, and the IF algorithm as a function of the total number of velocity channels simulated. Of critical importance in the context of the closing remarks of Section \ref{subsec:theory-algcomp-approx}, the samples simulated were not configured so as to develop wide bandwidth polarisation correlations; the SF algorithm therefore overtakes the full fidelity CTD algorithm when the distribution size exceeds about 640 total velocity channels.

\begin{figure}
    \centering
    \includegraphics[width=1.\columnwidth, trim=0cm 0cm 0cm 0cm, clip]{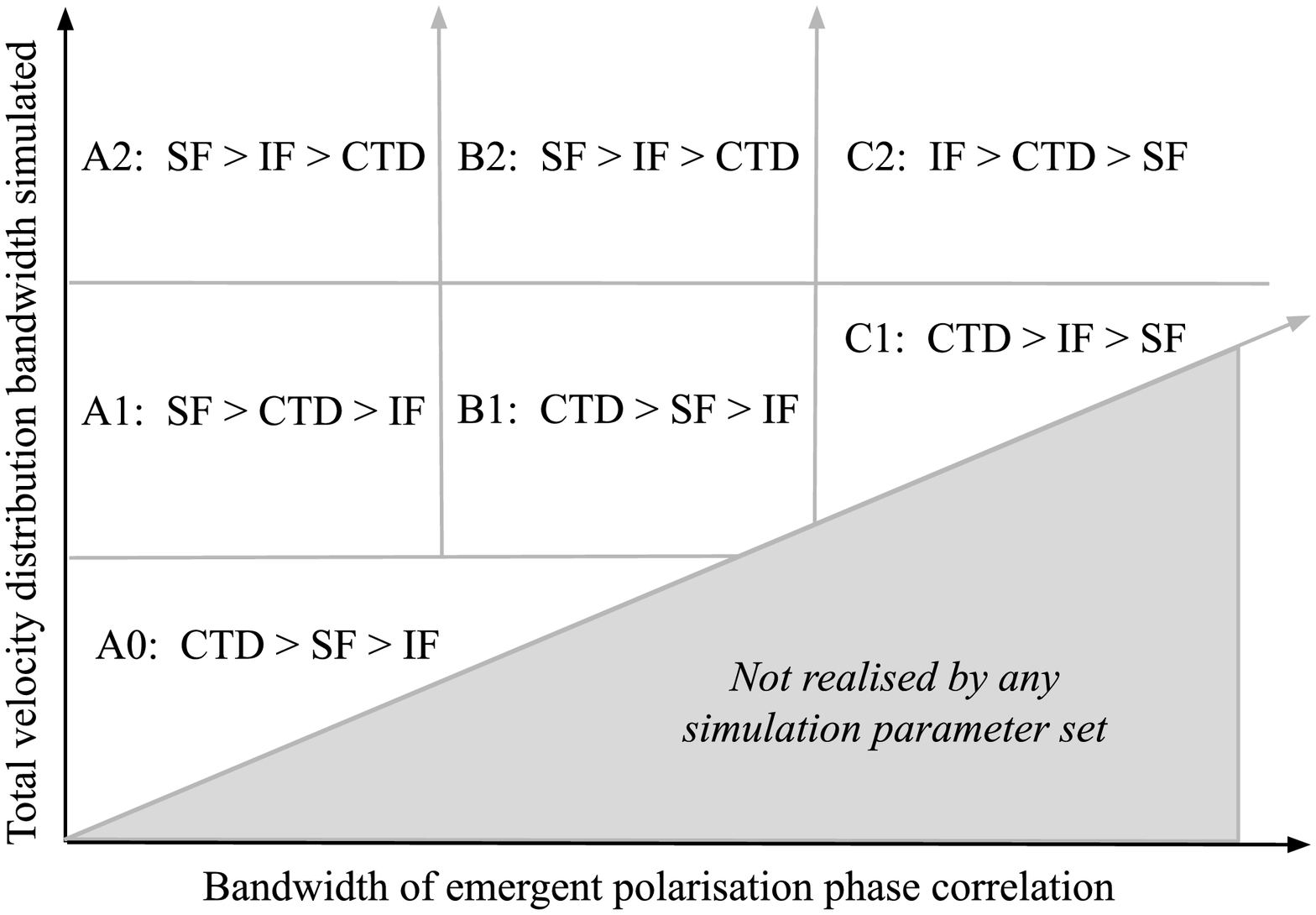}
    \caption{Schematic ranking algorithm efficiency (inequality key: faster > slower) across various sample configurations. Note that the schematic is only coarsely partitioned and does not precisely identify all inequality ordering transition boundaries.}
    \label{fig:alg_comparisons}
\end{figure}

Conversely, were the system to demand fine time stepping of the SF algorithm due to the development of wide bandwidth polarisation phase correlations, the SF complexity dependency would be quadratic and would plot along a line parallel to the CTD complexity dependency; the IF algorithm would then achieve the fastest execution times above about 1500 total velocity channels simulated. All of these results are summarised schematically in Figure \ref{fig:alg_comparisons}, where speed ranking of the three algorithms is shown for various distribution sizes and degrees of polarisation phase correlation development.

The simulations of Section \ref{sec:swept-disc} fall in region A0 of Figure \ref{fig:alg_comparisons} and are simulated exclusively with the full fidelity CTD algorithm. These simulations lay the intuitive foundations for broader velocity distributions simulated in Section \ref{sec:swept-cont}, which fall primarily in regions A1, A2 and B2 of Figure \ref{fig:alg_comparisons}. A few simulations will fall in regions B1, C1, and C2; however, Section \ref{sec:swept-cont} made exclusive use of the SF algorithm, as the regions B1, C1, and C2 will not be explored to such a depth as to render the SF algorithm dramatically inferior to either the CTD or the IF algorithms.

\section{Spurious Periodicity in the Comb Distribution Intensity Transients}\label{app:comb_spurious}

In Section \ref{subsubsec:swept-comb-hisat} we observed spurious peaks of periodicity $T/24$ in the intensity transient generated from a comb distribution of channel separation $24 dv$. We remarked that this structure implied the existence of at least some degree of correlation between neighbouring channels. To make this statement more precise, consider a reference signal $h(t)$ of duration $T$ that is generated by some physical process. Suppose that this signal is modulated by an array of carrier frequencies $f_n = n (M / T)$ for $n \in [-N, N]$, where $M$ is the number of fundamental frequencies between neighbouring comb channels; i.e., in the aforementioned example, $M=24$. Mathematically, we have a total signal
\begin{align}
    h_\textrm{tot}(t) &= \sum_{n=-N}^{N} h(t) e^{i \left[2 \pi (n M / T) t + \phi(n)\right]} \\
    &= h(t) \sum_{n=-N}^{N} e^{i \left[2 \pi (n M / T) t + \phi(n)\right]} \label{eq:spurious_superposn}
\end{align}
for $\phi(n)$ a (potentially varying) phase angle of the $n^\textrm{th}$ signal.

Suppose first that $\phi(n)$ does not vary across channels; say, $\phi(n) = 0$. Let $S$ denote the sum of equation \eqref{eq:spurious_superposn} which obeys the recurrence relation
\begin{equation}
    S = e^{i 2 \pi (M / T) t} S + e^{-N i 2 \pi (M / T) t} - e^{(N+1) i 2 \pi (M / T) t}
\end{equation}
such that
\begin{equation}
    S = \begin{cases}
    \frac{\sin\left[2 \pi (N+1/2) (M/T) t\right]}{\sin\left[\pi (M/T) t\right]} &t \neq \mbox{integer multiple of } T/m\\
    2 N + 1 &\mbox{otherwise}.
    \end{cases} \label{eq:s_cases}
\end{equation}
Notice that $S$ is periodic in $T/M$. Inspecting (without loss of generality) the first cycle of $S$ and assuming large $N$, for $t$ on the order of $T / [(N+1)M]$ we may Taylor expand the denominator in expression \eqref{eq:s_cases} to obtain
\begin{equation}
    S \approx (2N+1) \: \textrm{sinc}\left[(N+1/2) M t\right] \mbox{ for } t < T / [(N+1)M]
\end{equation}
which makes precise our statement that the spurious intensity spikes of periodicity $T/24$ are expected of a comb distribution possessing correlation between its channels.

At the other limit, suppose that the channel phases are now completely uncorrelated; i.e., that $\phi(n)=\textrm{Rand}[0,2\pi]$. In this case $S$ may be viewed as the DFT of periodicity $T/M$ of an array of unit norm and randomly phased elements $\textrm{exp}\left[i\phi(n)\right]$. By a familiar result of Fourier theory \citep{Bracewell1978}, the squared norm of such a DFT is a random signal; thus, if the $\phi(n)$ are completely uncorrelated, then $h_\textrm{tot}(t)$ will be the signal $h(t)$ modulated by random noise.

In-between the two limits of completely correlated and completely un-correlated channel phases, we may investigate the signal generated by partially correlated channels numerically. For this purpose, let us construct $\phi_l (n)$ of varying finite channel phase correlation length $l$ via convolution of a completely random phase angle array $\phi^\textrm{R}(n)$ against a Gaussian $g_l(n)$ of characteristic width $l$:
\begin{equation}
    \phi_l (n) = \left\{g_l \star \phi^\textrm{R} \right\}(n).
\end{equation}

\begin{figure}
    \centering
    \includegraphics[width=.45\columnwidth, trim=2cm 1.3cm 1.5cm 1.3cm, clip]{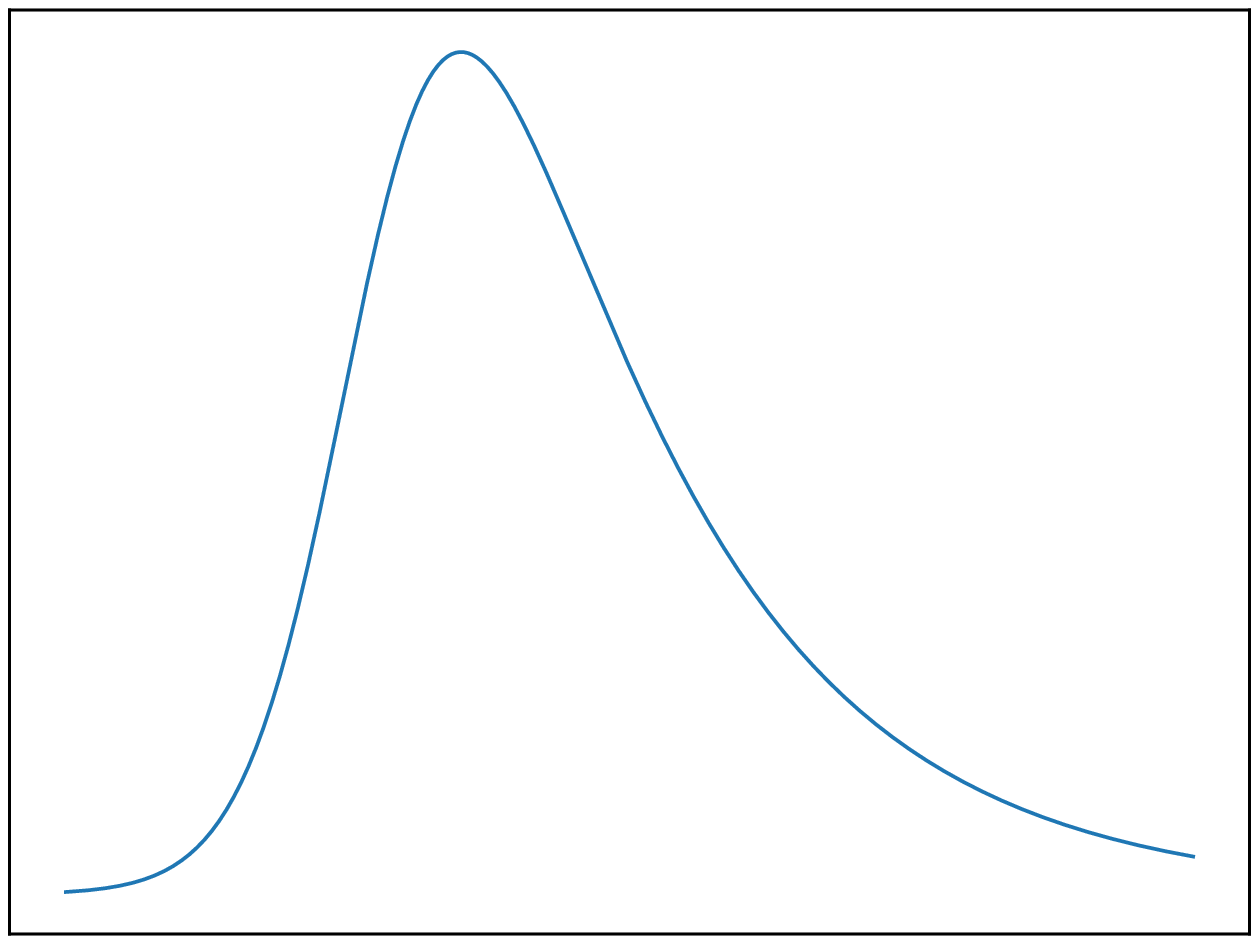}
    \includegraphics[width=.45\columnwidth, trim=2cm 1.3cm 1.5cm 1.3cm, clip]{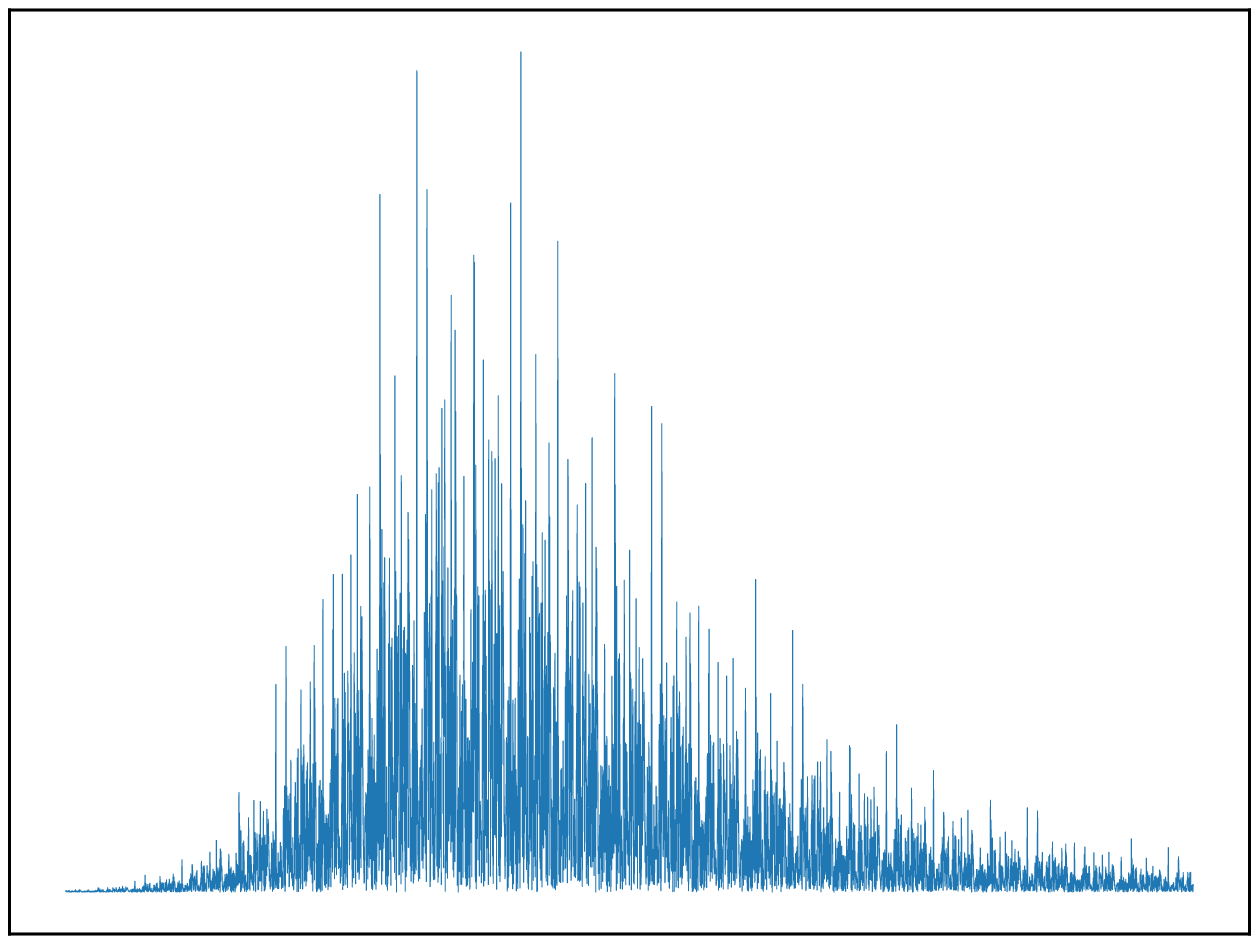}
    \includegraphics[width=.45\columnwidth, trim=2cm 1.3cm 1.5cm 1.3cm, clip]{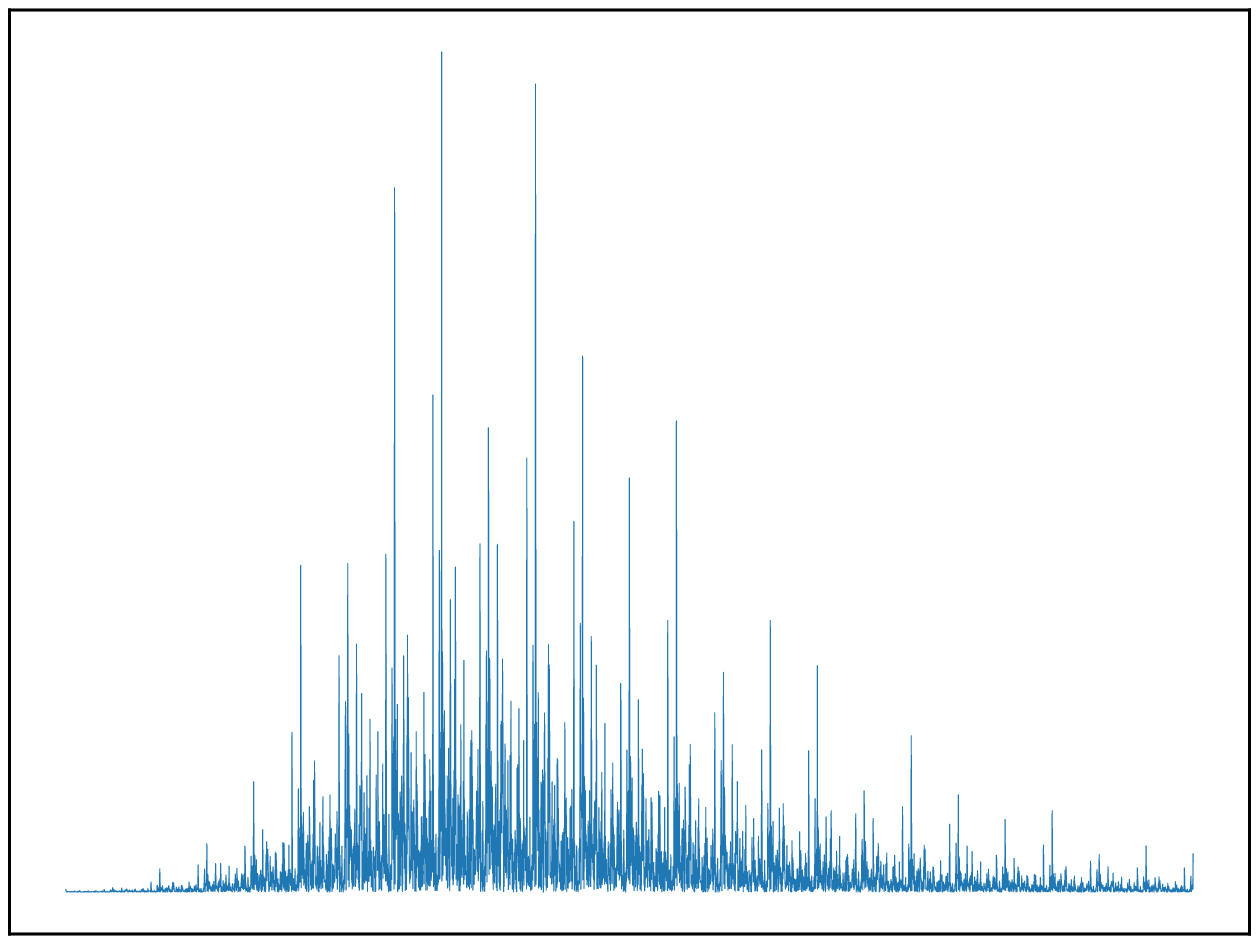}
    \includegraphics[width=.45\columnwidth, trim=2cm 1.3cm 1.5cm 1.3cm, clip]{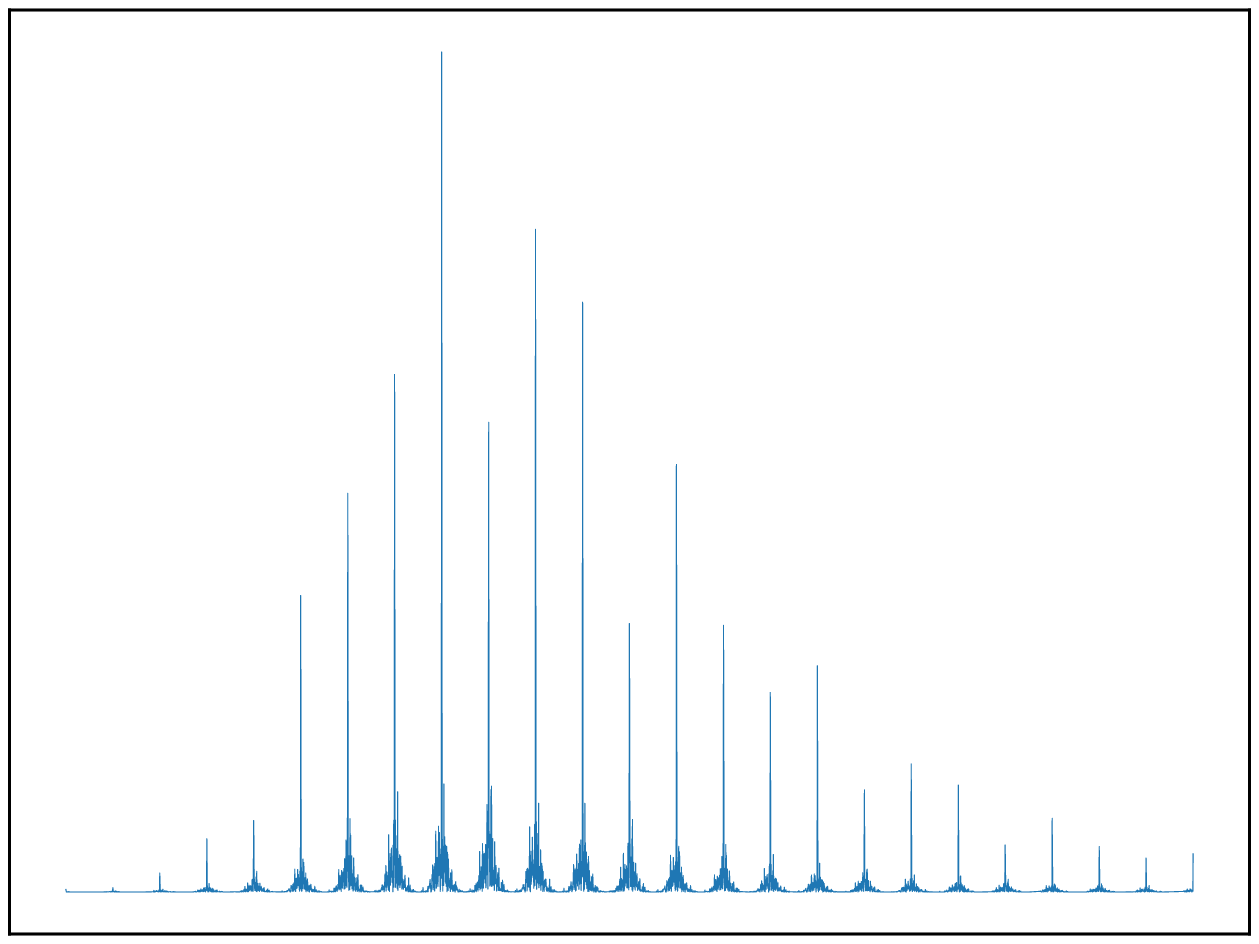}
    \caption{Demonstration of the effect of modulating a reference signal at offset frequencies possessing varying degrees of phase correlation. Top-left: reference signal. Top-right, bottom-left, bottom-right: phase correlation length $l=0.1,0.7,1.4$ (respectively).}
    \label{fig:app_sign_generation}
\end{figure}

We now consider, for example, the reference signal $h(t)$ pictured in the top-left panel of Figure \ref{fig:app_sign_generation}. We multiply $h(t)$ by the sum $S$ resulting from three different degrees of channel phase correlation; namely, $l=0.1$, $1.0$, and $2.0$ for $161$ channels separated by $24dv$ ($M=24$ in the present notation). The results, shown in the top-right, bottom-left, and bottom-right panels of Figure \ref{fig:app_sign_generation}, can be compared against Figure \ref{fig:comb_161channels_toothsepdecreasing}. Roughly speaking, the bottom-left panel with $l=0.7$ appears most similar in character to the spurious features of Figure \ref{fig:comb_161channels_toothsepdecreasing}, hence our qualitative assertion in section \ref{subsubsec:swept-comb-hisat} that those features were indicative of a small degree of correlation between neighbouring channels.

\bsp	
\label{lastpage}
\end{document}